\documentclass[11pt,twoside]{article}
\usepackage{amsmath,amsthm,amssymb,amsbsy,bm,calc,comment,fancyhdr,
            float,glossaries,ifthen,mathrsfs,textgreek,titlesec,tocloft,xy}

\PassOptionsToPackage{ctagsplt,Righttag}{amstex}
\usepackage[title]{appendix}
\usepackage{cite}
\usepackage[dvips,dvipsnames,usenames]{color} 
\usepackage[colorlinks=true,citecolor=MidnightBlue,linkcolor=BrickRed,breaklinks=true]{hyperref} 
\usepackage[usenames,dvipsnames]{xcolor}
\usepackage[super]{nth}
\usepackage{graphicx}
\usepackage{mathtools}
\usepackage{wrapfig}
\usepackage[font={footnotesize}]{caption}
\usepackage{tikz}
\usepackage{tikz-cd}
\usepackage{subcaption}
\usepackage[version=4]{mhchem}
\usepackage{svg}
\definecolor{darkgreen}{rgb}{0,0.65,0}
\definecolor{darkred}{rgb}{0.65,0,0}

\usepackage{bm}
\usepackage{mathrsfs}
\usepackage{sectsty}

\usepackage{mathdots}
\usepackage{yhmath}
\usepackage{cancel}
\usepackage{siunitx}
\usepackage{array}
\usepackage{multirow}
\usepackage{tabularx}
\usepackage{booktabs}
\usetikzlibrary{fadings}
\usetikzlibrary{patterns}
\usetikzlibrary{shadows.blur}
\usetikzlibrary{shapes}
\usepackage{stmaryrd}
\usepackage{braket}

\newcommand{\pder}[2]{\frac{\partial #1}{\partial #2}}

\newcommand{\jump}[1]{\left\llbracket #1 \right\rrbracket}
\newcommand{\bracfrac}[2]{\left(\frac{#1}{#2}\right)}
\newcommand{\LMavg}[1]{\braket{#1}_\text{LM}}

\newcommand{\trace}{\operatorname{tr}}
\newcommand{\grad}{\operatorname{grad}}

\newcommand{\divg}{\operatorname{div}}

\newcommand{\etal}{\emph{et al.}}

\newcommand{\ft}{\tilde{f}}
\newcommand{\Ft}{\tilde{F}}

\newcommand{\scrM}{\mathscr{M}}

\newcommand{\bbR}{\mathbb{R}}

\newcommand{\bbP}{\mathbb{P}}

\newcommand{\betab}{\bm{\beta}}

\newcommand{\ab}{\bm{a}}
\newcommand{\bb}{\bm{b}}
\newcommand{\cb}{\bm{c}}
\newcommand{\db}{\bm{d}}
\newcommand{\eb}{\bm{e}}
\newcommand{\fb}{\bm{f}}
\newcommand{\gb}{\bm{g}}
\newcommand{\hb}{\bm{h}}

\newcommand{\jb}{\bm{j}}

\newcommand{\mb}{\bm{m}}
\newcommand{\nb}{\bm{n}}
\newcommand{\tb}{\bm{t}}

\newcommand{\vb}{\bm{v}}

\newcommand{\xb}{\bm{x}}

\newcommand{\Ab}{\bm{A}}
\newcommand{\Bb}{\bm{B}}

\newcommand{\Fb}{\bm{F}}

\newcommand{\Ib}{\bm{I}}
\newcommand{\Jb}{\bm{J}}

\newcommand{\Lb}{\bm{L}}
\newcommand{\Mb}{\bm{M}}

\newcommand{\Tb}{\bm{T}}

\newcommand{\Vb}{\bm{V}}

\newcommand{\Xb}{\bm{X}}

\newcommand{\Bc}{{\mathcal B}}

\newcommand{\Jc}{{\mathcal J}}

\newcommand{\Mc}{{\mathcal M}}

\newcommand{\Pc}{{\mathcal P}}

\newcommand{\zerob}{{\boldsymbol{0}}}

\newcommand{\brm}{\mathrm{b}}

\newcommand{\taub}{\boldsymbol{\tau}}
\newcommand{\nub}{\boldsymbol{\nu}}
\newcommand{\chib}{\bm{\chi}}

\newcommand{\thetab}{\bm{\theta}}

\advance\textheight by \topskip
\begin{document}


\begin{center}
	{\textbf{\Large{The irreversible thermodynamics of curved lipid membranes II: Permeability and osmosis}}
	} \\
	\vspace{0.21in}

	Ahmad M. Alkadri$^{1, \ddag}$ and Kranthi K. Mandadapu$^{1, 2, \dag}$ \\
	\vspace{0.25in}

	\footnotesize{
		{
			$^1$ Department of Chemical \& Biomolecular Engineering, University of California at Berkeley,
			Berkeley, CA, 94720, USA \\[3pt]
			$^2$
			Chemical Sciences Division, Lawrence Berkeley National Laboratory, Berkeley, CA 94720, USA
		}
	}
\end{center}

\vspace{5pt}
%
%

\begin{abstract}
    We present a theory that combines the framework of irreversible thermodynamics with modified integral theorems to model arbitrarily curved and deforming membranes immersed in bulk fluid solutions. We study the coupling between the mechanics and permeability of a viscous and elastically-bendable membrane, and a multi-component bulk fluid solution. An equation for the internal entropy production for irreversibilities at the membrane is derived, determining 
    the generalized thermodynamic forces and fluxes, from which we identify the deviatoric stress as a novel driving force for permeability. 
    A complete set of equations of motion, constitutive laws, and boundary conditions to model the lipid membrane and bulk fluid system are provided.
\end{abstract}
\vspace{5pt}


\noindent\rule{4.6cm}{0.4pt}

\noindent \small{\ddag \, ahmadalkadri@berkeley.edu \\
	\dag \, kranthi@berkeley.edu
}

\vspace{15pt}

%
%

\small \tableofcontents
\vspace{08pt}



%
%

\section{Introduction} \label{sec:intro}

Lipid membranes act as separation structures in biological systems, forming boundaries that compartmentalize different regions of bulk fluid \cite{Watson2015}. In addition, these membranes perform a “gatekeeping” role by regulating the transport of chemical species between these compartments. This regulatory role is achieved through selective permeability: the ability to allow some chemical species to pass through the membrane while hindering the transport of others. This selective permeability is crucial for many of life's most vital processes, as it allows for the concentrations of salts, glucose, and other chemical species within the cell to exist at different concentrations from the aqueous environment surrounding the cell. Broadly, transport across lipid membranes takes place through endo/exo-cytosis, protein mediated transport, or direct permeation through the membrane. In this paper, we will focus on direct membrane permeability to non-electrolyte solutions, and how this permeability couples with the mechanics of curved, deforming, and flowing membranes.

One phenomenon of interest that arises in the presence of a selectively permeable membrane is osmosis. When two compartments of aqueous solution are separated by a membrane that is impermeable to a solute, water will typically flow from the solution of lower solute concentration into the solution of higher concentration \cite{koretsky2012engineering}. The additional pressure required to halt the flow of water across the membrane is known as the \textit{osmotic pressure}. Instances of osmotic transport phenomena are ubiquitous in biology, with examples ranging from water reabsorption in the kidneys, to maintaining turgor pressure in plant cells \cite{Marbach2019}. Cells find themselves in natural physiological concentrations on the order of $\SI{150}{mmol/L}$, yet a deviation of just $\SI{10}{mmol/L}$ yields an osmotic pressure of $\SI{25}{kPa}$ in an ideally selective membrane -- a large force for soft materials like lipid bilayers to withstand. Moreover, these osmotic pressures can also induce large curvature changes in the membrane \cite{liu2022vesicles}, so that osmotic transport couples with the mechanics of the membrane. 

Of practical interest to engineers, scientists, and medical practitioners is an equation for the flowrate of solute and solvent across a semi-permeable membrane. Rigorous identification of the relevant driving forces for osmotic transport, however, can be a non-trivial task. In 1896, Ernest H. Starling was studying the movement of fluid between  capillary membranes and the surrounding interstitial tissue, and hypothesized that:\\
\indent ``\textit{{[...]} at any given time, there must be a balance between the hydrostatic pressure of the blood in the capillaries and the osmotic attraction of the blood for the surrounding fluids. [...] Here then we have the balance of forces necessary to explain the speedy regulation of the quantity of circulating fluid.''}\cite{starling1896absorption}\\
Having identified the osmotic pressure and hydrostatic pressure inside and outside of a capillary membrane as driving forces for flow across a capillary (a phenomenon known as ``filtration''), these driving forces have come to be known as ``Starling forces''~\cite{levick2010microvascular}. Starling's hypothesis was later verified by the experimental observations of Landis~\cite{landis1927micro} and Pappenheimer and Soto-Rivera~\cite{pappenheimer1948effective}, with theoretical support emerging shortly thereafter. Beginning in 1951, Staverman used Onsager's irreversible thermodynamic framework~\cite{onsager1931a,onsager1931b} to introduce the \textit{reflection coefficient}, a parameter that reconciled discrepancies between theoretical and measured values for the osmotic pressure by accounting for the membrane's solute permeability~\cite{staverman1951theory}. Building on this work, Kedem and Katchalsky applied linear irreversible thermodynamics to provide a mathematical description for membrane permeability to non-electrolytes in discontinuous systems~\cite{kedem1958thermodynamic}. The seminal result of this work is the Starling equation, which expresses the filtration rate in terms of the Starling forces and Staverman's reflection coefficient (see Section~\ref{subsec:negligible-deviatoric-stresses}). 
Kedem and Katchalsky's analysis, however, does not take into account all of the mechanical forces that drive lipid membrane deformations.

Modelling the mechanical behaviour of lipid membranes presents a unique challenge, as they are soft materials with a thickness on the order of only $\SI{5}{nm}$, but have in-plane length scales on the order of micrometers. Moreover, lipid bilayers can assume arbitrarily curved and deforming configurations, and behave as in-plane fluids with out-of-plane bending elasticity. Continuum models of lipid membranes often make use of the small thickness dimension to approximate membranes as being two-dimensional surfaces embedded in three-dimensional space. In previous work \cite{sahu-mandadapu-pre-2017}, the irreversible thermodynamic framework of Prigogine \cite{prigogine1961introduction} and de Groot \& Mazur \cite{deGroot2013nonequilthermo} was developed for biological membranes modelled as two-dimensional fluids with Helfrich bending elasticity. Using tools from differential geometry, the governing equations are written in terms of an arbitrary local parametrization, so that they are valid for arbitrarily curved and deforming membranes. This model, however, does not include the surrounding bulk fluid, and therefore does not recapitulate the Starling forces or permeability. In this second part, our goal is to extend and combine this model with the works of Kedem and Katchalsky to develop an irreversible thermodynamic framework for the transport of non-electrolyte solutions across curved and deforming membranes.

\begin{figure}
    \centering
    \input{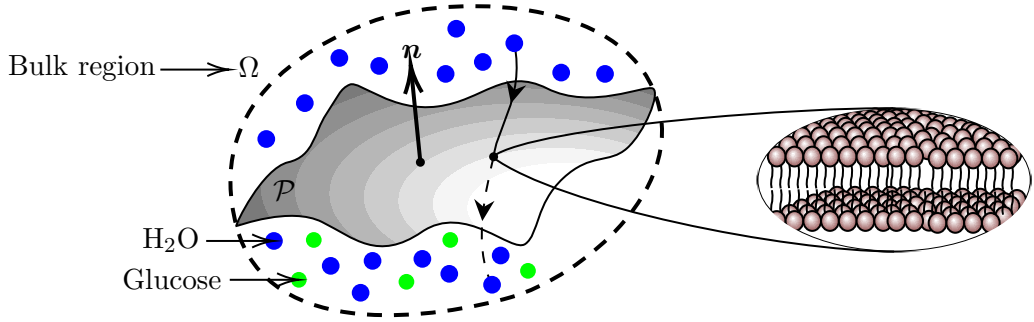}
    \vspace{-15pt}
    \caption{
		A lipid bilayer membrane modelled as a two-dimensional surface $\Pc$ immersed in a three-dimensional non-electrolyte solution $\Omega$. The membrane compartmentalizes the bulk region into two regions, and water is seen to permeate through the membrane into a glucose solution. On the right, we depict a zoomed-in view of the lipid bilayer.
	}
	\label{fig:membrane-bulk-non-electrolyte}
\end{figure}

While we shall continue to model the membrane as an ostensibly two-dimensional material, the bulk domain that the membrane is immersed in is three-dimensional (see Figure~\ref{fig:membrane-bulk-non-electrolyte}). Reducing the membrane to a two-dimensional object embedded in a three-dimensional fluid causes bulk quantities such as density, velocity, and shear stress to undergo a discontinuous ``jump'' at the membrane interface. The presence of these ``jumps'' renders the membrane interface a \textit{surface of discontinuity} (also sometimes called a \textit{singular surface}~\cite{gurtin2010mechanics}). This poses additional challenges in developing the irreversible thermodynamic framework, as the integral theorems used to write conservation laws need to be modified to account for finite jumps in the otherwise smooth bulk fields. Surfaces of discontinuity have previously been utilized by continuum mechanicians studying shock waves or other non-material interfaces~\cite{casey2011derivation,gurtin2010mechanics}. In this work, we show that these integral theorems generalize naturally to material surfaces: surfaces of a non-zero mass density, such as lipid bilayer membranes. 
The two modified integral theorems required are the modified divergence theorem and the modified Reynolds transport theorem. While the modified divergence theorem remains unchanged for a material surface of discontinuity, we derive a novel form of the modified Reynolds transport theorem that takes the material nature of the membrane into account. In addition, we argue that these theorems remain valid in the presence of tangential slip between the bulk fluid and the membrane.
The modified integral theorems are then used to separate out the local balance laws for the bulk fluid and the membrane, and to derive the form of the rate of internal entropy production at the membrane interface. The ensuing formalism yields more general thermodynamic forces than the ones obtained by Kedem and Katchalsky, and therefore gives rise to more generalized constitutive equations.

The remainder of the article is structured as follows. In Section~\ref{sec:mathematical-preliminaries} we set up the continuum mechanical and differential geometric frameworks used to describe the bulk fluid and the membrane. We also introduce the notation used to describe ``jumps'' across the membrane, and summarize the modified integral theorems derived in Appendix~\ref{sec:integral-theorem-proofs}. In Section~\ref{sec:balance-laws}, we apply the modified integral theorems to obtain local balances of mass, linear momentum, angular momentum, energy, and entropy in the bulk fluid and the membrane. Additionally, we also derive an equation for the internal entropy production at the membrane interface, thus elucidating the relevant thermodynamic forces and fluxes for mass transport across the membrane. In Section~\ref{sec:linear-irrev-thermo}, we propose more generalized linear constitutive laws for permeable membranes, and simplify them to certain special cases of interest in Section~\ref{sec:special-cases-for-permeability}. We end with a set of equations of motion for a lipid membrane in Section~\ref{sec:EOM-membrane-bulk-system}, together with a more generalized Starling equation that includes the deviatoric stresses from the bulk fluid as an additional driving force for osmotic transport.

\section{Mathematical Preliminaries} \label{sec:mathematical-preliminaries}

In this section, we consider a $2$-dimensional membrane surface $\Mc$ embedded in a $3$-dimensional bulk solution $\Bc$, and summarize the tools from differential geometry used to describe the evolution of the membrane. The bulk fluid is assumed to occupy a region in Euclidean space $\Bc \subseteq \bbR^3$, and we assume that the membrane is always completely immersed in this region (i.e., $\Mc \subset \Bc$). Additionally, since phospholipids typically spontaneously self-assemble into closed structures due to the large energetic cost of a free edge \cite{zelisko_determining_2017}, we will assume that the membrane does not have a (global) boundary (i.e., $\partial\Mc = \emptyset$). Consequently, while our theory cannot resolve transient processes where a free edge arises due to a topological change in the membrane---such as during pore formation and vesicle fission/fusion \cite{deserno_contact_2007,agrawal_boundary-value_2009,tu_compatibility_2010}---we can still model these processes immediately preceding and after the appearance of a free edge.

\subsection{Kinematics of the Bulk Fluid and Membrane} 

As the membrane $\Mc$ is embedded in three-dimensional Euclidean space $\bbR^3$, the vectors that describe a position $\xb^{(m)}$ on the membrane surface and a position $\xb$ in the bulk fluid $\Bc$ are drawn from the same vector space (see Figure~\ref{fig:membrane-immersion}). The coordinates used to describe each of the two position vectors, however, can differ. To describe the position of the membrane, we follow a similar notation to that in previous works \cite{Steigmann1999,sahu-mandadapu-pre-2017} and parametrize a patch $\Pc \subseteq \Mc$ of the membrane by 
\begin{equation}
    \xb^{(m)} = \xb^{(m)}(\theta^1,\theta^2,t) \,, \label{eq:surface-param}
\end{equation}
where $\theta^1$ and $\theta^2$ are referred to as \textit{surface coordinates} (also sometimes called \textit{local coordinates}). For the bulk fluid, we use Cartesian coordinates $(x^1,x^2,x^3)$ so that
\begin{equation}
    \xb = \xb(x^1,x^2,x^3,t) \,.
    \label{eq:bulk-Cartesian}
\end{equation}
For convenience, we will always use Greek indices to indicate components that span the set $\{1, 2\}$, and Roman letters to indicate components that span the set $\{1, 2, 3\}$. This means that Eq.~\eqref{eq:surface-param} can be rewritten as $\xb^{(m)} = \xb^{(m)}(\theta^\alpha,t)$, while Eq.~\eqref{eq:bulk-Cartesian} can be rewritten as $\xb = \xb(x^i,t)$. Additionally, partial differentiation with respect to the surface coordinates will be denoted by $(\,\bullet\,)_{,\alpha} := \pder{}{\theta^{\alpha}}(\,\bullet\,)$, and partial differentiation with respect to a Cartesian coordinate will be denoted by $(\,\bullet\,)_{,i} := \pder{}{x^{i}}(\,\bullet\,)$.

\begin{figure}
    \centering
    \input{membrane-immersion.tex}
    \caption{
            A patch of membrane $\Pc$ embedded in three-dimensional space within a bulk medium $\Bc$. The bulk and membrane position vectors, $\xb$ and $\xb^{(m)}$, are depicted as lying in the same vector space $\bbR^3$. The local surface parametrization $\xb^{(m)}(\theta^\alpha,t)$ induces the in-plane tangent vectors $\ab_1$ and $\ab_2$ that span the tangent space at the membrane, which the unit normal vector $\nb$ sits orthogonal to. At the patch boundary $\partial\Pc$, the unit vector $\taub$ lies tangent to the boundary curve, from which we define the boundary normal $\nub = \taub \times \nb$.
	}
	\label{fig:membrane-immersion}
\end{figure}

The surface parametrization (Eq.~\eqref{eq:surface-param}) induces many kinematic quantities that are useful in describing the evolution of the membrane. Here, we briefly summarize the necessary definitions and results for this work, and direct the reader to Ref.~\cite{sahu-mandadapu-pre-2017} for additional membrane-related quantities. The in-plane tangent vectors are given by $\ab_\alpha := \xb_{,\alpha}^{(m)}$, and subsequently the unit normal is $\nb := \frac{\ab_1 \times \ab_2}{||\ab_1 \times \ab_2||}$. This forms a local set of basis vectors $\{\ab_1,\ab_2,\nb\}$. The covariant metric tensor is $a_{\alpha\beta} := \ab_\alpha \cdot \ab_\beta$, and the contravariant metric tensor is computed from its inverse as $a^{\alpha\beta} := (a_{\alpha\beta})^{-1}$. The metric tensor may be used to raise and lower indices of a tensor. For instance, the contravariant tangent vectors are given by $\ab^\alpha = a^{\alpha\beta} \ab_\beta$, where we follow the Einstein summation convention in any term containing a pair of covariant and contravariant indices. Note that $\{\ab^1,\ab^2,\nb\}$ also forms a basis at each point on the membrane, so that a vector $\vb$ may be written as $\vb = v^\alpha \ab_\alpha + v\nb = v_\alpha \ab^\alpha + v\nb$. To incorporate spatial changes in the in-plane basis vectors automatically into derivatives of a vector $\vb$, we use the Christoffel symbols of the second kind $\Gamma_{\alpha\beta}^\gamma := \frac{1}{2}a^{\gamma\delta}(a_{\delta\alpha,\beta} + a_{\delta\beta,\alpha} - a_{\alpha\beta,\delta})$ and define the covariant derivative of contravariant components as $v_{;\alpha}^\beta := v_{,\alpha}^\beta + \Gamma_{\alpha\mu}^\beta v^\mu$ and covariant components as $v_{\beta;\alpha} := v_{\beta,\alpha} - \Gamma_{\beta\alpha}^\mu v_\mu$. This way, we have $(v^\beta\ab_\beta)_{,\alpha} \cdot \ab^\gamma = v^\gamma_{;\alpha}$ and $(v_\beta\ab^\beta)_{,\alpha} \cdot \ab_\gamma = v_{\gamma;\alpha}$.

To evaluate the change in area of the membrane, consider a reference patch $\Pc_0$ corresponding to the membrane patch $\Pc$ fixed at some time $t_0$. Let $A_{\alpha\beta}$ be the metric tensor of the reference surface. The relative contraction/dilation of the membrane is measured by the Jacobian of the map from the reference surface $\Pc_0$ to the current surface $\Pc$, given by $J = \sqrt{\det(a_{\alpha\beta})\big/\det(A_{\alpha\beta})}$. To compute quantities related to the principal curvatures of the membrane, we use the curvature tensor $b_{\alpha\beta} := \nb \cdot \xb_{,\alpha\beta}^{(m)}$. The mean and Gaussian curvatures are then given by $H := \frac{1}{2}a^{\alpha\beta}b_{\alpha\beta}$ and $K := \det(b_{\alpha\beta}) \big/ \det(a_{\alpha\beta})$, respectively. 

For the boundary of the membrane patch $\partial\Pc$, we describe a point $\xb_\brm^{(m)}$ using an arc-length parametrization 
\begin{equation}
    \xb_\brm^{(m)} = \xb^{(m)}(\theta_\brm^\alpha(s),t) \,, \label{eq:boundary-parametrization}
\end{equation}
where $s$ denotes an arc-length parameter, and $\theta_\brm^\alpha$ denotes the restriction of the surface coordinates to the boundary of their domain. This induces another basis at the boundary of the membrane that is more convenient for prescribing boundary conditions. Namely, we have the unit tangent $\taub := \ab_\alpha \frac{d\theta_\brm^\alpha}{ds}$ and the boundary normal $\nub := \taub \times \nb$ so that $\{\nub,\taub,\nb\}$ forms a local orthonormal basis. Note that implicit in the choice of a surface parametrization (Eq.~\eqref{eq:surface-param}) is a choice for the local orientation of the membrane as defined by the unit normal $\nb$. The orientation on the boundary $\partial\Pc$ is chosen so that the boundary normal $\nub$ is outward-pointing.

To track the rate of change of quantities in the bulk fluid, we define the material time derivative of a bulk quantity as 
\begin{align}
    \frac{d}{dt} \left( \bullet \right) &:= \left( \bullet \right)_{,t} + v^i \left( \bullet \right)_{,i} \,,
\end{align}
where $\left( \bullet \right)_{,t}$ denotes partial differentiation with respect to time, and $v^i$ are the components of the bulk velocity $\vb := \dfrac{d\xb}{dt}$. To track rates of change at the membrane, we similarly define the material time derivative of a membrane quantity as 
\begin{align}
    \frac{d}{dt}\left( \bullet \right)^{(m)} := \left( \bullet \right)^{(m)}_{,t} + v^\alpha \left( \bullet \right)^{(m)}_{,\alpha} \,,
\end{align}
where $v^\alpha$ are the components of the membrane velocity $\vb^{(m)}$, which is given by 
\begin{align}
    \vb^{(m)} := \pder{\xb^{(m)}}{t} + \pder{\theta^{\alpha}}{t} \ab_\alpha = v\nb + v^\alpha\ab_\alpha \,.
    \label{eq:membrane-velocity}
\end{align}
Note that in Eq.~\eqref{eq:membrane-velocity}, without loss of generality on the arbitrary shape of the membrane patch, we have assumed that the surface parametrization (Eq.~\eqref{eq:surface-param}) is chosen so that $\xb^{(m)}_{,t}$ is always normal to the membrane, i.e., $\xb^{(m)}_{,t} = v \nb$. For both the bulk fluid and the membrane, we will often denote the material time derivative with an overset dot, i.e., $\dot{\left( \bullet \right)} := \dfrac{d}{dt}\left( \bullet \right)$. The material time derivative of the metric tensor follows the kinematic identity~\cite{sahu-mandadapu-pre-2017} 
\begin{align}
\begin{split}
    \dot{a}_{\alpha\beta} &= \vb_{,\alpha}\cdot\ab_\beta + \vb_{,\beta}\cdot\ab_\alpha\\
    &= v_{\alpha;\beta} + v_{\beta;\alpha} - 2v b_{\alpha\beta} \,.
\end{split}
\end{align}
It follows from the above identity that the material time derivative of the surface Jacobian $J$ is given by
\begin{align}
    \frac{\dot{J}}{J} &= \frac{1}{2}\dot{a}_{\alpha\beta} a^{\alpha\beta} = v_{;\alpha}^\alpha - 2vH =: \divg_\text{S}\vb^{(m)} \,,
    \label{eq:J-dot}
\end{align}
where we have denoted the \textit{surface divergence} of the membrane velocity as $\divg_\text{S}\vb^{(m)}$.


\subsection{Surfaces of Discontinuity and the Modified Integral Theorems}
\label{subsec:modified-integral-theorems}

We now outline the mathematical tools used to integrate the membrane into a fluid system. Consider a region of fluid~$\Bc$ with a membrane patch~$\Pc$ immersed inside of it. 
Let~$\Omega$ denote any subregion of the bulk~$\Bc$ whose boundary $\partial\Omega$ contains the boundary of the membrane (i.e.,~$\Pc \subset \Omega$ and~$\partial\Pc \subset \partial\Omega$; 
\begin{wrapfigure}{r}{0.5\textwidth}
    \centering
    \tikzset{every picture/.style={line width=0.75pt}} 

\begin{tikzpicture}[x=0.75pt,y=0.75pt,yscale=-1,xscale=1]

\draw  [fill={rgb, 255:red, 0; green, 0; blue, 0 }  ,fill opacity=1 ] (196.58,190.99) .. controls (196.58,190.16) and (197.25,189.49) .. (198.08,189.49) .. controls (198.91,189.49) and (199.58,190.16) .. (199.58,190.99) .. controls (199.58,191.82) and (198.91,192.49) .. (198.08,192.49) .. controls (197.25,192.49) and (196.58,191.82) .. (196.58,190.99) -- cycle ;
\draw   (268.08,25.99) .. controls (340.42,32.74) and (355.92,93.07) .. (358.08,110.99) .. controls (360.25,128.91) and (370.08,168.99) .. (328.08,210.99) .. controls (286.08,252.99) and (177.25,225.24) .. (168.08,210.99) .. controls (158.92,196.74) and (150.08,174.99) .. (163.08,140.99) .. controls (176.08,106.99) and (139.58,74.33) .. (164.08,53.83) .. controls (188.58,33.33) and (195.75,19.24) .. (268.08,25.99) -- cycle ;
\draw    (163.08,140.99) .. controls (205.25,89.91) and (303.25,182.41) .. (358.08,110.99) ;
\draw [line width=1.5]    (214.08,125.56) -- (216.81,95.22) ;
\draw [shift={(217.08,92.23)}, rotate = 95.14] [color={rgb, 255:red, 0; green, 0; blue, 0 }  ][line width=1.5]    (14.21,-4.28) .. controls (9.04,-1.82) and (4.3,-0.39) .. (0,0) .. controls (4.3,0.39) and (9.04,1.82) .. (14.21,4.28)   ;
\draw  [fill={rgb, 255:red, 0; green, 0; blue, 0 }  ,fill opacity=1 ] (278.08,134.49) .. controls (278.08,133.66) and (278.75,132.99) .. (279.58,132.99) .. controls (280.41,132.99) and (281.08,133.66) .. (281.08,134.49) .. controls (281.08,135.32) and (280.41,135.99) .. (279.58,135.99) .. controls (278.75,135.99) and (278.08,135.32) .. (278.08,134.49) -- cycle ;
\draw  [fill={rgb, 255:red, 0; green, 0; blue, 0 }  ,fill opacity=1 ] (278.08,142.49) .. controls (278.08,141.66) and (278.75,140.99) .. (279.58,140.99) .. controls (280.41,140.99) and (281.08,141.66) .. (281.08,142.49) .. controls (281.08,143.32) and (280.41,143.99) .. (279.58,143.99) .. controls (278.75,143.99) and (278.08,143.32) .. (278.08,142.49) -- cycle ;
\draw    (268.08,25.99) .. controls (277.48,3.52) and (281.08,34.21) .. (296.28,14.27) ;
\draw    (288.08,70.99) .. controls (309.08,69.92) and (299.88,42.72) .. (333.08,37.92) ;
\draw    (333.22,132.06) .. controls (340.35,146.12) and (361.95,115.72) .. (376.42,139.26) ;
\draw    (136.19,257.44) -- (173.92,257.44) ;
\draw [shift={(175.92,257.44)}, rotate = 180] [color={rgb, 255:red, 0; green, 0; blue, 0 }  ][line width=0.75]    (10.93,-3.29) .. controls (6.95,-1.4) and (3.31,-0.3) .. (0,0) .. controls (3.31,0.3) and (6.95,1.4) .. (10.93,3.29)   ;
\draw    (136.19,257.44) -- (136.19,217.27) ;
\draw [shift={(136.19,215.27)}, rotate = 90] [color={rgb, 255:red, 0; green, 0; blue, 0 }  ][line width=0.75]    (10.93,-3.29) .. controls (6.95,-1.4) and (3.31,-0.3) .. (0,0) .. controls (3.31,0.3) and (6.95,1.4) .. (10.93,3.29)   ;

\draw    (136.19,257.44) -- (113.58,273.98) ;
\draw [shift={(111.96,275.16)}, rotate = 323.82] [color={rgb, 255:red, 0; green, 0; blue, 0 }  ][line width=0.75]    (10.93,-3.29) .. controls (6.95,-1.4) and (3.31,-0.3) .. (0,0) .. controls (3.31,0.3) and (6.95,1.4) .. (10.93,3.29)   ;

\draw    (136.19,257.44) -- (196.72,192.45) ;
\draw [shift={(198.08,190.99)}, rotate = 132.97] [color={rgb, 255:red, 0; green, 0; blue, 0 }  ][line width=0.75]    (10.93,-3.29) .. controls (6.95,-1.4) and (3.31,-0.3) .. (0,0) .. controls (3.31,0.3) and (6.95,1.4) .. (10.93,3.29)   ;
\draw    (328.08,210.99) .. controls (350.38,211.43) and (328.1,226) .. (350.1,223.43) ;
\draw    (300.67,213.58) .. controls (325.67,213.58) and (300.17,249.08) .. (321.17,246.08) ;

\draw (281.08,127.39) node [anchor=north west][inner sep=0.75pt]  [font=\scriptsize]  {$+$};
\draw (281.58,137.39) node [anchor=north west][inner sep=0.75pt]  [font=\scriptsize]  {$-$};
\draw (211.08,78.19) node [anchor=north west][inner sep=0.75pt]    {$\bm{n}$};
\draw (298.08,2.19) node [anchor=north west][inner sep=0.75pt]    {$\partial \Omega ^{( +)}$};
\draw (335.08,25.39) node [anchor=north west][inner sep=0.75pt]    {$\Omega ^{( +)}$};
\draw (374.08,136.39) node [anchor=north west][inner sep=0.75pt]    {$\mathcal{P}$};
\draw (242.08,148.39) node [anchor=north west][inner sep=0.75pt]    {$\llbracket f\rrbracket =f^{(+)} -f^{(-)}$};
\draw (97,272.4) node [anchor=north west][inner sep=0.75pt]    {$x_1$};
\draw (175.58,248.28) node [anchor=north west][inner sep=0.75pt]    {$x_2$};
\draw (130.71,200.17) node [anchor=north west][inner sep=0.75pt]    {$x_3$};
\draw (168.89,223.15) node [anchor=north west][inner sep=0.75pt]    {$\bm{x}$};
\draw (350.8,213.43) node [anchor=north west][inner sep=0.75pt]    {$\partial \Omega ^{( -)}$};
\draw (322.58,235.92) node [anchor=north west][inner sep=0.75pt]    {$\Omega ^{( -)}$};

\end{tikzpicture}
    \caption{
		The lipid membrane patch $\Pc$ as a surface of discontinuity. The bulk fluid neighbourhood $\Omega$ enclosing the patch is partitioned into two regions $\Omega^{(+)}$ and $\Omega^{(-)}$. The membrane normal $\nb$ points into the ``positive'' side of the membrane, so that a ``jump'' $\jump{f}$ at the membrane is given by the difference between the limiting value $f^{(+)}$ from the positive side and the limiting value $f^{(-)}$ from the negative side.
	}
	\label{fig:surface-of-discontinuity}
\end{wrapfigure}
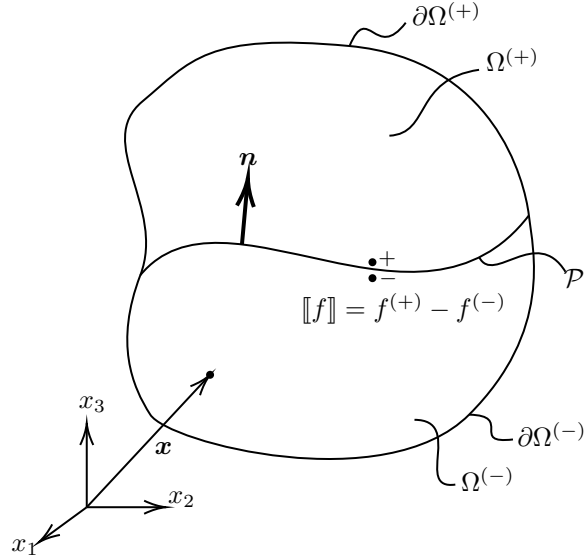
see Figure~\ref{fig:surface-of-discontinuity}). 
This choice of bulk region is required in the derivation of the modified integral theorems (see Appendix~\ref{sec:integral-theorem-proofs}), and will also enable us to localize the balance laws in Section~\ref{sec:balance-laws}. 
In particular, the membrane $\Pc$ partitions $\Omega$ into two open regions~$\Omega^{(+)}$ and~$\Omega^{(-)}$, where we choose the ``positive'' side~$\Omega^{(+)}$ to be the partition that the membrane normal~$\nb$ points into.
Furthermore, the membrane surface is allowed to act as a \textit{surface of discontinuity}, i.e., as one moves transversely through the membrane, quantities associated with the bulk fluid such as the density scalar field~$\rho$, the velocity vector field~$\vb$, and the stress tensor field~$\Tb$, may undergo a finite jump discontinuity at the membrane. 

The limiting values of any bulk quantity~$\fb$ as one approaches from the top/bottom of the membrane are defined as 
\begin{align}
    \fb^{(+)}(\xb^{(m)},t) &:= \lim_{\zeta \to 0^+} \fb(\xb^{(m)} + \zeta\nb,t) \, , \label{eq:f+}\\
    \fb^{(-)}(\xb^{(m)},t) &:= \lim_{\zeta \to 0^+} \fb(\xb^{(m)} - \zeta\nb,t) \, . \label{eq:f-}
\end{align}
Note that although the quantity $\fb$ itself is ill-defined at the membrane, the limiting quantities $\fb^{(+)}$ and $\fb^{(-)}$ are evaluated directly on the membrane. The difference between the two limits~\eqref{eq:f+} and~\eqref{eq:f-} is called the \textit{jump}, and is defined by 
\begin{equation}
    \jump{\fb(\xb^{(m)},t)} := \fb^{(+)}(\xb^{(m)},t) - \fb^{(-)}(\xb^{(m)},t) \,,
    \label{eq:jump-defn}
\end{equation}
while the \textit{average limiting value} is defined by 
\begin{equation}
    \braket{\fb(\xb^{(m)},t)} := \frac{\fb^{(+)}(\xb^{(m)},t) + \fb^{(-)}(\xb^{(m)},t)}{2} \,.
\end{equation}
From here on, we will often suppress the dependence of a quantity on $\xb^{(m)}$ and $t$. It is useful to note that if $\gb$ is another bulk quantity, then we have the identity
\begin{equation}
    \jump{\fb \cdot \gb} = \jump{\fb} \cdot \braket{\gb} + \braket{\fb} \cdot \jump{\gb} \,.
    \label{eq:jump-identity}
\end{equation}
In particular, if $\jump{\gb} = \zerob$ then Eq.~\eqref{eq:jump-identity} implies that $\jump{\fb\cdot\gb} = \jump{\fb}\cdot\gb$. Any surface quantity $\fb^{(m)}$ appearing within the jump brackets ``$\jump{\bullet}$'' is evaluated on the surface as usual, and is assumed to satisfy $\jump{\fb^{(m)}} = \zerob$.

Moving between the integral and local form of a balance law is typically done using the divergence and Reynolds transport theorems. When the system contains a surface of discontinuity, however, these integral theorems need to be modified to account for the jump across the surface. The physical significance of this modification will be expounded upon in Section~\ref{sec:balance-laws} when we derive the balance laws. Briefly, however, one can visualize that a conserved quantity (such as mass) may accumulate onto the membrane as it is being transported between the $\Omega^{(+)}$ and $\Omega^{(-)}$ compartments in Figure~\ref{fig:surface-of-discontinuity}. 
The ability of the membrane to act as an extra source/sink, as well as maintain finite jumps across the interface, is reflected in the modified integral theorems as additional integrals of jump quantities over the membrane surface of discontinuity. The formal details for the proof of the modified integral theorems are laid out in Appendix \ref{sec:integral-theorem-proofs}, with the main results summarized below.

Let $\varphi$ denote a scalar field, $\fb$ a vector field, and $\Fb$ a tensor field. Then the \textit{modified divergence theorem} for each of these fields can be stated as 
\begin{subequations}
\label{eq:jump-div-thm}
\begin{align}
    \int_{\partial\Omega} \varphi\nb \,da &= \int_\Omega \grad\varphi \,dv + \int_\Pc \jump{\varphi}\nb \,da \,, \label{eq:jump-div-thm-scalar}\\
    \int_{\partial\Omega} \fb\cdot\nb \,da &= \int_\Omega \divg \fb \,dv + \int_{\Pc} \llbracket \fb \rrbracket\cdot\nb \,da \,, \label{eq:jump-div-thm-vector}\\
    \int_{\partial\Omega} \Fb \nb \,da &= \int_\Omega \divg \Fb \,dv + \int_\Pc \jump{\Fb}\nb \,da \,. \label{eq:jump-div-thm-tensor}
\end{align}
\end{subequations}
Note that in the above equations, the surface normal $\nb$ is always taken with respect to the surface that is being integrated over. The orientation of the normal for the membrane is locally defined and is induced by the surface parametrization, whereas for the bulk boundary we choose the outward pointing normal.

In Section~\ref{sec:balance-laws}, the bulk quantities, $\varphi$, $\fb$, and~$\Fb$ represent intensive quantities (e.g., mass density, momentum density, and stress) defined in the bulk fluid. Since the membrane~$\Pc$ is a material interface, there are corresponding intensive scalar~$\varphi^{(m)}$, vector~$\fb^{(m)}$, and tensor~$\Fb^{(m)}$ fields that can be evaluated at points~$\xb^{(m)}$ on the membrane~$\Pc$. The modified Reynolds transport theorem gives an expression for the total rate of change of the associated extensive quantity~$\left(\int_{\Omega} \fb \,dv + \int_{\Pc} \fb^{(m)} \,da\right)$ in the presence of a surface of discontinuity. In Appendix~\ref{sec:integral-theorem-proofs}, we prove that this theorem generalizes to material surfaces of discontinuity, as well as the case where there is relative slip between the bulk fluid and the membrane. 
Recalling that the \textit{surface divergence} of the membrane velocity $\vb^{(m)} = (v^{(m)})^\alpha\ab_\alpha + v^{(m)}\nb$ is defined as $\divg_\text{S}\vb^{(m)} := (v^{(m)})_{;\alpha}^\alpha - 2v^{(m)} H$ \cite{sahu-mandadapu-pre-2017}, the \textit{modified} Reynolds transport theorems for the scalar, vector and tensor fields are then given by 
\begin{subequations}
\label{eq:jump-reynolds-transport}
\begin{align}
    \begin{split}
        \frac{d}{dt}\Bigg( \int_{\Omega} \varphi \,dv + \int_{\Pc} \varphi^{(m)} \,da \Bigg)
        = &\int_{\Omega} (\dot{\varphi} + \varphi\divg\vb) \,dv\\
        &\hspace{0.in} + \int_{\Pc} \left(\dot{\varphi}^{(m)} + \varphi^{(m)}\divg_\text{S} \vb^{(m)} + \jump{\varphi(\vb - \vb^{(m)})} \cdot \nb \right) \,da \,,
    \end{split} 
    \label{eq:jump-reynolds-transport-scalar}\\
    \begin{split}
    \frac{d}{dt}\Bigg( \int_{\Omega} \fb \,dv + \int_{\Pc} \fb^{(m)} \,da\Bigg) =  &\int_{\Omega} \left(\dot{\fb} + \fb\divg(\vb) \right)\,dv \\
    &\hspace{0.in} +
    \int_{\Pc} \left(\dot{\fb}^{(m)} + \fb^{(m)}\divg_\text{S} \vb^{(m)} + \jump{\fb \otimes \left(\vb - \vb^{(m)}\right) } \nb \right) \,da \,,
    \end{split}
    \label{eq:jump-reynolds-transport-vector}\\
    \begin{split}
        \frac{d}{dt}\Bigg( \int_{\Omega} \Fb \,dv + \int_{\Pc} \Fb^{(m)} \,da \Bigg)
        = &\int_{\Omega} (\dot{\Fb} + \Fb\divg\vb) \,dv\\
        &\hspace{0.in} + \int_{\Pc} \left(\dot{\Fb}^{(m)} + \Fb^{(m)}\divg_\text{S} \vb^{(m)} + \jump{\Fb \otimes (\vb - \vb^{(m)})} \nb \right) \,da \,.
    \end{split}
    \label{eq:jump-reynolds-transport-tensor}
\end{align}
\end{subequations}
The quantities inside the jump brackets in Eq.~\eqref{eq:jump-reynolds-transport} correspond to the flux of bulk quantities relative to the membrane.

While the above modified integral theorems are useful for converting integral balances over bulk regions to local balances, we still require two-dimensional analogues of the integral theorems to localize integral balances on the membrane surface. Allowing the membrane to act as a surface of discontinuity does not modify the surface integral theorems from previous work \cite{sahu-mandadapu-pre-2017}. With that in mind, let $\fb^{(m)}$ be a scalar, vector, or tensor field that can be evaluated at points $\xb^{(m)}$ on the membrane $\Pc$. Suppose that $\fb^{(m)}$ can be written in terms of the local basis as $\fb^{(m)} = (f^{(m)})^\alpha \ab_\alpha + f^{(m)}\nb$. Since the boundary normal $\nub$ is an in-plane quantity, it may be expanded as $\nub = \nu^\alpha\ab_\alpha = \nu_\alpha\ab^\alpha$. The \textit{surface divergence theorem} states that 
\begin{equation}
    \int_{\partial\Pc} (f^{(m)})^\alpha \nu_\alpha \,ds = \int_\Pc (f^{(m)})_{;\alpha}^\alpha \,da \,. 
    \label{eq:surface-div-thm}
\end{equation}
In the case where no accumulation takes place on the membrane, it is also possible to separately write a \textit{surface Reynolds transport theorem} \cite{sahu-mandadapu-pre-2017}
\begin{equation}
    \frac{d}{dt} \int_\Pc \fb^{(m)} \,da = \int_\Pc \left(\dot{\fb}^{(m)} + \fb^{(m)}\divg_\text{S} \vb^{(m)}\right) \,da \,.
    \label{eq:surface-reynolds-transport}
\end{equation}
Note, however, that Eq.~\eqref{eq:surface-reynolds-transport} is only a special case of Eq.~\eqref{eq:jump-reynolds-transport} that arises when the bulk fluid and the membrane do not interact. We refer the reader to Section~\ref{subsubsec:proof-of-modified-rtt} for a detailed discussion on when Eq.~\eqref{eq:jump-reynolds-transport} can be separated into a bulk modified Reynolds transport theorem, and a (membrane) surface Reynolds transport theorem.

\section{Balance Laws} \label{sec:balance-laws}

In this section, we formulate balance laws for mass, linear momentum, angular momentum, energy, and entropy for a system involving a membrane $\Pc$ immersed in a body of fluid $\Omega$. 
Using the integral theorems, we separate the balance laws for the fluid and the membrane. 
The membrane is allowed to act as a surface of discontinuity. Physically, this corresponds to mechanisms by which the membrane and bulk can exchange conserved quantities, modifying the membrane balance laws from previous work \cite{sahu-mandadapu-pre-2017}. 
We end by deriving an expression for the internal entropy production in the fluid and at the membrane, identifying the thermodynamic forces and fluxes.


\subsection{Mass Balance} \label{subsec:mass-balance}

Suppose that there are $N$ chemical species in the overall bulk fluid and membrane system. Species ``$i$'' can exist in the bulk fluid with mass density $\rho_i$ or in the membrane with mass density $\rho_i^{(m)}$. Let the average velocity of species ``$i$'' be $\vb_i$, then the mass-averaged velocity is given by $\vb := \sum_i (\rho_i/\rho)\vb_i$. The diffusive flux of species ``$i$'' is the mass flux of species ``$i$'' relative to the mass-averaged velocity, and is given by $\jb_i := \rho_i(\vb_i - \vb)$. By using the molar mass $\scrM_i$, we can convert to a molar basis, where the molar concentration is represented by $c_i := \rho_i/\scrM_i$ and the molar diffusive flux is given by $\Jb_i := \jb_i/\scrM_i$. On the membrane, similar quantities are defined as follows: $\rho^{(m)} := \sum_i \rho_i^{(m)},\ \vb^{(m)} :=  \sum_i (\rho_i^{(m)}/\rho^{(m)})\vb_i^{(m)},\, \jb_i^{(m)} := \rho_i^{(m)}(\vb_i^{(m)} - \vb^{(m)}),\ c_i^{(m)} := \rho_i^{(m)}/\scrM_i$, and $\Jb_i^{(m)} := \jb_i^{(m)}/\scrM_i$. 
With regards to notation, it's important to note that superscripts are used to distinguish between quantities specific to the membrane and those specific to the bulk fluid. Specifically, quantities that pertain to the membrane are decorated with a superscript $(\,\bullet\,)^{(m)}$, while quantities that relate to the bulk fluid are undecorated. As defined in Section~\ref{subsec:modified-integral-theorems}, quantities decorated with a superscript $(\,\bullet\,)^{(+)}$ or $(\,\bullet\,)^{(-)}$ are limiting values of bulk quantities, and are evaluated at points directly on the membrane.

In the absence of chemical reactions, the total mass of species ``$i$'' in the bulk fluid and membrane system $\Omega \oplus \Pc$ (see Figure~\ref{fig:surface-of-discontinuity}) changes only by way of diffusive flux through either the fluid boundary $\partial\Omega$ or the membrane patch boundary $\partial\Pc$:
\begin{equation}
    \frac{d}{dt} \left(\int_\Omega \rho_i \,dv + \int_\Pc \rho_i^{(m)} \,da \right) = -\int_{\partial\Omega} \jb_i \cdot \nb \,da - \int_{\partial\Pc} \jb_i^{(m)} \cdot \nub \,ds \,.
\end{equation}
Applying the modified Reynolds transport theorem (Eq.~\eqref{eq:jump-reynolds-transport-scalar}) and the divergence theorems (Eqs.~\eqref{eq:jump-div-thm-vector} and~\eqref{eq:surface-div-thm}) leads to 
\begin{multline}
    \int_\Omega \left(\dot{\rho}_i + \rho_i\divg\vb + \divg\jb_i \right) \,dv \\
    + \int_\Pc \left(\dot{\rho}_i^{(m)} + \rho_i^{(m)}\divg_\text{S}\vb^{(m)} + (j_i^{(m)})^{\alpha}_{;\alpha} + \jump{\jb_i - \rho_i(\vb^{(m)} - \vb)} \cdot \nb \right) \,da = 0 \,. \label{eq:global-mass-bal-intermediate-1}
\end{multline}
Using the relation $\jb_i - \rho_i(\vb^{(m)} - \vb) = \rho_i(\vb_i - \vb^{(m)})$, we can rewrite Eq.~\eqref{eq:global-mass-bal-intermediate-1} as 
\begin{multline}
    \int_\Omega \left(\dot{\rho}_i + \rho_i\divg\vb + \divg\jb_i \right) \,dv \\ 
    + \int_\Pc \left(\dot{\rho}_i^{(m)} + \rho_i^{(m)}\divg_\text{S}\vb^{(m)} + (j_i^{(m)})^{\alpha}_{;\alpha} + \jump{\rho_i(\vb_i - \vb^{(m)})} \cdot \nb \right) \,da = 0 \,. \label{eq:global-mass-bal-intermediate-2}
\end{multline}

Equation~\eqref{eq:global-mass-bal-intermediate-2} can be separated to yield localized forms of mass balances for the fluid and the membrane. By choosing a small enough neighbourhood $\Omega$ around any interior point $\xb$ in the bulk fluid and selecting a region of the bulk fluid that does not intersect the surface of discontinuity, Eq.~\eqref{eq:global-mass-bal-intermediate-2} reduces to 
\begin{equation}
    \int_\Omega \left(\dot{\rho}_i + \rho_i\divg\vb + \divg\jb_i \right) \,dv \,.
    \label{eq:global-mass-bal-intermediate-3}
\end{equation}
Since this is true for any arbitrary neighbourhood $\Omega$, Eq.~\eqref{eq:global-mass-bal-intermediate-3} localizes to 
\begin{equation}
    \dot{\rho}_i + \rho_i\divg\vb + \divg\jb_i = 0 \,. \label{eq:local-bulk-mass-balance-spec-i}
\end{equation}
We refer to the above equation as the \textit{bulk mass balance on species ``$i$''}. Substituting Eq.~\eqref{eq:local-bulk-mass-balance-spec-i} back into the global balance reduces Eq.~\eqref{eq:global-mass-bal-intermediate-2} to 
\begin{equation}
    \int_\Pc \left(\dot{\rho}_i^{(m)} + \rho_i^{(m)}\divg_\text{S}\vb^{(m)} + (j_i^{(m)})^{\alpha}_{;\alpha} + \jump{\rho_i(\vb_i - \vb^{(m)})} \cdot \nb \right) \,da = 0 \,.
\end{equation}
Since the surface patch $\Pc$ is arbitrary, the local \textit{membrane mass balance on species ``$i$''} is then given by 
\begin{equation}
    \dot{\rho}_i^{(m)} + \rho_i^{(m)}\divg_\text{S}\vb^{(m)} + (j_i^{(m)})^{\alpha}_{;\alpha} + \jump{\rho_i(\vb_i - \vb^{(m)})} \cdot \nb = 0 \,.
    \label{eq:local-membrane-mass-balance-spec-i}
\end{equation}
Dividing Eqs.~\eqref{eq:local-bulk-mass-balance-spec-i} and~\eqref{eq:local-membrane-mass-balance-spec-i} by the molar mass $\scrM_i$, gives the species balance on a molar basis, i.e., the concentration balances, as 
\begin{subequations}
\label{eq:local-mole-balance-spec-i}
\begin{align}
    \dot{c}_i + c_i\divg\vb + \divg\Jb_i &= 0 \label{eq:local-bulk-mole-balance-spec-i} \,, \\
    \dot{c}_i^{(m)} + c_i^{(m)}\divg_\text{S}\vb^{(m)} + (J_i^{(m)})^{\alpha}_{;\alpha} + \jump{c_i(\vb_i - \vb^{(m)})} \cdot \nb &= 0 \,.
    \label{eq:local-membrane-mole-balance-spec-i}
\end{align}
\end{subequations}
Summing Eqs.~\eqref{eq:local-bulk-mass-balance-spec-i} and~\eqref{eq:local-membrane-mass-balance-spec-i} over all species, and noting that $\sum_i \jb_i = \zerob$ and $\sum_i \jb_i^{(m)} = \zerob$, yields the overall mass balances for the bulk and membrane
\begin{subequations}
\label{eq:local-mass-balance}
\begin{align}
    \dot{\rho} + \rho\divg\vb &= 0 \,, 
    \label{eq:local-bulk-mass-balance}\\
    \dot{\rho}^{(m)} + \rho^{(m)}\divg_\text{S}\vb^{(m)} + \jump{\rho(\vb - \vb^{(m)})} \cdot \nb &= 0 \,.
    \label{eq:local-membrane-mass-balance}
\end{align}
\end{subequations}

A special case of interest is when a particular species does not accumulate on the membrane. We impose this condition by requiring that the total mass of the species in a material region of the surface $\Pc$ does not change, which leads to the following global balance 
\begin{equation}
    \frac{d}{dt} \int_\Pc \rho_i^{(m)} \,da = - \int_{\partial\Pc} \jb_i^{(m)} \cdot \nub \,ds \,.
\end{equation}
Applying Eqs.~\eqref{eq:surface-reynolds-transport} and~\eqref{eq:surface-div-thm} then localizing yields 
\begin{equation}
    \dot{\rho}_i^{(m)} + \rho_i^{(m)}\divg_\text{S}\vb^{(m)} + (j_i^{(m)})^{\alpha}_{;\alpha} = 0 \,,
    \label{eq:local-membrane-mass-balance-spec-i-no-acc}
\end{equation}
which can also be written in terms of concentration as 
\begin{equation}
    \dot{c}_i^{(m)} + c_i^{(m)}\divg_\text{S}\vb^{(m)} + (J_i^{(m)})^{\alpha}_{;\alpha} = 0 \,.
    \label{eq:local-membrane-mole-balance-spec-i-no-acc}
\end{equation}
Substituting Eqs.~\eqref{eq:local-membrane-mass-balance-spec-i-no-acc} and~\eqref{eq:local-membrane-mole-balance-spec-i-no-acc} into Eqs.~\eqref{eq:local-membrane-mass-balance-spec-i} and~\eqref{eq:local-membrane-mole-balance-spec-i}, respectively, yields the jump conditions 
\begin{equation}
    \jump{\rho_i(\vb_i - \vb^{(m)})} \cdot \nb = 0 
    \quad \text{and} \quad 
    \jump{c_i(\vb_i - \vb^{(m)})} \cdot \nb = 0 
    \,. \label{eq:local-membrane-mass-bal-spec-i-noacc}
\end{equation}
The quantities in the jump brackets above carry the physical significance of being the mass (or concentration) flux of species ``$i$'' relative to the membrane, which we will be interested in developing constitutive laws for in Section~\ref{sec:linear-irrev-thermo}. 
Defining the relative mass fluxes~$\dot{n}_i^{(m)} := \rho_i (\vb_i - \vb^{(m)}) \cdot \nb$ and~$\dot{n}^{(m)} := \rho (\vb - \vb^{(m)}) \cdot \nb$, 
or~$\dot{N}_i^{(m)} := c_i (\vb_i - \vb^{(m)}) \cdot \nb$ and~$\dot{N}^{(m)} := c (\vb - \vb^{(m)}) \cdot \nb$ in a molar basis, Eq.~\eqref{eq:local-membrane-mass-bal-spec-i-noacc} implies that whenever there's no accumulation of a species on the membrane, the limiting value of the flux of that component must be equal on both sides of the membrane. In this case, it is not necessary to decorate the quantities $\dot{n}_i^{(m)}$ or $\dot{N}_i^{(m)}$ with a superscript $(\, \bullet \,)^{(\pm)}$. 

\subsection{Linear Momentum Balance} \label{subsec:linear-momentum-balance}

\begin{figure}
    \centering
    \includegraphics[width=1.0\linewidth]{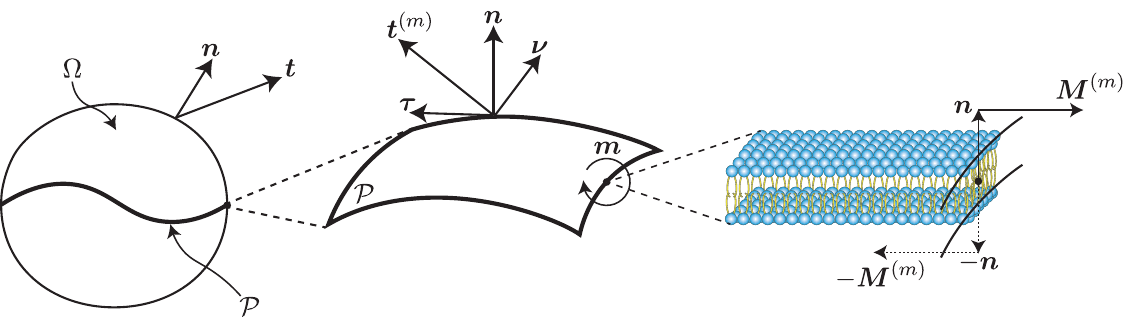}
    \caption{A traction $\tb$ acting on the bulk boundary, and a membrane traction $\tb^{(m)}$ acting on the membrane boundary. The membrane also supports a moment $\mb = \nb \times \Mb^{(m)}$, which is resolved as a director traction $\Mb^{(m)}$ acting on a moment arm with length corresponding to the thickness of the membrane and orientation in the direction of the surface normal $\nb$.}
    \label{fig:force-balance}
\end{figure}

The rate of change of the total momentum of the membrane and bulk fluid system must be balanced by the net force acting on the system. Within both the bulk fluid and the membrane, we allow for two types of forces: body forces $\bb$ and tractions $\tb$ on the boundary\footnote{We do not include electric or magnetic forces, but readers interested in the treatment of these forces in continuous systems should consult \cite{fong2020transport}.} (see Figure~\ref{fig:force-balance}). The integral formulation of the balance of linear momentum for the region occupied by the bulk fluid $\Omega$ with membrane $\Pc$ (see Figure~\ref{fig:surface-of-discontinuity}) is then\footnote{%
Note that since the membrane has measure zero in the bulk fluid we do not need to explicitly remove it from the bulk fluid integrals, 
i.e., 
\[
    \int_{\Omega} (\,\cdots) \,dv = \int_{\Omega\setminus\Pc} (\,\cdots) \,dv \quad \text{and} \quad \int_{\partial\Omega} (\,\cdots) \,da = \int_{\partial\Omega\setminus\partial\Pc} (\,\cdots) \,da \,.
\]
}
\begin{equation}
    \frac{d}{dt}\left(\int_\Omega \rho \vb \,dv + \int_\Pc \rho^{(m)} \vb^{(m)} \, da\right) = \left(\int_\Omega \rho\bb \,dv + \int_{\partial\Omega} \tb \,da \right) + \left(\int_\Pc \rho^{(m)} \bb^{(m)} \,da + \int_{\partial\Pc}  \tb^{(m)} \,ds\right) \,. \label{eq:global-linear-momentum-balance}
\end{equation}
Using Cauchy's tetrahedron argument \cite{gurtin2010mechanics}, we can decompose the tractions $\tb(\xb,t;\nb)$ at a point $\xb_\brm$ on the boundary of the bulk domain as 
\begin{equation}
    \tb(\xb_\brm,t;\nb) = \tb^i(\xb_\brm,t) n_i = \Tb\nb\,,
    \label{eq:bulk-traction-decomposition}
\end{equation}
where $\tb^i(\xb_\brm,t)$ are the principal tractions, and $n_i := \nb \cdot \eb_i$ are the Cartesian components of the normal vector with $\eb_i$ being the Cartesian basis vectors. The Cauchy stress tensor in Eq.~\eqref{eq:bulk-traction-decomposition} is given by $\Tb = \tb^i \otimes \eb_i$ so that the traction on the bulk boundary is $\tb = \Tb\nb$. A similar curvilinear triangle argument by Naghdi~\cite{Naghdi1973} can be employed to show that the tractions on the boundary of a surface $\tb^{(m)}(\xb_\brm^{(m)},t;\nub)$ can be decomposed as 
\begin{equation}
    \tb^{(m)}(\xb_\brm^{(m)},t; \nub) = (\Tb^{(m)})^\alpha(\xb_\brm^{(m)},t) \nu_\alpha
    \,,
    \label{eq:membrane-traction-decomposition}
\end{equation}
where $(\Tb^{(m)})^\alpha(\xb_\brm^{(m)},t)$ are the principal membrane traction vectors \cite{Rangamani2013,sahu-mandadapu-pre-2017}. 

With Eqs.~\eqref{eq:bulk-traction-decomposition} and~\eqref{eq:membrane-traction-decomposition}, the global balance of linear momentum (Eq.~\eqref{eq:global-linear-momentum-balance}) becomes 
\begin{align}
\begin{split}
    \frac{d}{dt}\left(\int_\Omega \rho \vb \,dv + \int_\Pc \rho^{(m)} \vb^{(m)} \, da\right) &= \left(\int_\Omega \rho\bb \,dv + \int_{\partial\Omega} \Tb\nb \,da \right)\\
    &\hspace{0.75in} + \left(\int_\Pc \rho^{(m)} \bb^{(m)} \,da + \int_{\partial\Pc}  (\Tb^{(m)})^\alpha \nu_\alpha \,ds\right) \,. 
\end{split}
    \label{eq:global-linear-momentum-balance-2}
\end{align}
Applying the integral theorems of Section~\ref{subsec:modified-integral-theorems} and localizing yields 
\begin{subequations}
\label{eq:local-linear-momentum-balance}
\begin{align}
    \rho\dot{\vb} &= \rho\bb + \divg \Tb \,,
    \label{eq:local-bulk-linear-momentum-balance}\\
    \rho^{(m)}\dot{\vb}^{(m)} &= \rho^{(m)}\bb^{(m)} + (\Tb^{(m)})^\alpha_{;\alpha} + \jump{\Tb - \rho (\vb - \vb^{(m)}) \otimes (\vb - \vb^{(m)})}\nb \,.
    \label{eq:local-membrane-linear-momentum-balance}
\end{align}
\end{subequations}
The jump term in Eq.~\eqref{eq:local-membrane-linear-momentum-balance} corresponds to the additional forces
from the bulk fluid that act on the membrane, which arise from bulk traction forces and relative momentum exchanges. Typically, for low Reynolds number flows---as is common for many biological systems \cite{purcell1977life,stone2015low}%
---the kinetic contributions to the jump term in Eq.~\eqref{eq:local-membrane-linear-momentum-balance} are negligible in comparison to the traction forces.

\subsection{Angular Momentum Balance} \label{subsec:angular-momentum-balance}

The net torque acting on the system is obtained by integrating the torque densities due to the body forces ($\rho\xb\times\bb$ and $\rho^{(m)}\xb^{(m)}\times\bb^{(m)}$) and tractions ($\rho\xb\times\tb$ and $\rho^{(m)}\xb^{(m)}\times\tb^{(m)}$) that act on the bulk fluid and membrane. In addition, it is known that lipid membranes can hold additional moments $\mb$ due to their representation as two-dimensional shells \cite{sahu-mandadapu-pre-2017} (see Figure~\ref{fig:force-balance}). The moment per unit length $\mb$ can be decomposed as the cross product $\db^{(m)} \times \Mb^{(m)}$ between a director field $\db^{(m)}$ that acts as the moment arm, and a director traction $\Mb^{(m)}$ that applies an equal and opposite force to the moment arm to give rise to the couple-stress. The director field $\db^{(m)}$ indicates the orientation of the lipid molecules that constitute the membrane, and in general does not need to be aligned with the surface normal $\nb$ due to the presence of phenomena such as lipid tilt. We neglect such effects here\footnote{We refer the reader to \cite{sahu-mandadapu-pre-2017,steigmann2013model} for a more thorough discussion on when including the director field $\db^{(m)}$ may be necessary. In such cases, an additional balance law is needed to describe the evolution $\dot{\db}^{(m)}$ of the director field~\cite{Naghdi1973,steigmann2013model}.}, however, and assume that the moment arm that $\Mb^{(m)}$ acts on is always aligned with the membrane normal so that
\begin{equation}
    \mb = \nb \times \Mb^{(m)}\,.
    \label{eq:moment-couple-decomposition}
\end{equation}
Equation~\eqref{eq:moment-couple-decomposition} shows that without loss of generality, we can take $\Mb^{(m)}$ to lie in the plane of the membrane. Once again, using the curvilinear triangle argument~\cite{Naghdi1973}, the director traction $\Mb^{(m)}$ can be decomposed as 
\begin{equation}
    \Mb^{(m)}(\xb_\brm^{(m)},t; \nub) = (\Mb^{(m)})^\alpha(\xb_\brm^{(m)},t) \nu_\alpha \,,
\end{equation}
where $(\Mb^{(m)})^\alpha$ are the principal couple-stress vectors. 
The rate of change of angular momentum of the system can then be expressed as 
\begin{multline}
    \frac{d}{dt}\left(\int_\Omega \rho \xb \times \vb \,dv + \int_\Pc \rho^{(m)} \xb^{(m)} \times \vb^{(m)} \,da\right) = \int_\Omega \rho \xb \times \bb \,dv + \int_{\partial\Omega} \xb \times \tb \,da \\ 
    + \int_\Pc \rho^{(m)} \xb^{(m)} \times \bb^{(m)} \,da + \int_{\partial\Pc} \xb^{(m)} \times \tb^{(m)} \,ds + \int_{\partial\Pc} \nb \times \Mb^{(m)} \,ds \,. \label{eq:overall-ang-momentum-bal}
\end{multline}

Applying the integral theorems of Section~\ref{subsec:modified-integral-theorems} to Eq.~\eqref{eq:overall-ang-momentum-bal}, incorporating the local forms of the momentum balance (Eq.~\eqref{eq:local-linear-momentum-balance}), and localizing yields 
\begin{subequations}
\label{eq:local-ang-momentum-balance}
\begin{gather}
    \Tb^T = \Tb \,, 
    \label{eq:local-bulk-ang-momentum-balance}\\
    \ab_\alpha \times (\Tb^{(m)})^\alpha - b_\alpha^\beta\ab_\beta \times (\Mb^{(m)})^\alpha + \nb \times (\Mb^{(m)})_{;\alpha}^\alpha = 0 \,.
    \label{eq:local-membrane-ang-momentum-balance}
\end{gather}
\end{subequations}
Equation~\eqref{eq:local-bulk-ang-momentum-balance} is the standard symmetry condition for the bulk stress tensor from classical continuum mechanics, while Eq.~\eqref{eq:local-membrane-ang-momentum-balance} is identical to the local balance of angular momentum for a membrane that appears in standalone membrane theories \cite{sahu-mandadapu-pre-2017}. To state this result differently, no new mechanisms for angular momentum transport are introduced by considering the composite bulk and membrane system with the membrane acting as a surface of discontinuity. It is not self-evident, however, that depositing angular momentum onto a permeable membrane should be impossible. Indeed, this mathematical result may stem from our choice of representation for the permeable membrane as a strictly two-dimensional interface, and other theories that account for the finite thickness of the membrane \cite{yannick2024part1} could yield new mechanisms for angular momentum transport.

Equation~\eqref{eq:local-membrane-ang-momentum-balance} also imposes restrictions on the membrane stress tensor
(see \cite{sahu-mandadapu-pre-2017} for a detailed derivation). That is, without loss of generality, expanding the principal stress vectors using the in-plane basis as $(\Tb^{(m)})^\alpha = N^{\alpha\beta} \ab_\beta + S^\alpha \nb$ and the couple stress vectors as $(\Mb^{(m)})^\alpha = -M^{\alpha\beta} \ab_\beta$, we obtain the following restrictions:
\begin{subequations}
\label{eq:ang-mom-consequences}
\begin{align}
    \Tb &\text{ is symmetric} \,,\\
    \sigma^{\alpha\beta} &:= (N^{\alpha\beta} - b_\mu^\beta M^{\mu\alpha}) \text{ is symmetric} \,, 
    \label{eq:membrane-ang-momentum-consequence-1}\\
    S^\alpha &= -M_{;\beta}^{\beta\alpha} \,.
    \label{eq:membrane-ang-mom-consequence-2}
\end{align}
\end{subequations}
In Eq.~\eqref{eq:ang-mom-consequences}, $N^{\alpha\beta}$ and $S^\alpha$ are the in-plane and normal membrane traction vector components, $M^{\alpha\beta}$ are the in-plane couple-stress vector components, and $\sigma^{\alpha\beta}$ are the moment-free contributions to the in-plane membrane tractions. 
Note that since components with Greek indices in Eqs.~\eqref{eq:membrane-ang-momentum-consequence-1} and~\eqref{eq:membrane-ang-mom-consequence-2} are all membrane quantities, for notational simplicity we omit
the superscript $(\, \bullet \,)^{(m)}$. As notational convenience, whenever there is no ambiguity, we will henceforth also expand membrane quantities as $\fb^{(m)} = f^\alpha \ab_\alpha + f\nb$ in lieu of $\fb^{(m)} = (f^{(m)})^\alpha \ab_\alpha + f^{(m)}\nb$.

\subsection{Energy Balance} \label{subsec:energy-balance}

According to the first law of thermodynamics, the total energy of a system may change by way of work done on it and heat flow into it. 
The overall balance of energy for the bulk and membrane system can be written as 
\begin{multline}
    \frac{d}{dt}\left(\int_\Omega \rho e \,dv + \int_\Pc \rho^{(m)} e^{(m)} \,da \right) = \int_\Omega \left( \rho\bb\cdot\vb + \rho r \right)\,dv + \int_{\partial\Omega} \left( h + \tb\cdot\vb \right)\,da \\ 
    + \int_\Pc \left( \rho^{(m)}\bb^{(m)}\cdot\vb^{(m)} + \rho^{(m)} r^{(m)} \right) \,da + \int_{\partial\Pc} \left( h^{(m)} + \tb^{(m)} \cdot \vb^{(m)} + \dot{\nb}\cdot\Mb^{(m)} \right)\,ds \,,
    \label{eq:energy-rate-of-change-1}
\end{multline}
where $e$ and $e^{(m)}$ denote the bulk and membrane energies per unit mass, $\bb$ and $\bb^{(m)}$ are the bulk and membrane body forces per unit mass, $h$ and $h^{(m)}$ are the bulk and membrane boundary heat fluxes, and $r$ and $r^{(m)}$ are the local heat per unit mass supplied to the bulk and membrane bodies.
Additionally, in Eq.~\eqref{eq:energy-rate-of-change-1}, $\tb\cdot\vb$ and $\tb^{(m)} \cdot \vb^{(m)}$ represent the work done by the bulk fluid and membrane tractions, $\rho\bb\cdot\vb$ and $\rho^{(m)}\bb^{(m)}\cdot\vb^{(m)}$ represent the work done by the external body forces on the bulk fluid and membrane, and $\dot{\nb}\cdot\Mb^{(m)}$ represents the work due to moments acting on the membrane. 
Using again the Cauchy Tetrahedron Argument and its curvilinear triangle analogue~\cite{Naghdi1973}, we can rewrite the bulk and membrane heat fluxes as 
\begin{align}
    h(\xb_\brm,t;\nb) &= -\Jb_q(\xb_\brm,t;\nb) \cdot \nb \,,
    \label{eq:bulk-heat-flux-decomp}\\
    h^{(m)}(\xb_\brm^{(m)},t;\nub) &= -\Jb_q^{(m)}(\xb_\brm^{(m)},t) \cdot \nub \,,
    \label{eq:membrane-heat-flux-decomp}
\end{align}
where $\Jb_q$ and $\Jb_q^{(m)}$ are the corresponding heat flux vectors. Given Eqs.~\eqref{eq:bulk-heat-flux-decomp} and~\eqref{eq:membrane-heat-flux-decomp}, as well as the bulk and membrane traction decompositions~\eqref{eq:bulk-traction-decomposition} and~\eqref{eq:membrane-traction-decomposition}, we apply the divergence theorems~\eqref{eq:jump-div-thm-vector} and~\eqref{eq:surface-div-thm} to re-express Eq.~\eqref{eq:energy-rate-of-change-1} as 
\begin{multline}
    \frac{d}{dt}\left(\int_\Omega \rho e \,dv + \int_\Pc \rho^{(m)} e^{(m)} \,da \right) = \int_\Omega \left( \rho\bb \cdot \vb + \divg(\Tb^T\vb) + \rho r - \divg\Jb_q \right) \,dv \\
    + \int_\Pc \left( \rho^{(m)}\bb^{(m)} \cdot \vb^{(m)} + \left( 
    (\Tb^{(m)})^\alpha \cdot \vb^{(m)}
    \right)_{;\alpha}
    + \rho^{(m)} r^{(m)} - (J_q^{(m)})_{;\alpha}^{\alpha}\right) \,da
    \hspace{0.in} \\
    + \int_\Pc \left( \dot{\nb} \cdot (\Mb^{(m)})^\alpha \right)_{;\alpha} \,da
    + \int_\Pc \jump{\Tb^T\vb - \Jb_q}\cdot\nb \,da \,.
    \label{eq:energy-rate-of-change-2}
\end{multline}

The total energy $e$ and $e^{(m)}$ of each body can be decomposed into a kinetic part\footnote{Strictly speaking, the total kinetic energy is $\sum_{i=1}^N \frac{1}{2} \rho_i ||\vb_i||^2 = \frac{1}{2} \rho ||\vb||^2 + \sum_{i=1}^N \frac{1}{2} \rho_i ||\vb_i - \vb||^2$. Here, we neglect the inertia of each species relative to the barycentric velocity; see \cite{deGroot2013nonequilthermo} for examples of where inertial terms may matter.
}
$\frac{1}{2}\rho \vb\cdot\vb$ and $\frac{1}{2}\rho^{(m)} \vb^{(m)}\cdot\vb^{(m)}$ and an internal part $\rho u$ and $\rho^{(m)} u^{(m)}$ as
\begin{align}
    \rho e &= \rho u + \frac{1}{2}\rho \vb \cdot \vb \,,\\
    \rho^{(m)} e^{(m)} &= \rho^{(m)} u^{(m)} + \frac{1}{2}\rho^{(m)} \vb^{(m)} \cdot \vb^{(m)} \,.
\end{align}
Using this decomposition, the Reynolds transport theorem~\eqref{eq:jump-reynolds-transport-scalar}, and the local mass balances~\eqref{eq:local-mass-balance}, we can express the rate of change of the system's energy (i.e., the left-hand side of Eq.~\eqref{eq:energy-rate-of-change-1}) as 
\begin{multline}
    \frac{d}{dt}\left(\int_\Omega \rho e \,dv + \int_\Pc \rho^{(m)} e^{(m)} \,da \right) = \int_\Omega \left(\rho \dot{u} + \rho \dot{\vb}\cdot\vb\right) \,dv + \int_\Pc \left( \rho^{(m)} \dot{u}^{(m)} + \rho^{(m)} \dot{\vb}^{(m)}\cdot\vb^{(m)} \right) \,da \\
    + \int_\Pc \jump{\rho(u - u^{(m)})(\vb - \vb^{(m)}) + \frac{1}{2}\rho(\vb\cdot\vb - \vb^{(m)}\cdot\vb^{(m)})(\vb - \vb^{(m)})}\cdot\nb \,da \,.
    \label{eq:energy-rate-of-change-3}
\end{multline}

Using the balance of linear momentum, we can show that the rate of change of the kinetic energy of the system is balanced by the work done by the body forces and internal stresses. To this end, taking the inner product of the bulk linear momentum balance~\eqref{eq:local-bulk-linear-momentum-balance} with the velocity $\vb$ and integrating over the region of fluid $\Omega$, we obtain the mechanical energy balance 
\begin{align}
    \int_\Omega \rho\dot{\vb}\cdot\vb \,dv &= \int_\Omega \left(\rho\bb\cdot\vb + \divg\Tb\cdot\vb \right) \,dv \,.
    \label{eq:local-bulk-mechanical-energy-balance}
\end{align}
Similarly, taking the inner product of the membrane linear momentum balance~\eqref{eq:local-membrane-linear-momentum-balance} and the membrane velocity $\vb^{(m)}$, and integrating over the membrane surface gives the membrane mechanical energy balance 
\begin{multline}
    \int_\Pc \rho^{(m)}\dot{\vb}^{(m)}\cdot\vb^{(m)} \,da = \int_\Pc \left(\rho^{(m)}\bb^{(m)} \cdot \vb^{(m)} + (\Tb^{(m)})_{;\alpha}^\alpha \cdot\vb^{(m)} \right) \,da \\
    + \int_\Pc \left( \jump{\Tb^T\vb^{(m)} - \rho(\vb\cdot\vb^{(m)} - \vb^{(m)}\cdot\vb^{(m)})(\vb - \vb^{(m)})}\cdot\nb \right) \,da \,.
    \label{eq:local-membrane-mechanical-energy-balance}
\end{multline}
Combining the bulk and membrane mechanical energy balances (Eqs.~\eqref{eq:local-bulk-mechanical-energy-balance} and~\eqref{eq:local-membrane-mechanical-energy-balance}) with Eq.~\eqref{eq:energy-rate-of-change-3}, we obtain the following expression 
\begin{multline}
    \frac{d}{dt}\left(\int_\Omega \rho e \,dv + \int_\Pc \rho^{(m)} e^{(m)} \,da \right) = \int_\Omega \left(\rho \dot{u} + \rho\bb\cdot\vb + \divg\Tb\cdot\vb\right) \,dv\\
    + \int_\Pc \left( \rho^{(m)} \dot{u}^{(m)} + \rho^{(m)}\bb^{(m)} \cdot \vb^{(m)} + (\Tb^{(m)})_{;\alpha}^\alpha \cdot\vb^{(m)} \right) \,da \\
    + \int_\Pc \jump{\rho(u - u^{(m)})(\vb - \vb^{(m)}) + \Tb^T\vb^{(m)} + \frac{1}{2}\rho||\vb - \vb^{(m)}||^2(\vb - \vb^{(m)})}\cdot\nb \,da \,.
    \label{eq:energy-rate-of-change-4}
\end{multline}
Combining Eq.~\eqref{eq:energy-rate-of-change-4} with Eq.~\eqref{eq:energy-rate-of-change-2} yields a balance for the internal energy as
\begin{multline}
    \int_\Omega \rho \dot{u} \,dv + \int_\Pc \rho^{(m)} \dot{u}^{(m)} \,da = \int_\Omega \left( \rho r - \divg\Jb_q + \Tb:\grad\vb \right) \,dv \\
    + \int_\Pc \left(\rho^{(m)} r^{(m)}- \jump{\Jb_q}\cdot\nb - (J_q^{(m)})_{;\alpha}^{\alpha} + (\Tb^{(m)})^\alpha \cdot \vb^{(m)}_{,\alpha} + (\dot{\nb} \cdot (\Mb^{(m)})^\alpha)_{,\alpha} \right) \,da \\
    + \int_{\Pc} \jump{-\rho(u - u^{(m)})(\vb - \vb^{(m)}) + \Tb^T(\vb - \vb^{(m)}) - \frac{1}{2}\rho||\vb - \vb^{(m)}||^2(\vb - \vb^{(m)})}\cdot\nb \,da \,.
    \label{eq:energy-rate-of-change-5}
\end{multline}

Since the local angular momentum balance of the membrane~\eqref{eq:local-membrane-ang-momentum-balance} is unchanged from that of a standalone membrane, the last two terms term in the integrand of the second term on the right-hand side of Eq.~\eqref{eq:energy-rate-of-change-2} can be evaluated using the following identity 
\begin{equation}
    (\Tb^{(m)})^\alpha \cdot \vb^{(m)}_{,\alpha} + (\dot{\nb} \cdot (\Mb^{(m)})^\alpha)_{,\alpha} = \frac{1}{2}\sigma^{\alpha\beta}\dot{a}_{\alpha\beta} + M^{\alpha\beta}\dot{b}_{\alpha\beta} \,,
    \label{eq:traction-couple-work-identity}
\end{equation}
which follows from the symmetry of the tensor $\sigma^{\alpha\beta}$ (Eq.~\eqref{eq:membrane-ang-momentum-consequence-1}) and other kinematic identities (see~\cite{sahu-mandadapu-pre-2017} for a detailed derivation).
Incorporating Eq.~\eqref{eq:traction-couple-work-identity} into Eq.~\eqref{eq:energy-rate-of-change-5} and then localizing to the bulk and membrane gives 
\begin{subequations}
\label{eq:local-energy-balance}
\begin{align}
    \rho \dot{u} &= \Tb:\grad\vb + \rho r - \divg\Jb_q \,,
    \label{eq:local-bulk-energy-balance}\\
    \begin{split}
        \rho^{(m)} \dot{u}^{(m)} &= \rho^{(m)} r^{(m)}- \jump{\Jb_q}\cdot\nb - (J_q^{(m)})_{;\alpha}^{\alpha} + \frac{1}{2}\sigma^{\alpha\beta}\dot{a}_{\alpha\beta} + M^{\alpha\beta}\dot{b}_{\alpha\beta}\\
    &\quad + \jump{-\rho(u - u^{(m)})(\vb - \vb^{(m)}) + \Tb^T(\vb - \vb^{(m)}) - \frac{1}{2}\rho||\vb - \vb^{(m)}||^2(\vb - \vb^{(m)})}\cdot\nb \,.
    \end{split}
    \label{eq:local-membrane-energy-balance}
\end{align}
\end{subequations}
Equation~\eqref{eq:local-bulk-energy-balance} is the usual balance of energy for bulk fluid media \cite{gurtin2010mechanics}. Equation~\eqref{eq:local-membrane-energy-balance}, however, differs from the balance of energy for a standalone membrane in previous works \cite{sahu-mandadapu-pre-2017} through the appearance of new jump terms that act as sources/sinks of energy. The term $- \jump{\Jb_q}\cdot\nb = \Jb_q^{(+)}\cdot(-\nb) + \Jb_q^{(-)} \cdot \nb$ represents the net heat flux from the surrounding bulk fluid into the membrane. 
Similarly, the term $\jump{\Tb^T(\vb - \vb^{(m)})}\cdot\nb = \Tb^{(+)}\nb \cdot (\vb^{(+)} - \vb^{(m)}) + \Tb^{(-)}(-\nb) \cdot (\vb^{(-)} - \vb^{(m)})$ is the net work done on the membrane by surrounding bulk fluid.

Using Eq.~\eqref{eq:jump-identity}, the term $\jump{-\rho(u - u^{(m)})( \vb - \vb^{(m)} )} \cdot \nb$ in Eq.~\eqref{eq:local-membrane-energy-balance} can be written as 
\begin{align}
    \jump{-\rho(u - u^{(m)})(\vb - \vb^{(m)})}\cdot\nb &= \jump{-u} \braket{\dot{n}^{(m)}} + \left( \braket{u} - u^{(m)} \right) \jump{-\dot{n}^{(m)}} \,,
    \label{eq:relative-internal-energy-flux-expansion}
\end{align}
where the first term on the right-hand side of Eq.~\eqref{eq:relative-internal-energy-flux-expansion} corresponds to the rate of energy exchange between the membrane and the (net) mass flux $\braket{\dot{n}^{(m)}}$ through the membrane, and the second term corresponds to the rate of energy change that accompanies the rate of mass deposition $\jump{-\dot{n}^{(m)}}$ onto the membrane. 
Observe that when mass is deposited onto the membrane, it is the relative internal energy $\braket{u} - u^{(m)}$, and not just the bulk internal energy $\braket{u}$ that contributes to changes in the internal energy of the membrane $\dot{u}^{(m)}$. This is because if material with the same internal energy per unit mass as that of the membrane is deposited, then the intensive (per unit mass) quantity $u^{(m)}$ should not change.
Similarly, the last term  $\jump{-\frac{1}{2}\rho||\vb - \vb^{(m)}||^2(\vb - \vb^{(m)})}\cdot\nb$ in Eq.~\eqref{eq:local-membrane-energy-balance} accounts for both the change in the kinetic energy of the material permeating through the membrane, as well as the additional kinetic energy contributed by material deposited onto the membrane.

\subsection{Entropy Balance} \label{subsec:entropy-balance}

In what follows, we consider the entropy balance of the combined system by extending the works of de Groot \& Mazur \cite{deGroot2013nonequilthermo} and Sahu \textit{et al.}\cite{sahu-mandadapu-pre-2017} to a combined membrane and bulk system, with the membrane acting as a material surface of discontinuity.
We begin with a general entropy balance, and propose a second law of  thermodynamics corresponding to the internal entropy production. We introduce the thermodynamic potentials and derive the form of the internal entropy production that drives irreversibilities in the composite system in terms of the thermodynamic forces and fluxes. We conclude with a discussion of the new terms in the internal entropy production at the membrane, which will give rise to the results presented in Section~\ref{sec:linear-irrev-thermo}.

For the combined membrane and bulk fluid system, we postulate the total rate of change of entropy as 
\begin{multline}
    \frac{d}{dt}\left(\int_\Omega \rho s \,dv + \int_\Pc \rho^{(m)} s^{(m)} \,da \right) = \int_{\partial\Omega} \rho h_s \,da + \int_\Omega \left( \rho \eta_e + \rho \eta_i \right) \,dv
    \\ 
    + \int_{\partial\Pc} \rho^{(m)} h_s^{(m)} \,ds + \int_\Pc \left( \rho^{(m)} \eta_e^{(m)} + \rho^{(m)} \eta_i^{(m)} \right) \,da \,,
    \label{eq:entropy-rate-of-change-1}
\end{multline}
where $s$ and $s^{(m)}$ are the bulk and membrane entropy per unit mass, $h_s$ and $h_s^{(m)}$ are the bulk and membrane entropy fluxes into the system, $\eta_e$ and $\eta_e^{(m)}$ are the bulk and membrane entropy supplies from external sources, and $\eta_i$ and $\eta_i^{(m)}$ are the rates of internal entropy production in the bulk and at the membrane. 
Equation~\eqref{eq:entropy-rate-of-change-1} presents a generic statement of the balance of entropy in terms of external and internal sources. We postulate the second law of thermodynamics with the statement that internal entropy production must be non-negative both in the bulk fluid and at the membrane \cite{deGroot2013nonequilthermo}, i.e., 
\begin{subequations}
\begin{align}
    \eta_i &\ge 0 \,,
    \label{eq:bulk-second-law-inequality}\\
    \eta_i^{(m)} &\ge 0 \,.
    \label{eq:membrane-second-law-inequality}
\end{align}
\end{subequations}

The tetrahedron and curvilinear triangle arguments yield the following relations for the entropy fluxes
\begin{align}
    h_s(\xb_\brm,t;\nb) &= -\Jb_s(\xb_\brm,t;\nb) \cdot \nb \,,\\
    h_s^{(m)}(\xb_\brm^{(m)},t;\nub) &= -\Jb_s^{(m)}(\xb_\brm^{(m)},t) \cdot \nub \,,
\end{align}
where $\Jb_s$ and $\Jb_s^{(m)}$ are the bulk and membrane entropy flux vectors. Using these relations and the divergence theorems~\eqref{eq:jump-div-thm-scalar} and~\eqref{eq:surface-div-thm}, Eq.~\eqref{eq:entropy-rate-of-change-1} becomes
\begin{multline}
    \frac{d}{dt}\left(\int_\Omega \rho s \,dv + \int_\Pc \rho^{(m)} s^{(m)} \,da \right) = \int_\Omega \left( -\divg\Jb_s + \rho \eta_e + \rho \eta_i \right) \,dv \\ 
    + \int_\Pc \left( -(J_s^{(m)})_{;\alpha}^{\alpha} + \rho^{(m)} \eta_e^{(m)} + \rho^{(m)} \eta_i^{(m)} - \jump{\Jb_s}\cdot \nb\right) \,da \,.
    \label{eq:entropy-rate-of-change-2}
\end{multline}
With the Reynolds transport theorem~\eqref{eq:jump-reynolds-transport-scalar}, and the mass balances~\eqref{eq:local-bulk-mass-balance} and~\eqref{eq:local-membrane-mass-balance},
we obtain the local balances of entropy

\begin{subequations}
\label{eq:local-entropy-balance}
\begin{gather}
    \rho\dot{s} = -\divg\Jb_s + \rho \eta_e + \rho \eta_i \,,
    \label{eq:local-bulk-entropy-balance}\\
    \rho^{(m)}\dot{s}^{(m)} + \jump{\Jb_s + \rho(s^{(m)} - s)(\vb^{(m)} - \vb)}\cdot\nb = -(J_s^{(m)})_{;\alpha}^{\alpha} + \rho^{(m)} \eta_e^{(m)} + \rho^{(m)} \eta_i^{(m)} \,.
    \label{eq:local-membrane-entropy-balance}
\end{gather}
\end{subequations}
Equation~\eqref{eq:local-bulk-entropy-balance} is the typical entropy balance for a bulk fluid medium \cite{deGroot2013nonequilthermo}, which attributes the rate of change of entropy to entropy fluxes $\Jb_s$, external supplies of entropy $\eta_e$, and internal sources of entropy entropy production $\eta_i$. 
Equation~\eqref{eq:local-membrane-entropy-balance} presents the same general entropy balance as that of a standalone membrane in previous work \cite{sahu-mandadapu-pre-2017}, but with the addition of a source of entropy flux~$\jump{\Jb_s}$ from the bulk fluid, as well as entropy exchange with the bulk fluid $\jump{\rho(s^{(m)} - s)(\vb^{(m)} - \vb)}\cdot\nb$ due to material permeating through and depositing onto the membrane. Essentially, the left-hand side of Eq.~\eqref{eq:local-membrane-entropy-balance} states that the rate of change of entropy at the membrane can be due to phenomena that take place within the membrane itself (i.e., $\rho\dot{s}^{(m)}$), or due to the membrane's interaction with the surrounding bulk fluid. In the spirit of Eq.~\eqref{eq:relative-internal-energy-flux-expansion}, we can expand the jump term that describes these interactions as 
\begin{align}
    \jump{-\rho(s - s^{(m)})(\vb - \vb^{(m)})}\cdot\nb &= \jump{-s} \braket{\dot{n}^{(m)}} + \left( \braket{s} - s^{(m)} \right) \jump{-\dot{n}^{(m)}} \,.
    \label{eq:relative-entropy-flux-expansion}
\end{align}
The first term on the right-hand side of Eq.~\eqref{eq:relative-entropy-flux-expansion} corresponds to the entropy exchanged between the membrane and the flux of mass~$\braket{\dot{n}^{(m)}}$ permeating through the membrane, while the second term on the right-hand side corresponds to the entropy change accompanying the rate of mass deposition~$\jump{-\dot{n}^{(m)}}$.

Note that although the jump terms in Eq.~\eqref{eq:local-membrane-entropy-balance} correspond to entropy production at the membrane due to interactions between the bulk fluid and the membrane, these terms are not directly a part of the inequality in Eq.~\eqref{eq:membrane-second-law-inequality}. Instead, we must identify the contributions to the internal entropy production~$\eta_i^{(m)}$, the external entropy production~$\eta_e^{(m)}$, and the entropy flux~$\Jb_s^{(m)}$. We do so in the following sections by introducing a thermodynamic potential for each continuum body and expanding the left-hand sides of Eq.~\eqref{eq:local-entropy-balance}.

\subsubsection{Entropy Production in the Fluid}\label{subsubsec:bulk-entropy-production}

Recall that the bulk fluid is a liquid-phase mixture consisting of $N$ uncharged chemical species. We will present the thermodynamics of this multi-component mixture in an analogous manner to \cite{fong2020transport}. The most natural choice of thermodynamic potential is the Helmholtz free energy per unit volume $\ft(T,c_1,\dots,c_N)$, which is given by 
\begin{equation}
    \ft := \rho u - \rho T s \,.
    \label{eq:bulk-Helmholtz-defn}
\end{equation}
Taking the material time derivative, and incorporating the mass balance~\eqref{eq:local-bulk-mass-balance} and the energy balance~\eqref{eq:local-bulk-energy-balance}, we can rearrange Eq.~\eqref{eq:bulk-Helmholtz-defn} to obtain
\begin{equation}
    \rho\dot{s} = \frac{1}{T}\left(\Tb : \grad\vb + \rho r - \divg \Jb_q - \rho\dot{T}s - \dot{\ft} - \ft\divg\vb \right) \,.
    \label{eq:entropy-production-bulk-1}
\end{equation}
Now assuming that we have local equilibrium at each point in the fluid, we can use the fundamental relation \cite{callen1960thermodynamics}
\begin{equation}
    d\ft = -\rho s \,dT + \sum_{i=1}^N \mu_i \,dc_i \,, \label{eq:bulk-Helmholtz-fundamental-relation}
\end{equation}
where $\mu_i$ is the chemical potential of species ``$i$''. Notice that Eq.~\eqref{eq:bulk-Helmholtz-fundamental-relation} is tantamount to defining  
\begin{equation}
    s := -\frac{1}{\rho}\left(\pder{\ft}{T}\right)_{c_j} \quad \text{and} \quad \mu_i := \left(\pder{\ft}{c_i}\right)_{T,c_{j \ne i}} \,.
    \label{eq:membrane-entropy-mu-defn}
\end{equation}
Using Euler's homogeneous function theorem, the local equilibrium assumption also gives the Euler equation 
\begin{equation}
    \ft = -p + \sum_{i=1}^N \mu_i c_i \,,
    \label{eq:bulk-Euler-equation}
\end{equation}
where $p$ is the thermodynamic pressure, as well as the Gibbs-Duhem equation
\begin{equation}
    0 = \rho s \,dT - dp + \sum_{i=1}^N c_i \,d\mu_i \,.
    \label{eq:Gibbs-Duhem-1}
\end{equation}

From Eq.~\eqref{eq:bulk-Helmholtz-fundamental-relation}, we have that the material time time derivative of the Helmholtz free energy is 
\begin{equation}
    \dot{\ft} = -\rho s \dot{T} + \sum_{i=1}^N \mu_i \dot{c}_i \,.
    \label{eq:bulk-Helmholtz-dot}
\end{equation}
Substituting Eq.~\eqref{eq:bulk-Helmholtz-dot} into Eq.~\eqref{eq:entropy-production-bulk-1}, incorporating the mole balance (Eq.~\eqref{eq:local-bulk-mole-balance-spec-i}), and then simplifying using Eq.~\eqref{eq:bulk-Euler-equation}, we have that the rate of entropy production is 
\begin{equation}
    \rho\dot{s} = \frac{1}{T}\left[\rho r + \left(\Tb + p\Ib\right):\grad\vb + \sum_{i=1}^N \left(\mu_i \cdot \divg\Jb_i\right) - \divg\Jb_q\right] \,. 
    \label{eq:entropy-production-bulk-2}
\end{equation}

Now, to resolve the entropy production (Eq.~\eqref{eq:entropy-production-bulk-2}) into the form given by the right-hand-side of Eq.~\eqref{eq:local-bulk-entropy-balance}, we rewrite Eq.~\eqref{eq:entropy-production-bulk-2} as 
\begin{equation}
     \rho\dot{s} = \left(\frac{\rho r}{T}\right) + \left[\frac{1}{T}(\taub:\grad\vb) - \sum_{i=1}^N \grad\bracfrac{\mu_i}{T} \cdot \Jb_i - \frac{\Jb_q \cdot \grad T}{T^2}\right] - \divg\left(\frac{\Jb_q - \sum_{i=1}^N \mu_i \Jb_i}{T}\right) \,,
     \label{eq:entropy-production-bulk-3}
\end{equation}
where we have defined the stress $\taub := \Tb + p\Ib$, with $p$ being the thermodynamic pressure. Comparing the terms in Eq.~\eqref{eq:entropy-production-bulk-3} with the terms in Eq.~\eqref{eq:local-bulk-entropy-balance}, we can identify the relations
\begin{align}
    \Jb_s &= \frac{\Jb_q - \sum_{i=1}^N \mu_i \Jb_i}{T} \,,
    \label{eq:Js-bulk}\\
    \rho\eta_{\text{ext}} &= \frac{\rho r}{T} \,,
    \label{eq:eta-e-bulk}\\
    \rho \eta_{\text{int}} &= \frac{1}{T}\left(\taub:\grad\vb\right) - \sum_{i=1}^N \grad\bracfrac{\mu_i}{T} \cdot \Jb_i - \frac{\Jb_q \cdot \grad T}{T^2} \,.
    \label{eq:eta-i-bulk}
\end{align}
Equation~\eqref{eq:eta-i-bulk} shows that irreversibilities in the bulk fluid arise from stress fluxes $\taub$ across velocity gradients $\grad\vb$, from species mass fluxes $\Jb_i$ across chemical potential gradients $\grad\bracfrac{\mu_i}{T}$, and from heat fluxes $\Jb_q$ across temperature gradients $\grad\bracfrac{1}{T}$.

\subsubsection{Entropy Production at the Membrane}
\label{subsubsec:membrane-entropy-production}

While entropy production in the bulk fluid takes place solely due to processes associated with the bulk fluid, entropy production at the membrane surface of discontinuity results from both bulk and membrane processes. 
Mathematically, this additional entropy production due to discontinuities in bulk quantities appears as the jump term in Eq.~\eqref{eq:local-membrane-entropy-balance}. 

In analogy to the bulk fluid in Section~\ref{subsubsec:bulk-entropy-production}, we choose the Helmholtz free energy per unit area of the membrane $\ft^{(m)}$ as the thermodynamic potential, which is defined as
\begin{align}
    \ft^{(m)} &:= \rho^{(m)}u^{(m)} - \rho^{(m)}Ts^{(m)} \,. \label{eq:membrane-Helmholtz-defn}
\end{align}
Note that the Helmholtz free energy per unit area is related to the Helmholtz free energy per unit mass $f^{(m)}$ through $\ft^{(m)} := \rho^{(m)} f^{(m)}$.
For simplicity, in what follows, we have assumed that the bulk fluid and the membrane are always in thermal equilibrium (i.e., $T = T^{(+)} = T^{(m)} = T^{(-)}$) so that the temperature field is continuous everywhere. 
Given the multicomponent nature of the membrane, and that work is required in to bend or stretch the membrane, 
in order to completely specify its thermodynamic state we must specify its: stretch and curvature, temperature, and chemical composition. 
This gives rise to a functional dependence of the form 
\begin{align}
    \ft^{(m)} &= \ft^{(m)}(a_{\alpha\beta}, b_{\alpha\beta}, T, c_1^{(m)}, \dots, c_N^{(m)}) \,,
    \label{eq:membrane-Helmholtz-functional-dependence}
\end{align}
with the understanding that $\ft^{(m)}$ is independent of the particular choice of coordinates system (i.e., $a_{\alpha\beta}$ and $b_{\alpha\beta}$ are paired with appropriate contravariant factors). 
Paralleling our assumption on the bulk fluid in Section~\ref{subsubsec:bulk-entropy-production}, we assume that the membrane satisfies local equilibrium. At equilibrium, Eq.~\eqref{eq:membrane-Helmholtz-functional-dependence} then allows us to define the surface chemical potential $\mu_i^{(m)}$ of species ``$i$'' in the membrane as
\begin{equation}
    \mu_i^{(m)} := \bracfrac{\partial  \ft^{(m)}}{\partial  c_i^{(m)}}_{a_{\alpha\beta},b_{\alpha\beta}, c_{j\ne i}^{(m)}} \,,
    \label{eq:membrane-chemical-potential-defn}
\end{equation}
and the entropy per unit volume of the membrane as
\begin{equation}
    \rho^{(m)} s^{(m)} := -\bracfrac{\partial  \ft^{(m)}}{\partial T}_{a_{\alpha\beta},b_{\alpha\beta},c_1^{(m)},\dots,c_k^{(m)}} \,. 
    \label{eq:membrane-entropy-defn}
\end{equation}

Taking the material time derivative, we can rearrange Eq.~\eqref{eq:membrane-Helmholtz-defn} as 
\begin{equation}
    \rho^{(m)} T \dot{s}^{(m)} = \dot{\rho}^{(m)} f^{(m)} + \rho^{(m)} \dot{u}^{(m)} - \rho^{(m)} \dot{T} s^{(m)} - \dot{\ft}^{(m)} \,.
    \label{eq:membrane-helmholtz-dot-0}
\end{equation} 
Additionally, taking the material time derivative of Eq.~\eqref{eq:membrane-Helmholtz-functional-dependence}, we have 
\begin{align}
    \dot{\ft}^{(m)} &= \frac{1}{2}\left(\frac{\partial \ft^{(m)}}{\partial a_{\alpha\beta}} + \frac{\partial \ft^{(m)}}{\partial a_{\beta\alpha}}\right) \dot{a}_{\alpha\beta} + \frac{1}{2}\left(\frac{\partial \ft^{(m)}}{\partial b_{\alpha\beta}} + \frac{\partial \ft^{(m)}}{\partial b_{\beta\alpha}}\right) \dot{b}_{\alpha\beta} + \frac{\partial  \ft^{(m)}}{\partial T} \dot{T} + \sum_{i=1}^N \mu_i^{(m)} \dot{c}_i^{(m)} \,. 
    \label{eq:membrane-helmholtz-dot-1}
\end{align}
In Eq.~\eqref{eq:membrane-helmholtz-dot-1}, it is important to symmetrize the derivative with respect to the symmetric tensors $a_{\alpha\beta}$ and $b_{\alpha\beta}$ as not all of their components can change independently. 
Using Eqs.~\eqref{eq:membrane-helmholtz-dot-1}, \eqref{eq:membrane-chemical-potential-defn}, and~\eqref{eq:membrane-entropy-defn}, Eq.~\eqref{eq:membrane-helmholtz-dot-0} leads to 
\begin{align}
\begin{split}
    \rho^{(m)} T \dot{s}^{(m)} &= \dot{\rho}^{(m)} f^{(m)} + \rho^{(m)}\dot{u}^{(m)} \\
    &\qquad - \left[ \frac{1}{2} \left(\frac{\partial \ft^{(m)}}{\partial a_{\alpha\beta}} + \frac{\partial \ft^{(m)}}{\partial a_{\beta\alpha}}\right) \dot{a}_{\alpha\beta} + \frac{1}{2} \left(\frac{\partial \ft^{(m)}}{\partial b_{\alpha\beta}} + \frac{\partial \ft^{(m)}}{\partial b_{\beta\alpha}}\right) \dot{b}_{\alpha\beta} + \sum_{i=1}^N \mu_i^{(m)} \dot{c}_i^{(m)}  \right]\,.
\end{split}
\end{align}
Substituting the local membrane species balance (Eq.~\eqref{eq:local-mole-balance-spec-i}), mass balance (Eq.~\eqref{eq:local-membrane-mass-balance}), and energy balance (Eq.~\eqref{eq:local-membrane-energy-balance}) into the above equation, and incorporating Eq.~\eqref{eq:J-dot} gives the rate of entropy production of the membrane as
\begin{align}
\begin{split}
    \rho^{(m)} T \dot{s}^{(m)} &= \rho^{(m)} r^{(m)} - \jump{\Jb_q}\cdot\nb - (J_q^{(m)})_{;\alpha}^{\alpha}
    + \sum_{i=1}^N \mu_i^{(m)} \left( (J_i^{(m)})^{\alpha}_{;\alpha} + \jump{c_i(\vb_i - \vb^{(m)})} \cdot \nb \right)
    \\
    & 
    \quad + \left[ \frac{1}{2}\sigma^{\alpha\beta} - \frac{1}{2} \left(\frac{\partial \ft^{(m)}}{\partial a_{\alpha\beta}} + \frac{\partial \ft^{(m)}}{\partial a_{\beta\alpha}}\right) + \frac{1}{2}\left(-\ft^{(m)} + \sum_{i=1}^N \mu_i^{(m)} c_i^{(m)}\right) a^{\alpha\beta}  \right]\dot{a}_{\alpha\beta} \\
    &\qquad + \left[ M^{\alpha\beta} - \frac{1}{2} \left(\frac{\partial \ft^{(m)}}{\partial b_{\alpha\beta}} + \frac{\partial \ft^{(m)}}{\partial b_{\beta\alpha}}\right) \right]\dot{b}_{\alpha\beta} \\
    &\quad\qquad - \jump{\rho(u - u^{(m)})(\vb - \vb^{(m)}) + \rho f^{(m)}(\vb - \vb^{(m)}) }\cdot\nb\\
    &\qquad\qquad + \jump{-\frac{1}{2}\rho||\vb - \vb^{(m)}||^2(\vb - \vb^{(m)}) + \Tb^T(\vb - \vb^{(m)})}\cdot\nb
    \,.
\end{split}
\label{eq:local-membrane-entropy-balance-2}
\end{align}%
In Eq.~\eqref{eq:local-membrane-entropy-balance-2}, we recognize the appearance of the surface tension $\gamma^{(m)}$ through the Euler relation~\eqref{eq:surface-Euler-relation} derived in Appendix~\ref{sec:thermodynamic-relations-curved-surfaces}. There, we also discuss the physical meaning of $\gamma^{(m)}$ by considering the Helmholtz free energy per unit mass $f^{(m)}$.

What remains to be done is to associate terms with the entropy balance postulate (Eq.~\eqref{eq:local-membrane-entropy-balance}). 
Equation~\eqref{eq:local-membrane-entropy-balance-2} gives an expression for the first term on the left-hand side of Eq.~\eqref{eq:local-membrane-entropy-balance}, while for the second term we can use the bulk entropy flux (Eq.~\eqref{eq:Js-bulk}) and the definition of the bulk Helmholtz free energy (Eq.~\eqref{eq:bulk-Helmholtz-defn}) to write 
\begin{multline}
    T\jump{\Jb_s + \rho(s - s^{(m)})(\vb - \vb^{(m)})}\cdot\nb 
    \ = \\
    = \jump{ \rho(u - u^{(m)})(\vb - \vb^{(m)}) - \left(\ft - \rho f^{(m)}\right)(\vb - \vb^{(m)})
    + \Jb_q - \sum_{i=1}^N \mu_i \Jb_i } \cdot \nb \,. 
    \label{eq:membrane-entropy-jump}
\end{multline}
Furthermore, decomposing the stress tensor as $\taub := \Tb + p\Ib$, and using the Euler equation (Eq.~\eqref{eq:bulk-Euler-equation}), we have the identity  
\begin{align}
    \jump{-\ft (\vb - \vb^{(m)}) + \Tb^T(\vb - \vb^{(m)})}\cdot\nb = \jump{-\left( \sum_{i=1}^N \mu_i c_i\right)(\vb - \vb^{(m)}) + \taub^T(\vb - \vb^{(m)})}\cdot\nb \,. 
    \label{eq:jump-relative-Helmholtz}
\end{align}
Adding Eqs.~\eqref{eq:local-membrane-entropy-balance-2} and~\eqref{eq:membrane-entropy-jump} together, incorporating Eqs.~\eqref{eq:jump-relative-Helmholtz} and~\eqref{eq:surface-Euler-relation}, and using the definitions of the mass density $\rho := \sum_i \rho_i = \sum_i \scrM_i c_i$ and the mass-averaged velocity $\vb := \sum_i (\rho_i/\rho)\vb_i$, we have
\begin{align}
\begin{split}
    &\rho^{(m)} \dot{s}^{(m)} + \jump{\Jb_s + \rho(s^{(m)} - s)(\vb^{(m)} - \vb)}\cdot\nb \\
    &= - \left[\frac{\left(J_q^{(m)}\right)^\alpha - \sum_{i=1}^N \mu_i^{(m)}  (J_i^{(m)})^{\alpha}}{T} \right]_{;\alpha} 
    + \frac{\rho^{(m)} r^{(m)}}{T}
    - \frac{(J_q^{(m)})^{\alpha}}{T^2} T_{,\alpha} - \sum_{i=1}^N \bracfrac{\mu_i^{(m)}}{T}_{,\alpha}  (J_i^{(m)})^{\alpha} \\
    &\quad + \frac{1}{T}\Bigg\{ \left[ \sigma^{\alpha\beta} - \left(\frac{\partial \ft^{(m)}}{\partial a_{\alpha\beta}} + \frac{\partial \ft^{(m)}}{\partial a_{\beta\alpha}}\right) - \gamma^{(m)}a^{\alpha\beta} \right]\frac{1}{2}\dot{a}_{\alpha\beta}
    \\ &\hspace{2in}
    + \left[ M^{\alpha\beta} - \frac{1}{2} \left(\frac{\partial \ft^{(m)}}{\partial b_{\alpha\beta}} + \frac{\partial \ft^{(m)}}{\partial b_{\beta\alpha}}\right) \right]\dot{b}_{\alpha\beta} \Bigg\} \\
    &\qquad+ \sum_{i=1}^N \frac{1}{T}\jump{-\left(\mu_i - \mu_i^{(m)} + \frac{1}{2}\scrM_i ||\vb - \vb^{(m)}||^2 \right) c_i(\vb_i - \vb^{(m)}) +\frac{\scrM_i}{\rho}\taub^T c_i(\vb_i - \vb^{(m)}) }\cdot\nb \,.
\end{split}
\label{eq:local-membrane-entropy-balance-4}
\end{align}
The jump terms in Eq.~\eqref{eq:local-membrane-entropy-balance-4} correspond to entropy produced due to both the permeation and accumulation of species ``$i$'' on the membrane. The irreversibilities that arise from species permeation can be attributed to internal entropy production at the membrane, while deposition can be viewed as external entropy supplied from the bulk fluid (within the bulk fluid, this will appear as entropy flux out of the system). In this article, we will assume that no accumulation takes place at the membrane so that all membrane-bulk interactions correspond to internal entropy production at the membrane.

A term-by-term comparison of Eq.~\eqref{eq:local-membrane-entropy-balance-4} leads us to identify
\begin{align}
    \Jb_s^{(m)} &= \frac{\Jb_q^{(m)} - \sum_{i=1}^N \mu_i^{(m)} \Jb_i^{(m)}}{T} \,, 
    \label{eq:Js-membrane}\\
    \rho^{(m)}\eta_{\text{ext}}^{(m)} &= \frac{\rho^{(m)} r^{(m)}}{T} \,, 
    \label{eq:eta-e-membrane}\\
    \begin{split}
    \rho^{(m)} \eta_{\text{int}}^{(m)} &= -\frac{(J_q^{(m)})^{\alpha}}{T^2} T_{,\alpha} - \sum_{i=1}^N \bracfrac{\mu_i^{(m)}}{T}_{,\alpha}  (J_i^{(m)})^{\alpha} \\
    &\ + \frac{1}{T}\Bigg\{\left[\sigma^{\alpha\beta} - \left(\frac{\partial \ft^{(m)}}{\partial a_{\alpha\beta}} + \frac{\partial \ft^{(m)}}{\partial a_{\beta\alpha}}\right) - \gamma^{(m)} a^{\alpha\beta} \right]\frac{1}{2}\dot{a}_{\alpha\beta}
    \\ &\hspace{2in}
    + \left[ M^{\alpha\beta} - \frac{1}{2} \left(\frac{\partial \ft^{(m)}}{\partial b_{\alpha\beta}} + \frac{\partial \ft^{(m)}}{\partial b_{\beta\alpha}}\right) \right]\dot{b}_{\alpha\beta} \Bigg\} \\
    &\quad + \sum_{i=1}^N \frac{1}{T}\jump{-\left(\mu_i - \mu_i^{(m)} + \frac{1}{2}\scrM_i ||\vb - \vb^{(m)}||^2 \right) c_i(\vb_i - \vb^{(m)}) +\frac{\scrM_i}{\rho}\taub^T c_i(\vb_i - \vb^{(m)}) }\cdot\nb \,.
    \end{split}
    \label{eq:eta-i-membrane}
\end{align}
Equation~\eqref{eq:eta-i-membrane} shows that all the irreversibilities that arise in previous work for a standalone membrane~\cite{sahu-mandadapu-pre-2017} also arise at a permeable and deforming membrane. These include irreversibilities arising from heat fluxes across temperature gradients, diffusive fluxes of a chemical species across chemical potential gradients, stress fluxes in the presence of areal changes, and moment fluxes in the presence of curvature changes. In addition, Eq.~\eqref{eq:eta-i-membrane} states that irreversibilities may also arise from chemical species permeation and accumulation across chemical potential jumps at the membrane, and chemical species permeation and slip in the presence of normal stress jumps $\jump{\taub\nb}$ at the membrane.

\section{Linear Irreversible Thermodynamics} \label{sec:linear-irrev-thermo}

In this section, we present a set of linear constitutive relations that describe osmotic transport across the membrane following the framework of linear irreversible thermodynamics~\cite{prigogine1961introduction,deGroot2013nonequilthermo}. 
To that end, we rewrite the rate of internal entropy production at the membrane as a contraction between thermodynamic forces and corresponding fluxes. This allows us to propose linear constitutive relations between these forces and fluxes. 
We begin by deriving some well-known constitutive laws for the bulk fluid, and proceed to develop the constitutive relations for the membrane. 
We end by establishing the general form of the constitutive relations for osmotic transport.

\subsection{Constitutive Laws for the Bulk Fluid} 
\label{subsec:bulk-constitutive-laws}

The rate of internal entropy production (Eq.~\eqref{eq:eta-i-bulk}) can be written as a contraction of thermodynamic fluxes $\Jb_\alpha$ and conjugate thermodynamic forces $\Xb_\alpha$, i.e., 
\begin{equation}
    \rho T \eta_\text{int} = \sum_\alpha \Jb_\alpha \cdot \Xb_\alpha \ge 0 \,,
    \label{eq:eta-int-flux-force}
\end{equation}
with the inequality following 
from the second law of thermodynamics. Here, the forces $\Xb_\alpha$ correspond to gradients in chemical potential, velocity, and temperature, which correspondingly drive the fluxes $\Jb_\alpha$ of species diffusion, stress, and heat. The most general linear relationship between a thermodynamic flux $\Jb_\alpha$ and a thermodynamic force $\Xb_\alpha$ that may be posed is 
\begin{equation}
    \Jb_\alpha = \sum_\beta \Lb^{\alpha\beta} \Xb_\beta \,,
    \label{eq:phenom-rels}
\end{equation}
where $\Lb^{\alpha\beta}$ are referred to as \textit{phenomenological transport coefficients}. The phenomenological coefficients for which $\alpha \ne \beta$ correspond to what is known as \textit{cross-phenomena}. 
In the absence of a magnetic field, the cross-phenomenological coefficients are symmetric via the Onsager reciprocal relations~\cite{onsager1931a,onsager1931b}, i.e., 
\begin{equation}
    \Lb^{\alpha\beta} = \Lb^{\beta\alpha} \,.
\end{equation}
Substituting Eq.~\eqref{eq:phenom-rels} into Eq.~\eqref{eq:eta-int-flux-force} yields 
\begin{equation}
    \rho T \eta_\text{int} = \sum_{\alpha} \sum_{\beta} \Xb_\alpha \cdot \Lb^{\alpha\beta} \Xb_\beta \ge 0 \,,
\end{equation}
so that the matrix of phenomenological coefficients $\pmb{\mathscr{L}} = \begin{bmatrix} \Lb^{\alpha\beta} \end{bmatrix}$ is a symmetric positive semi-definite matrix, placing the following restrictions 
\begin{equation}
    \trace\pmb{\mathscr{L}} \ge 0 \quad , \quad 
    \bm{1} \cdot \pmb{\mathscr{L}} \bm{1} \ge 0 \quad , \quad \text{and} \quad 
    \det\pmb{\mathscr{L}} \ge 0 \,,
    \label{eq:Ljk-PSD-consequences}
\end{equation}
where $\bm{1} := \begin{bmatrix} 1 & \cdots & 1\end{bmatrix}^T$ is a vector of all ones (with the same size as the number of rows of $\pmb{\mathscr{L}}$).
The application of this linear irreversible thermodynamic framework on fluids has already been carried out numerous times \cite{deGroot2013nonequilthermo,curtiss1999multicomponent}, but we will summarize it here again for multicomponent bulk solutions as posed in~\cite{fong2020transport}, as it will aid in illustrating the use of the framework for membranes in the next section.

We begin by noting that since $\sum_{i=1}^N \scrM_i \Jb_i = \zerob$, only $N-1$ of the $N$ diffusive fluxes $\Jb_i$ in Eq.~\eqref{eq:eta-i-bulk} are independent. Thus, there are only $N-1$ independent conjugate thermodynamic driving forces\footnote{To see this more clearly, first observe that we can write the Gibbs-Duhem relation (Eq.~\eqref{eq:Gibbs-Duhem-1}) as 
\begin{equation}
    \sum_{i=1}^N c_i T \,d\left(\frac{\mu_i}{T}\right) = -\frac{\Tilde{h}}{T} \,dT +\,dp  \,,
    \label{eq:Gibbs-Duhem-2}
\end{equation}
where $\Tilde{h} := \rho u + p$ is the enthalpy per unit volume. This implies that 
\begin{equation}
    \sum_{i=1}^N c_i T \grad\left(\frac{\mu_i}{T}\right) = -\frac{\Tilde{h}}{T} \grad T +\grad p  \,.
    \label{eq:Gibbs-Duhem-3}
\end{equation}
By considering $\grad T$ and $\grad p$ as independent driving forces, it is then sufficient to examine the remaining forces under isothermal and isobaric conditions. This yields $\sum_{i=1}^N c_i \grad\left(\frac{\mu_i}{T}\right) = 0$, which implies that $\grad\bracfrac{\mu_1}{T} = -\sum_{i=2}^N \frac{c_i}{c_1}\grad\bracfrac{\mu_i}{T}$, thus demonstrating that not all of the $\grad\bracfrac{\mu_i}{T}$ driving forces are independent.}.
Using $i=1$ to denote the major constituent in the fluid (i.e., the solvent), we have that its diffusive flux is $\scrM_1\Jb_1 = -\sum_{i \ne 1} \scrM_i\Jb_i$.
This reduces Eq.~\eqref{eq:eta-i-bulk} to
\begin{align}
    \rho \eta_{\text{int}} &= \frac{1}{T}\left(\taub:\grad\vb\right) - \sum_{i=2}^N \grad\left(\frac{\mu_i - \frac{\scrM_i}{\scrM_1} \mu_1}{T} \right)\cdot \Jb_i - \frac{\Jb_q \cdot \grad T}{T^2} \,.
    \label{eq:eta-i-bulk-2}
\end{align}
For an isotropic fluid, posing linear relationships between the thermodynamic forces and fluxes in Eq.~\eqref{eq:eta-i-bulk-2} then leads to~\cite{fong2020transport}:
\begin{gather}
    \Jb_q^{(h)} = \frac{L^{00}}{T} \grad T + \sum_{j=2}^N L^{0j} \left(\grad\bracfrac{\mu_i}{T} - \frac{\scrM_i}{\scrM_1} \grad\bracfrac{\mu_1}{T} \right) \,,\\
    \Jb_i = -\sum_{j=2}^N L^{ij} \left(\grad\bracfrac{\mu_i}{T} - \frac{\scrM_i}{\scrM_1} \grad\bracfrac{\mu_1}{T} \right) + \frac{L^{i0}}{T} \grad T \,,
    \label{eq:generalized-Ficks-law}\\
    \taub = \eta\left(\grad\vb + \grad\vb^T\right) + \left(\kappa - \frac{2}{3}\eta\right)\divg\vb \,,
    \label{eq:bulk-const-law-tau-1}
\end{gather}
where $L^{ij}$ are diffusive phenomenological coefficients, $\eta$ is the dynamic viscosity, and $\kappa$ is the dilatational viscosity, with the restriction from the second law of thermodynamics $\kappa,\eta \ge 0$. Note that the above constitutive equations are generally valid for concentrated systems. However, considering dilute solutions and assuming no cross phenomena and dilute concentrations, we may take $\grad \mu_i = \dfrac{k_B T}{c_i} \grad c_i$, $L^{ij} = 0$ ($j\neq i$), and $L^{ii} = \dfrac{D_i c_i}{k_B T}$ from the Green-Kubo relations~\cite{fong2020transport}, where $D_i$ is the self-diffusion coefficient and $k_B$ is the Boltzmann constant. The above constitutive relations for the bulk then simplify to 
\begin{gather}
    \Jb_q^{(h)} = -\kappa_\mathrm{T} \grad T \,,
    \label{eq:bulk-Fouriers-law} \\
    \Jb_i = -D_i \grad c_i \,,
    \label{eq:bulk-Ficks-law}\\
    \taub = \eta\left(\grad\vb + \grad\vb^T\right) + \left(\kappa - \frac{2}{3}\eta\right)\divg\vb \,,
    \label{eq:bulk-Newtons-law-of-viscosity}
\end{gather}
where $\kappa_\mathrm{T}$ is the thermal conductivity. Equations~\eqref{eq:bulk-Fouriers-law},~\eqref{eq:bulk-Ficks-law}, and~\eqref{eq:bulk-Newtons-law-of-viscosity} are the usual Fourier's law of conductivity, Fick's law of diffusion, and Newton's law of viscosity that appear in transport phenomena.

\subsection{Constitutive Laws for the Membrane}

We now develop constitutive laws governing the permeation of non-electrolyte solutes and solvents across mechanically deforming membranes. We will limit our discussion to cases where species from the bulk fluid don't accumulate on the membrane. The problem of simultaneous permeation and binding/unbinding of bulk chemical species can be addressed by incorporating chemical reactions into the balance laws developed in Section~\ref{sec:balance-laws}, but we will exclude them from the scope of this article. 

Recall from Eq.~\eqref{eq:local-membrane-mass-bal-spec-i-noacc} that 
the condition of no accumulation of species ``$i$'' on the membrane is equivalent to the jump condition on the relative mass flux $\jump{\dot{n}_i^{(m)}} = 0$. Since most processes involving biological membranes take place at low Reynolds number \cite{purcell1977life,salac2012reynolds,sahu2020geometry}, we may also neglect the jump in the relative kinetic energy-like terms between the membrane and the bulk fluid.
The internal entropy production at the membrane (Eq.~\eqref{eq:eta-i-membrane}) then simplifies to 
\begin{align}
    \begin{split}
    \rho^{(m)} T \eta_{\text{int}}^{(m)} &= -\frac{(J_q^{(m)})^{\alpha} T_{,\alpha}}{T} - T \sum_{i=1}^N \bracfrac{\mu_i^{(m)}}{T}_{,\alpha}  (J_i^{(m)})^{\alpha} \\
    &\ + \left[\sigma^{\alpha\beta} - \left(\frac{\partial \ft^{(m)}}{\partial a_{\alpha\beta}} + \frac{\partial \ft^{(m)}}{\partial a_{\beta\alpha}}\right) - \gamma^{(m)} a^{\alpha\beta} \right]\frac{1}{2}\dot{a}_{\alpha\beta}
    + \left[ M^{\alpha\beta} - \frac{1}{2} \left(\frac{\partial \ft^{(m)}}{\partial b_{\alpha\beta}} + \frac{\partial \ft^{(m)}}{\partial b_{\beta\alpha}}\right) \right]\dot{b}_{\alpha\beta}\\
    &\quad + \sum_{i=1}^N \left( \jump{- \frac{\mu_i}{\scrM_i} } \dot{n}_i^{(m)} + \jump{ \frac{\taub\nb}{\rho} \cdot \rho_i(\vb_i - \vb^{(m)}) } \right) \ge 0
    \,.
    \end{split}
    \label{eq:eta-i-membrane-2}
\end{align}

Since we are interested in the transport of species through the membrane, using
the three-dimensional identity tensor $\Ib = \ab_\alpha \otimes \ab^\alpha + \nb\otimes\nb$, we may further decompose the last term of Eq.~\eqref{eq:eta-i-membrane-2} into tangential and normal parts as  
\begin{align}
    \sum_{i=1}^N \jump{ \frac{\taub\nb}{\rho} \cdot \rho_i(\vb_i - \vb^{(m)}) } &= 
    \jump{ \bbP^{(m)} \taub\nb \cdot \bbP^{(m)}(\vb - \vb^{(m)})} + \sum_{i=1}^N \jump{ \frac{\nb \cdot \taub\nb}{\rho} } \dot{n}_i^{(m)} \,.
    \label{eq:stress-jump-decomp-2}
\end{align}
The tensor $\bbP^{(m)} := \ab_\alpha\otimes\ab^\alpha$ functions as both the identity tensor on the space of surface tensors, as well as the projection operator onto the tangent plane of the membrane surface, as it has the property $\bbP^{(m)} \bbP^{(m)} = \bbP^{(m)}$. Note that $\taub^{(\pm)}(\pm\nb)$ represents the deviatoric part of the traction acting on the membrane, and so the jump $\jump{\taub\cdot\nb} = \taub^{(+)}\nb + \taub^{(-)}(-\nb)$ is actually the net traction due to deviatoric stresses acting from both sides of the membrane. Incorporating Eq.~\eqref{eq:stress-jump-decomp-2} into Eq.~\eqref{eq:eta-i-membrane-2} gives the final form for entropy production at the membrane:
\begin{align}
    \begin{split}
    \rho^{(m)} T \eta_{\text{int}}^{(m)} &= 
    -\frac{(J_q^{(m)})^{\alpha} T_{,\alpha}}{T} - T \sum_{i=1}^N \bracfrac{\mu_i^{(m)}}{T}_{,\alpha}  (J_i^{(m)})^{\alpha} \\
    &\quad+ \left[\sigma^{\alpha\beta} - \left(\frac{\partial \ft^{(m)}}{\partial a_{\alpha\beta}} + \frac{\partial \ft^{(m)}}{\partial a_{\beta\alpha}}\right) - \gamma^{(m)} a^{\alpha\beta} \right]\frac{1}{2}\dot{a}_{\alpha\beta}
    \\ &\qquad\quad
    + \left[ M^{\alpha\beta} - \frac{1}{2} \left(\frac{\partial \ft^{(m)}}{\partial b_{\alpha\beta}} + \frac{\partial \ft^{(m)}}{\partial b_{\beta\alpha}}\right) \right]\dot{b}_{\alpha\beta} \\
    &\qquad\qquad + \jump{ \bbP^{(m)} \taub\nb \cdot \bbP^{(m)}(\vb - \vb^{(m)})} + \sum_{i=1}^N \jump{- \frac{\mu_i}{\scrM_i} + \frac{\nb \cdot \taub\nb}{\rho} } \dot{n}_i^{(m)} \ge 0
    \,.
    \end{split}
    \label{eq:eta-i-membrane-3}
\end{align}
In the absence of the jump terms, Eq.~\eqref{eq:eta-i-membrane-3} recovers the same result as previous work on arbitrarily curved and deforming viscous membranes \cite{sahu-mandadapu-pre-2017}. 
The first jump term in Eq.~\eqref{eq:eta-i-membrane-3} corresponds to tangential slip of the bulk fluid on the membrane, and the remaining jump terms correspond to individual species permeation through the membrane.
To the best of the authors' knowledge, among the jump terms in Eq.~\eqref{eq:eta-i-membrane-3}, only the chemical potential jump $\jump{-\frac{\mu_i}{\scrM_i}}$ has been previously considered as a driving force for the flux $\dot{n}_i^{(m)}$ of a solute across a selectively permeable membrane~\cite{kedem1958thermodynamic}. Herein, we find that the jump term $\jump{\frac{\nb \cdot \taub\nb}{\rho} }$ contributes as an additional driving force for osmosis and permeability.

In what follows, we develop constitutive laws for the membrane stresses~$\sigma^{\alpha\beta}$ and couple-stresses~$M^{\alpha\beta}$, as well as permeability and slip laws for a membrane with no deposition from the bulk fluid. In doing so, we neglect the effects of cross-phenomena, and only couple thermodynamic fluxes with their conjugate thermodynamic driving forces.

\subsubsection{Stresses and Helmholtz Free Energy}

From Eq.~\eqref{eq:eta-i-membrane-3}, we identify the thermodynamic fluxes conjugate to the areal contractions/dilations $\dot{a}_{\alpha\beta}$ and the rate of change of curvature $\dot{b}_{\alpha\beta}$ as 
\begin{align}
    \pi^{\alpha\beta} &:= \sigma^{\alpha\beta} - \left(\frac{\partial \ft^{(m)}}{\partial a_{\alpha\beta}} + \frac{\partial \ft^{(m)}}{\partial a_{\beta\alpha}}\right) - \gamma^{(m)} a^{\alpha\beta} \,,
    \label{eq:pi-alpha-beta-defn}\\
    \omega^{\alpha\beta} &:= M^{\alpha\beta} - \frac{1}{2} \left(\frac{\partial \ft^{(m)}}{\partial b_{\alpha\beta}} + \frac{\partial \ft^{(m)}}{\partial b_{\beta\alpha}}\right) \,,
    \label{eq:omega-alpha-beta-defn}
\end{align}
which, respectively, correspond to in-plane and out-of-plane viscous dissipation.
In what follows, we assume that we have perfectly elastic bending, i.e., $\omega^{\alpha\beta} \equiv 0$. Following Ref.~\cite{sahu-mandadapu-pre-2017}, posing a linear relationship between $\pi^{\alpha\beta}$ and $\dot{a}_{\alpha\beta}$ for an isotropic membrane (with no cross-coupling to the bending forces $\dot{b}_{\alpha\beta}$) leads to 
\begin{align}
    \pi^{\alpha\beta} &= \zeta a^{\alpha\gamma} a^{\beta\mu} \dot{a}_{\gamma\mu} + \lambda a^{\alpha\beta} \divg_\text{S}\vb^{(m)} \,,
    \label{eq:pi-ab-constitutive-law}
\end{align}
where $\zeta$ is the membrane viscosity, and $\lambda$ is the membrane dilatational viscosity. To compute the membrane stresses, it remains to specify the Helmholtz free energy $\ft^{(m)}$, from which the membrane stresses and moments can be computed using Eqs.~\eqref{eq:pi-alpha-beta-defn} and~\eqref{eq:omega-alpha-beta-defn} as 
\begin{align}
    \sigma^{\alpha\beta} &= \left(\frac{\partial \ft^{(m)}}{\partial a_{\alpha\beta}} + \frac{\partial \ft^{(m)}}{\partial a_{\beta\alpha}}\right) + \gamma^{(m)} a^{\alpha\beta} + \pi^{\alpha\beta} \,,
    \label{eq:sigma-ab-constitutive-law}\\
    M^{\alpha\beta} &= \frac{1}{2} \left(\frac{\partial \ft^{(m)}}{\partial b_{\alpha\beta}} + \frac{\partial \ft^{(m)}}{\partial b_{\beta\alpha}}\right) \,.
    \label{eq:M-ab-constitutive-law}
\end{align}

Let us consider the functional form of the Helmholtz free energy per unit area $\ft^{(m)}$ for the membrane given by Eq.~\eqref{eq:membrane-Helmholtz-functional-dependence}. 
Since $\ft^{(m)}$ is a scalar field it must be invariant to Galilean transformations, and so cannot assume arbitrary functional dependence on the (coordinate system dependent) components of the metric and curvature tensors. For a lipid bilayer membrane with curvature elasticity, we will assume that the energy is dependent only on the invariants of the curvature tensor, whence $\ft^{(m)}$ is expressible as~\cite{Steigmann1999}
\begin{align}
    \ft^{(m)}(a_{\alpha\beta}, b_{\alpha\beta}, T, c_1^{(m)}, \dots, c_N^{(m)})
    &= \tilde{\mathfrak{f}}^{(m)}(H,K,T,c_1^{(m)}, \dots, c_N^{(m)})
    \,.
    \label{eq:Helmholtz-change-of-variables}
\end{align}
We will use the Canham-Helfrich model \cite{canham1970minimum,helfrich1973elastic} to capture the curvature elasticity of the lipid membrane, 
for which the curvature free energy density (per unit area) $w_h$ takes the form
\begin{equation}
    w_h = k_b(H - C)^2 + k_g K \,,
\end{equation}
where $k_b$ is the bending modulus, $C$ is the spontaneous curvature that the membrane prefers to conform to (e.g., due to the presence of BAR domain proteins~\cite{simunovic2015physics}), and $k_g$ is the saddle splay modulus. 
With regards to the energetic cost of compression/dilation of the membrane surface, we follow the arguments of~\cite[Chapter 6.3]{safran2018statistical}. A locally flat membrane will prefer to saturate at some area area per lipid $1/c_0$ that minimizes the free energy. 
The membrane density may fluctuate about this local minimum with an energetic cost (per unit area) $w_c$, which we expand to second order as\footnote{In Eq.~\eqref{eq:membrane-compression-free-energy}, we multiply by a factor of $c$ in order to convert the free energy per unit lipid into a free energy per unit area.}
\begin{align}
    w_c &= k_c\, c c_0 \left(\frac{1}{c} - \frac{1}{c_0} \right)^2 \,,
    \label{eq:membrane-compression-free-energy}
\end{align}
where $k_c$ is the compression modulus. Altogether, we posit that the Helmholtz free energy per unit area is given by 
\begin{align}
    \tilde{\mathfrak{f}}^{(m)} &= w_h + w_c + w_s(c_1,\dots,c_N) + g(T)\,,
    \label{eq:membrane-Helmholtz-general-form}
\end{align}
where $w_s(c_1,\dots,c_N)$ is a model for the change in free energy due to the presence of each chemical species $c_i$ (and determines the chemical potential $\mu_i^{(m)}$), and $g(T)$ is a temperature-dependent function that determines the entropy of the membrane (see Eq.~\eqref{eq:membrane-entropy-mu-defn}). 
The species-dependent free energy $w_s$ is required in order to capture phenomena such as phase separation~\cite{baumgart2003imaging,veatch2003separation,zimmermann2019isogeometric} and curvature induced by protein binding~\cite{stachowiak2012membrane,brannigan2007contributions,tribet2008flexible}.
In this work, we will assume that we have a single-component lipid membrane with no deposition from the bulk, in which case we neglect the contribution from the $w_s$ term in Eq.~\eqref{eq:membrane-Helmholtz-general-form}. The Helmholtz free energy of the membrane then takes the form 
\begin{align}
    \tilde{\mathfrak{f}}^{(m)} &= k_b(H - C)^2 + k_g K + k_c\, c c_0 \left(\frac{1}{c} - \frac{1}{c_0} \right)^2 + g(T) \,.
    \label{eq:membrane-Helmholtz-free-energy-final-form}
\end{align}

Equation~\eqref{eq:membrane-Helmholtz-free-energy-final-form} can be used to compute the membrane stresses and moments in Eqs.~\eqref{eq:sigma-ab-constitutive-law} and~\eqref{eq:M-ab-constitutive-law}. To that end,
the partial derivatives $\dfrac{\partial \ft^{(m)}}{\partial a_{\alpha\beta}}$ and $\dfrac{\partial \ft^{(m)}}{\partial b_{\alpha\beta}}$ can be expressed using the chain rule on Eq.~\eqref{eq:Helmholtz-change-of-variables} to obtain
\begin{align}
    \frac{\partial \ft^{(m)}}{\partial a_{\alpha\beta}} 
    &= \tilde{\mathfrak{f}}^{(m)}_{,H} \frac{\partial H}{\partial a_{\alpha\beta}} + \tilde{\mathfrak{f}}^{(m)}_{,K} \frac{\partial K}{\partial a_{\alpha\beta}}
    = -\frac{1}{2} \tilde{\mathfrak{f}}^{(m)}_{,H} b^{\alpha\beta} - K \tilde{\mathfrak{f}}^{(m)}_{,K} a^{\alpha\beta} \,,
    \label{eq:f-metric-chain-rule}\\
    \frac{\partial \ft^{(m)}}{\partial b_{\alpha\beta}}
    &= \tilde{\mathfrak{f}}^{(m)}_{,H} \frac{\partial H}{\partial b_{\alpha\beta}} + \tilde{\mathfrak{f}}^{(m)}_{,K} \frac{\partial K}{\partial b_{\alpha\beta}}
    = \frac{1}{2} \tilde{\mathfrak{f}}^{(m)}_{,H} a^{\alpha\beta} + \tilde{\mathfrak{f}}^{(m)}_{,K} \Bar{b}^{\alpha\beta} \,,
    \label{eq:f-curvature-chain-rule}
\end{align}
where $(\bullet)_{,H}$ and $(\bullet)_{,K}$ denote partial differentiation with respect to $H$ and $K$, and $\Bar{b}^{\alpha\beta}$ is the cofactor of curvature, which may be calculated from the Cayley-Hamilton theorem as $\Bar{b}^{\alpha\beta} = 2H a^{\alpha\beta} - b^{\alpha\beta}$.
With the aid of Eqs.~\eqref{eq:f-metric-chain-rule}, \eqref{eq:f-curvature-chain-rule}, and~\eqref{eq:surface-Euler-relation}, 
we then have 
\begin{align}
    \begin{split}
        \sigma^{\alpha\beta} &= k_b \left[ (-3H^2 + 2HC + C^2) a^{\alpha\beta} + 2(H - C) \Bar{b}^{\alpha\beta} \right] \\
        &\qquad - k_g K a^{\alpha\beta} + 2k_c \left( \frac{c_0}{c} - 1 \right)a^{\alpha\beta} + \pi^{\alpha\beta} \,,
    \end{split}
    \\
    M^{\alpha\beta} &= k_b (H - C) a^{\alpha\beta} + k_g \Bar{b}^{\alpha\beta} \,,
\end{align}
where the viscous stresses $\pi^{\alpha\beta}$ are given by Eq.~\eqref{eq:pi-ab-constitutive-law}. Note that in the case where there is no deposition of mass from the bulk onto the membrane, the change in concentration is related to the areal change of the membrane through $J = \dfrac{c_0}{c}$, in which case the above stresses reduce to the result found in Ref.~\cite{sahu-mandadapu-pre-2017}.

\subsubsection{Incompressible Membranes}

Since lipid bilayer membranes can only tolerate an areal strain of $4$--$6$\% before rupturing \cite{janshoff2015mechanics}, they are often treated as area incompressible. When combined with the assumption of no mass deposition on the membrane, incompressibility is equivalent to the condition that the molar density of the membrane $c$ remains constant at some reference density $c_0$. We incorporate incompressibility into the membrane model by appending a Lagrange multiplier field $\lambda^{(m)}$ that enforces the constraint $\dfrac{c}{c_0}= 1$ on the membrane Helmholtz free energy $\tilde{\mathfrak{f}}^{(m)}$ in Eq.~\eqref{eq:membrane-Helmholtz-general-form} in lieu of the compressive energy density $w_c$. This yields the Helmholtz free energy
\begin{align}
	\tilde{\mathfrak{f}}^{(m)} &= w_h(H,K) + w_s(c_1,\dots,c_N) + g(T) + \lambda^{(m)} \cdot \left(\frac{c}{c_0} - 1 \right)
	\,.
	\label{eq:membrane-Helmholtz-incompressible-form}
\end{align}

For a single-component membrane, we drop the dependence of $\tilde{\mathfrak{f}}^{(m)} $ on $w_s(c_1,\dots,c_N)$ in Eq.~\eqref{eq:membrane-Helmholtz-incompressible-form}. From Eqs.~\eqref{eq:surface-Euler-relation} and~\eqref{eq:membrane-chemical-potential-defn}, the surface tension can then be calculated as 
\begin{align}
	\gamma^{(m)} &= -c^2 (\tilde{\mathfrak{f}}^{(m)}/c)_{,c}  \,,
\end{align}
where $(\bullet)_{,c}$ denotes partial differentiation with respect to $c$. In the case of Eq.~\eqref{eq:membrane-Helmholtz-incompressible-form} for a single-component membrane, this yields
\begin{align}
	\gamma^{(m)} &= w_h(H,K) + g(T) + \lambda^{(m)} \,,
	\label{eq:incompressible-surface-tension}
\end{align}
so that $\lambda^{(m)}$ represents an additional contribution to the surface tension that acts to keep the density of the membrane constant. 
Since we are considering isothermal systems, the contribution from $g(T)$ to both the free energy and the surface tension is constant, and so we may absorb it into the Lagrange multiplier $\lambda^{(m)}$. 
Substituting Eq.~\eqref{eq:membrane-Helmholtz-incompressible-form} into Eqs.~\eqref{eq:sigma-ab-constitutive-law} and~\eqref{eq:M-ab-constitutive-law}, and incorporating Eq.~\eqref{eq:incompressible-surface-tension} yields the membrane stresses and moments
\begin{align}
	\sigma^{\alpha\beta} &= k_b \left[ (-3H^2 + 2HC + C^2) a^{\alpha\beta} + 2(H - C) \Bar{b}^{\alpha\beta} \right]
    - k_g K a^{\alpha\beta} + \lambda^{(m)} a^{\alpha\beta} + \pi^{\alpha\beta} \,,
    \label{eq:sigma-alpha-beta-incompressible}
	\\
	M^{\alpha\beta} &= k_b (H - C) a^{\alpha\beta} + k_g \Bar{b}^{\alpha\beta} \,.
    \label{eq:M-alpha-beta-incompressible}
\end{align}

\subsubsection{Permeability and Slip Laws}
\label{subsec:permeability-and-slip-laws}

We now proceed to develop constitutive laws by proposing linear relations between the thermodynamic forces and fluxes corresponding to the entropy production at the membrane. Beginning with linear laws between the flux of solute and its conjugate thermodynamic driving force, we have the generalized constitutive relationship for permeability 
\begin{equation}
    \dot{N}_i^{(m)} = \sum_{j=1}^N L^{ij} \jump{-{\mu_j} + {\scrM_j}\bracfrac{\nb \cdot \taub\nb}{\rho} } \,,
    \label{eq:general-permeability-linear-law}
\end{equation}
where $L^{ij}$ are phenomenological permeability coefficients\footnote{Here, we reuse the same symbols $L^{ij}$ for the phenomenological coefficients as those that appear for the bulk fluid in Eq.~\eqref{eq:phenom-rels}. 
}.
Since the entropy production (Eq.~\eqref{eq:eta-i-membrane-3}) is non-negative by the second law of thermodynamics, the conditions in Eq.~\eqref{eq:Ljk-PSD-consequences} also apply to the coefficients $L^{ij}$ in Eq.~\eqref{eq:general-permeability-linear-law}.
At this stage, it is useful to express the chemical potential in terms of the activity $a_i$, 
which is defined via the relation~\cite{koretsky2012engineering}
\begin{equation}
    \mu_i - \mu_i^\circ = RT \ln a_i \,,
    \label{eq:activity-defn}
\end{equation}
where $\mu_i^\circ$ is the chemical potential of species ``$i$'' at a reference state that depends only on the temperature $T$. Taking the temperature $T$, pressure $p$, and composition $x_i$ as independent variables, we can write the total differential of the chemical potential $\mu_i = \mu_i(T,p,x_j)$ using the Maxwell relations from equilibrium thermodynamics~\cite{koretsky2012engineering} as
\begin{equation}
    T \,d\left(\frac{\mu_i}{T}\right) = -\frac{1}{T}\overline{H}_i \,dT + \overline{V}_i \,dp + \left(\pder{\mu_i}{x_j}\right)_{T,p,x_{j \ne i}} dx_j \,,
    \label{eq:chemical-potential-differential-form-1}
\end{equation}
where $\overline{H}_i$ is the partial molar enthalpy and $\overline{V}_i$ is the partial molar volume. Taking the reference state for the chemical potential to be independent of composition, we can incorporate Eq.~\eqref{eq:activity-defn} into Eq.~\eqref{eq:chemical-potential-differential-form-1} to obtain
\begin{equation}
    T \,d\left(\frac{\mu_i}{T}\right) = -\frac{1}{T}\overline{H}_i \,dT + \overline{V}_i \,dp + RT (\,d\ln a_i)_{T,p} \,,
    \label{eq:activity-defn-differential-form-3}
\end{equation}
where the notation $(d\ln a_i)_{T,p}$ indicates that the differential is taken at constant $T$ and $p$\footnote{We can expand $\displaystyle (d\ln a_i)_{T,p} = \sum_{i=1}^N \frac{1}{x_j} \left(\frac{\partial \ln a_i}{\partial \ln x_j}\right)_{T,p,x_{k \ne j}} \,dx_j$. Note that although only $N-1$ of the $N$ compositions $x_j$ are independent, the functional expression for $a_i$ may still involve all of the $x_j$. Upon integrating this expression, however, one needs to ensure that the constraint $\displaystyle \sum_j x_j = 1$ is satisfied.}.
Since we assume the temperature $T$ to be continuous across the membrane, Eq.~\eqref{eq:activity-defn-differential-form-3} simplifies to 
\begin{equation}
    d\mu_i = \overline{V}_i \,dp + RT \,(d\ln a_i)_{T,p} \,.
    \label{eq:const-T-mui}
\end{equation}
Assuming that the partial molar volume $\overline{V}_i$ is constant over our pressure range of interest, we can integrate Eq.~\eqref{eq:const-T-mui} from the $(-)$ state to the $(+)$ state to obtain 
\begin{equation}
    \jump{\mu_i} = \overline{V}_i \jump{p} + RT \jump{\ln a_i} \,.
    \label{eq:const-T-mui-jump}
\end{equation}
The activity $a_i$ can be calculated using an activity coefficient model $\gamma_i$, with the definition 
\begin{equation}
    a_i = x_i \gamma_i \,,
    \label{eq:activity-coefficient-defn}
\end{equation}
where the reference state for the activity is taken to be pure species ``$i$'' at the saturation pressure (i.e., the Lewis-Randall reference state~\cite{koretsky2012engineering}).
This representation is useful in many instances where the activity coefficient does not change significantly for small perturbations of the mole fraction $x_i$, and is typically~$\approx 1$ for ideal conditions. 
With Eqs.~\eqref{eq:const-T-mui-jump} and~\eqref{eq:activity-coefficient-defn}, we obtain  
an alternate form for the permeability law Eq.~\eqref{eq:general-permeability-linear-law} as 
\begin{equation}
    \dot{N}_i^{(m)} = \sum_{j=1}^N L^{ij} \left( \overline{V}_j \jump{-p} - RT \jump{\ln(x_j \gamma_j)} + \scrM_j\jump{ \frac{\nb \cdot \taub\nb}{\rho} } \right) \,,
    \label{eq:general-permeability-linear-law-2}
\end{equation}
indicating that the solute flux is driven by differences in concentration, pressure, and bulk stresses.

A no-slip assumption between the lipid bilayer and the surrounding bulk fluid is somewhat ubiquitous in the literature \cite{Seifert1999,Jahl2020}, but experiments have shown that slip phenomena can occur at lipid-water interfaces in the context of
supported lipid bilayers, such as fluid DOPC (dioleoylphosphatidylcholine) bilayers \cite{leroy2009probing,Olsen2021}. 
In such cases, the form of the entropy production in Eq.~\eqref{eq:eta-i-membrane-3} allows us to propose slip laws as linear constitutive relations between the relative tangential velocities and their conjugate in-plane tractions.
These tangential slip laws for both sides of the membrane can be posed as
\begin{equation}
    \bbP^{(m)}(\vb^{(\pm)} - \vb^{(m)}) = \pm \ \betab_\text{sl}^{(\pm)} \bbP^{(m)} \left(\taub^{(\pm)} \nb\right) \,,
    \label{eq:slip-law-1}
\end{equation}
where $\betab_\text{sl}^{(\pm)}$ is a second-order tensor of phenomenological slip coefficients. For an isotropic system, the tensor of slip coefficients can be represented as $\betab_\text{sl}^{(\pm)} = \beta_\text{sl}^{(\pm)} \bbP^{(m)}$ for some scalar $\beta_\text{sl}^{(\pm)}$. Equation~\eqref{eq:slip-law-1} written in local coordinates then becomes 
\begin{equation}
    \ab^\alpha \cdot (\vb^{(\pm)} - \vb^{(m)}) = \pm \beta_\text{sl}^{(\pm)} \ab^{\alpha} \cdot (\taub^{(\pm)}\nb) \,,
    \label{eq:slip-law-2}
\end{equation}
which is the well-known Navier-slip condition, and has been utilized to understand slip in high molecular weight fluids, particulate suspensions, and flows with large tangential stresses \cite{leal2007advanced,Lauga2007,Fraggedakis2016}. The coefficient $\beta_\text{sl}$ controls the ``slip length'' of the fluid---as for a Newtonian bulk fluid with viscosity $\eta$, the quantity $\eta \beta_\text{sl}$ represents the fictitious length below a planar surface under pure shear where the velocity profile becomes zero~\cite{Lauga2007}. Notice that when $\beta_\text{sl} = 0$, Eq.~\eqref{eq:slip-law-2} becomes the usual no-slip boundary condition~\cite{narsimhan2015pearling}.

\section{Special Cases for the Permeability Laws} 
\label{sec:special-cases-for-permeability}

In this section, we simplify the general constitutive relations developed in Section~\ref{subsec:permeability-and-slip-laws} to a few specific scenarios of membrane and bulk systems. In particular, we explore binary ideal systems, ideally selective membranes, and the case of pure mechanical permeability. In the case of negligible deviatoric stresses, we also recover the results of Kedem and Katchalsky~\cite{kedem1958thermodynamic}, and provide a linear map between our phenomenological coefficients and those of Kedem and Katchalsky.

\subsection{Binary Ideal System}
\label{subsec:binary-ideal-system}

Consider a two-component solution made up of a solvent `$w$' (typically aqueous water), and a solute `$s$'. Additionally, suppose that the solute is dilute and the solution is ideal.
By definition, an ideal solution is one for which the activity coefficient of each species $\gamma_i \approx 1$. Equation~\eqref{eq:const-T-mui-jump}, for each species, then leads to 
\begin{align}
    \jump{\mu_s} &= \overline{V}_s \jump{p} + RT\jump{\ln x_s} \label{eq:mu-s-jump-1} \,,\\
    \jump{\mu_w} &= \overline{V}_w \jump{p} + RT\jump{\ln x_w} \label{eq:mu-w-jump-1} \,.
\end{align}
Defining the volume fraction of each species `$i$' as 
\begin{equation}
    \varphi_i := c_i \overline{V}_i 
    \,,
\end{equation}
with $\overline{V}_i$ being its partial molar volume, 
we consider the component `$s$' to be dilute when $\varphi_s \ll 1$.
In this case, the total solution concentration $c \approx \overline{V}_w^{-1} \approx c_w$, and the jump in the solute mole fraction is
\begin{align}
    \jump{\ln x_s} &
    \approx \jump{\ln\left( c_s \overline{V}_w \right)} 
    = \jump{\ln c_s} = \frac{\jump{c_s}}{\LMavg{c_s}}\,,
    \label{eq:jump-lnxs-1}
\end{align}
where $\LMavg{c_i} := \dfrac{\jump{c_i}}{\jump{\ln c_i}}$ is the log-mean concentration difference.
For the solvent mole fraction, a Taylor expansion gives
\begin{align}
    \jump{\ln x_w} &= \jump{\ln(1 - x_s)} \approx -\jump{x_s} \approx -\jump{c_s}\overline{V}_w \,.
    \label{eq:jump-lnxw-1}
\end{align}
Incorporating the approximations in Eqs.~\eqref{eq:jump-lnxs-1} and~\eqref{eq:jump-lnxw-1} into Eqs.~\eqref{eq:mu-s-jump-1} and~\eqref{eq:mu-w-jump-1} yields 
\begin{align}
    \jump{\mu_s} &= \overline{V}_s \jump{p} + RT\frac{\jump{c_s}}{\LMavg{c_s}} \label{eq:mu-s-jump-2} \,,\\
    \jump{\mu_w} &= \overline{V}_w \jump{p} - RT\jump{c_s}\overline{V}_w \label{eq:mu-w-jump-2} \,.
\end{align}
With these approximations, the linear laws (Eq.~\eqref{eq:general-permeability-linear-law}) become 
\begin{align}
    \begin{split}
    \dot{N}_s^{(m)} & = \left( L^{ss}\overline{V}_s +  L^{sw}\overline{V}_w \right) \jump{-p} + \left( \frac{L^{ss}}{\LMavg{c_s}} - L^{sw}\overline{V}_w \right) RT\jump{-c_s}\\
    &\hspace{0.5in}  + \left( L^{ss} \scrM_s + L^{sw} \scrM_w \right) \jump{ \frac{\nb \cdot \taub\nb}{\rho} }\,,
    \end{split} \label{eq:binary-law-s} 
    \\
    \begin{split}
    \dot{N}_w^{(m)} &= \left( L^{sw}\overline{V}_s + L^{ww}\overline{V}_w \right) \jump{-p} + \left( \frac{L^{sw}}{\LMavg{c_s}} - L^{ww}\overline{V}_w \right) RT\jump{-c_s}\\
    &\hspace{0.5in} + \left( L^{sw} \scrM_s + L^{ww} \scrM_w \right) \jump{ \frac{\nb \cdot \taub\nb}{\rho} }
    \,,
    \end{split} \label{eq:binary-law-w} 
\end{align}
where we've applied Onsager Reciprocity to equate the ``cross-species'' phenomenological coefficients, i.e.,
\begin{equation}
    L^{sw} = L^{ws} \,. 
    \label{eq:membrane-Onsager-reciprocity}
\end{equation}

\subsection{The Normal Deviatoric Stress for a Newtonian Fluid}
\label{subsec:normal-deviatoric-stress}

We analyze the driving forces arising from the jump term $\jump{ \dfrac{\nb \cdot \taub\nb}{\rho} }$ for the special case when the outer and inner fluids behave as incompressible Newtonian fluids with viscosities $\eta^{(+)}$ and $\eta^{(-)}$, respectively.
With both the bulk mass density $\rho$ and the membrane mass density $\rho^{(m)}$ constant, the mass balances \eqref{eq:local-bulk-mass-balance} and \eqref{eq:local-membrane-mass-balance} simplify to 
\begin{subequations}
\label{eq:incompressible-mass-balance}
\begin{align}
    \divg\vb &= 0 \,, 
    \label{eq:incompressible-bulk-mass-balance}\\
    \divg_\text{S}\vb^{(m)}  &= 0 \,,
    \label{eq:incompressible-membrane-mass-balance}
\end{align}
\end{subequations}
where we've used the no accumulation condition $\jump{\rho(\vb - \vb^{(m)})} \cdot \nb = 0$. Expanding the divergence operator in Eq.~\eqref{eq:incompressible-bulk-mass-balance} into a surface and normal part as
\begin{align}
    \divg\vb &= \Ib : \grad\vb = (\bbP^{(m)} + \nb\otimes\nb) : \grad \vb = \divg_\text{S}\vb + \nb \cdot (\grad\vb) \nb \,,
    \label{eq:div-v-intermediate}
\end{align}
and using Eq.~\eqref{eq:incompressible-membrane-mass-balance},  we have  
\begin{align}
    \nb \cdot (\grad\vb) \nb &= -\divg_\text{S} ( \vb - \vb^{(m)} )\\
    &= 2H (\vb - \vb^{(m)}) \cdot \nb - \divg_\text{S} \cdot \bbP^{(m)} (\vb - \vb^{(m)}) \,.
    \label{eq:dvdn}
\end{align}
The deviatoric stress term $\dfrac{\nb\cdot\taub\nb}{\rho}$ for an incompressible Newtonian fluid from Eq.~\eqref{eq:bulk-const-law-tau-1} is 
\begin{equation}
    \frac{\nb \cdot \taub\nb}{\rho} = \frac{2\eta}{\rho} \left[ \nb \cdot (\grad\vb)\nb\right] \,.
    \label{eq:ntaun-local-coords}
\end{equation}
Combining Eqs.~\eqref{eq:dvdn} and ~\eqref{eq:ntaun-local-coords} and evaluating the jump across the membrane, we then have  
\begin{equation}
    \jump{\frac{\nb \cdot \taub\nb}{\rho}} = 4H \jump{\frac{\eta}{\rho} (\vb - \vb^{(m)})} \cdot \nb 
    - 2\jump{\frac{\eta}{\rho} \divg_\text{S} \left[ \mathbb{P} \left( \vb - \vb^{(m)} \right) \right] } \,.
    \label{eq:n-tau-n-1}
\end{equation} 

For the special case when there is no tangential slip on the membrane, and the assumption of no mass accumulation, Eq.~\eqref{eq:n-tau-n-1} can be reduced to 
\begin{equation}
    \jump{\frac{\nb \cdot \taub\nb}{\rho}} = 4H \jump{\frac{\eta}{\rho^2}} \dot{n}^{(m)}
    \,,
    \label{eq:n-tau-n-2-kinematic}
\end{equation}
where $\dot{n}^{(m)} := \rho^{(\pm)} (\vb^{(\pm)} - \vb^{(m)}) \cdot \nb$. In particular, this means that the driving force $\jump{\dfrac{\nb \cdot \taub\nb}{\rho}}$ only manifests in curved geometries (i.e., when the mean curvature $H$ of the membrane is nonzero) with some viscosity contrast between the inside and outside of the membrane. For dilute, binary systems, we can take the total mass density to be constant at the solvent density, i.e., $\rho \approx \rho_w$. In this case, Eq.~\eqref{eq:n-tau-n-2-kinematic} simplifies to 
\begin{equation}
    \jump{\frac{\nb \cdot \taub\nb}{\rho}} = 4H \jump{\frac{\eta}{\rho}}(\vb - \vb^{(m)}) \cdot \nb \,,
    \label{eq:n-tau-n-2-kinematic-dilute}
\end{equation}
in which the \textit{kinematic viscosity} $\eta/\rho$ appears.
Additionally, the condition of no mass accumulation on the membrane simplifies to $\jump{\vb\cdot\nb} = 0$, which enforces continuity on the normal component of the bulk velocity.

\subsection{Negligible Deviatoric Stresses and the Kedem and Katchalsky Relations}
\label{subsec:negligible-deviatoric-stresses}

In cases where the net stress $\jump{ \nb \cdot \taub\nb }$ is negligible (e.g., when viscous forces are negligible), we shall see that Eqs.~\eqref{eq:binary-law-s} and~\eqref{eq:binary-law-w} simplify to the constitutive laws presented by Kedem and Katchalsky \cite{kedem1958thermodynamic}. Following their notation, we define the ``volume flow per unit area'' as 
\begin{equation}
    J_v = \overline{V}_s \dot{N}_s^{(m)} + \overline{V}_w \dot{N}_w^{(m)} \,,
    \label{eq:Jv-defn}
\end{equation}
and the ``exchange flow'' as 
\begin{equation}
    J_D = \frac{\dot{N}_s^{(m)}}{\LMavg{c_s}} - \frac{\dot{N}_w^{(m)}}{c_w} \,.
    \label{eq:JD-defn}
\end{equation}
Note that the quantity $J_v$ has dimensions of length per time, and is better interpreted as the net velocity of the fluid flowing through the membrane. The quantity $J_D$ also has dimensions of length per time, and is a diffusion-like quantity that quantifies the velocity of the solute relative to the solvent. 
Kedem and Katchalsky propose linear relations between the fluxes $J_v$ and $J_D$ and the driving forces $\jump{-p}$ and $RT\jump{-c_s}$ as
\begin{align}
    J_v &= L^p \jump{-p} + L^{pD} RT\jump{-c_s} \,, \label{eq:JV-linear-law}\\
    J_D &= L^{pD} \jump{-p} + L^D RT\jump{-c_s} \,, \label{eq:JD-linear-law}
\end{align}
where $L^p$, $L^{pD}$, and $L^{D}$ are phenomenological coefficients. 

Setting the driving force $\jump{ \nb \cdot \taub\nb } \equiv 0$, Eqs.~\eqref{eq:binary-law-s} and~\eqref{eq:binary-law-w} provide linear relationships between the fluxes $\dot{N}_s^{(m)}$ and $\dot{N}_w^{(m)}$ and the driving forces $\jump{-p}$ and $RT\jump{-c_s}$. 
Within the linear irreversible thermodynamic framework, the choice of thermodynamic fluxes and the conjugate forces is arbitrary up to linear combination, and so Kedem and Katchalsky's phenomenological coefficients must be related to $L^{ss}$, $L^{sw}$, and $L^{sw}$ via an invertible linear transformation.
Substituting Eqs.~\eqref{eq:binary-law-s} and~\eqref{eq:binary-law-w} into Eqs.~\eqref{eq:Jv-defn} and~\eqref{eq:JD-defn} and comparing the coefficients of the driving forces, we obtain 
\begin{equation}
    \begin{pmatrix}
    L^p \\ L^{pD} \\ L^D
    \end{pmatrix} 
    = \begin{pmatrix}
        \overline{V}_s^2 & 2\overline{V}_s\overline{V}_w & \overline{V}_w^2
        \\[5pt]
        \frac{\overline{V}_s}{\LMavg{c_s}} & \frac{\overline{V}_w}{\LMavg{c_s}} - \overline{V}_s\overline{V}_w & - \overline{V}_w^2
        \\[5pt]
        \frac{1}{\LMavg{c_s}^2} & -2\frac{\overline{V}_w}{\LMavg{c_s}} & \overline{V}_w^2
    \end{pmatrix}
    \begin{pmatrix}
    L^{ss} \\ L^{sw} \\ L^{ww}
    \end{pmatrix} \,. \label{eq:to-KK-lin-alg}
\end{equation}

Kedem and Katchalsky~\cite{kedem1958thermodynamic} further transform their constitutive laws into a form that's more amenable for comparison with experimental data by writing 
\begin{gather}
    J_v = L^p \left(\jump{-p} - \sigma RT \jump{-c_s} \right) \,, \label{eq:Starling-eqn}\\
    \dot{N}_s^{(m)} = \omega RT \jump{-c_s} + (1 - \sigma) \LMavg{c_s} J_v \,,
    \label{eq:KK-solute-flux}
\end{gather}
where 
\begin{align}
    \sigma &:= -\frac{L^{pD}}{L^p} 
    = 1 - \frac{L^{ss}\overline{V}_s + L^{sw}\overline{V}_w}{\LMavg{c_s}(L^{ss}\overline{V}_s^2 + 2L^{sw}\overline{V}_s\overline{V}_w + L^{ww}\overline{V}_w^2)}
    & \left(\varphi_s \ll 1\right)
    \,,
    \label{eq:reflection-coefficient-defn}
\end{align}
is Staverman's reflection coefficient, and 
\begin{align}
    \omega &:= \frac{L^pL^D - (L^{pD})^2}{L^p}\LMavg{c_s}
    = \frac{\overline{V}_w^2 \left( L^{ss}L^{ww} - (L^{sw})^2 \right)}{\LMavg{c_s} \left( L^{ss}\overline{V}_s^2 + 2L^{sw}\overline{V}_s\overline{V}_w + L^{ww}\overline{V}_w^2 \right)}
    & \left(\varphi_s \ll 1\right)
    \,,
    \label{eq:omega-mobility-defn}
\end{align}
is the mobility. Equation~\eqref{eq:Starling-eqn} is known as the \textbf{Starling equation}, wherein $L^p$ is known as the filtration coefficient. The reflection coefficient $\sigma$ is a measurement of the membrane's ability to prevent the solute from permeating across it, a property known as \textit{selectivity}.
As we will discuss in the following sections, we have $\sigma = 0$ for a non-selective membrane, and $\sigma = 1$ for an ideally selective membrane.

\subsection{The Modified Starling Equation}
\label{subsec:modified-Starling-eqn}

When the bulk fluid on both sides of the membrane consists of a dilute, incompressible Newtonian fluid solution of solvent `$w$' and solute `$s$', we can combine the results of Sections~\ref{subsec:binary-ideal-system} and~\ref{subsec:normal-deviatoric-stress} to obtain more amenable forms of the constitutive relations for practical calculation. Our goal is to obtain an equation like that of Starling's (Eq.~\eqref{eq:Starling-eqn}), but modified appropriately to capture the additional viscous driving forces.

For a dilute system, we may take the total mass density to be constant at $\rho \approx \rho_w$.
The volume flux $J_v$ is then related to the total mass flux through the membrane by 
\begin{align}
     J_v := (\vb - \vb^{(m)})\cdot\nb =  \dot{n}^{(m)}/\rho 
     \,.
    \label{eq:volume-flux-to-mass-flux}
\end{align}
Using the identity $\dot{n}^{(m)} = \dot{n}_s^{(m)} + \dot{n}_w^{(m)}$ and the constitutive laws in Eqs.~\eqref{eq:binary-law-s} and~\eqref{eq:binary-law-w}, the total mass flux can be written as 
\begin{equation}
\begin{split}
    \dot{n}^{(m)} 
    &= \left( \frac{ L^{ss} \scrM_s + L^{sw} \scrM_w }{ 1 / \overline{V}_s} + \frac{L^{sw} \scrM_s + L^{ww} \scrM_w}{ 1 / \overline{V}_w} \right) \jump{-p}\\
    &\qquad + \left( \frac{L^{ss} \scrM_s + L^{sw} \scrM_w}{\LMavg{c_s}} - \frac{L^{sw} \scrM_s + L^{ww} \scrM_w}{1 / \overline{V}_w} \right) RT\jump{-c_s}\\
    &\qquad\qquad+ \left(L^{ss} \scrM_s^2 + 2L^{sw} \scrM_s\scrM_w + L^{ww} \scrM_w^2\right) \jump{\frac{\nb \cdot \taub\nb}{\rho}} 
    \,.
\end{split}
    \label{eq:total-mass-flux-result}
\end{equation}
Comparing Eq.~\eqref{eq:total-mass-flux-result} to Eqs.~\eqref{eq:Starling-eqn} and~\eqref{eq:volume-flux-to-mass-flux} motivates the definitions 
\begin{align}
    L_\text{mod}^p &:= \frac{\overline{V}_w}{\scrM_w} \left( \frac{ L^{ss} \scrM_s + L^{sw} \scrM_w }{ 1 / \overline{V}_s} + \frac{L^{sw} \scrM_s + L^{ww} \scrM_w}{ 1 / \overline{V}_w} \right)
    \,, \label{eq:Lp-mod} \\
    L_\text{mod}^{pD} &:= \frac{\overline{V}_w}{\scrM_w} \left( \frac{L^{ss} \scrM_s + L^{sw} \scrM_w}{\LMavg{c_s}} - \frac{L^{sw} \scrM_s + L^{ww} \scrM_w}{1 / \overline{V}_w} \right)
    \,. \label{eq:LpD-mod}
\end{align}
The constant solution density approximation in the dilute limit also yields $\overline{V}_s \approx \left({\scrM_s}/{\scrM_w}\right) \overline{V}_w$, so that Eqs.~\eqref{eq:Lp-mod} and~\eqref{eq:LpD-mod} along with Eq.~\eqref{eq:to-KK-lin-alg} result in $L_\text{mod}^p \equiv L^p$ and $L_\text{mod}^{pD} \equiv L^{pD}$. 
With the above definitions of the modified phenomenological coefficients, and the viscous driving force in Eq.~\eqref{eq:n-tau-n-2-kinematic}, we obtain the total volume flux $J_v$ as
\begin{align}
    J_v &= \frac{L^{p}}{1 - 4 H \jump{\eta} L^{p}} \left(\jump{-p} - \sigma RT \jump{-c_s} \right) \,.
    \label{eq:modified-Starling-equation}
\end{align}
We call Eq.~\eqref{eq:modified-Starling-equation} the \textbf{Modified Starling Equation}. Here, we see that the presence of membrane curvature and a viscosity contrast between the bulk fluids modifies the permeability of the membrane\footnote{Note that although the sign of the mean curvature $H$ is dependent upon orientation, the product $H \jump{\eta}$ is orientation-invariant since both factors will reverse in sign upon parity transformation.}. In particular, the permeability of a fluid flowing from a solution of higher viscosity into a solution of lower viscosity across a curved interface is enhanced. Note that the equation for the molar solute flux $\dot{N}_s^{(m)}$ is still given by Eq.~\eqref{eq:KK-solute-flux}, except that the volume flux $J_v$ should be calculated using the modified Starling equation (Eq.~\eqref{eq:modified-Starling-equation}).

\subsection{Ideally Selective and Non-Selective Membranes}
\label{subsec:ideally-selective-non-selective}

For a two-component system, an ideally selective membrane is one that allows for the passage of the solvent through the membrane, but perfectly resists the passage of the solute. In other words, an ideally selective membrane is characterized by the equation 
\begin{equation}
    \dot{N}_s^{(m)} \equiv 0 \,. \label{eq:ideal-selectivity-criterion-1}
\end{equation}
Since this relation must hold for arbitrary pressure differences~$\jump{-p}$ and concentration differences~$\jump{-c_s}$, Eq.~\eqref{eq:binary-law-s} implies that 
\begin{equation}
    L^{ss} = 0 = L^{sw} \,, \label{eq:ideal-selectivity-criterion-2}
\end{equation}
which is assumed to hold across all temperature, pressure, and concentration ranges where the membrane exhibits ideal behaviour.
Using the criterion~\eqref{eq:ideal-selectivity-criterion-2}, the modified phenomenological coefficients (Eq.~\eqref{eq:to-KK-lin-alg}) become 
\begin{align}
    L^p &= L^{ww} \overline{V}_w^2 = -L^{pD} \,,
\end{align}
leading to the reflection coefficient $\sigma=1$.
Substituting Eq.~\eqref{eq:ideal-selectivity-criterion-2} into Eq.~\eqref{eq:binary-law-w}, we find that the flux of solvent (which is equivalent to the total molar flux) for an ideally selective membrane is 
\begin{equation}
    \dot{N}_w^{(m)} = L^{ww}\overline{V}_w \left( \jump{-p} - RT\jump{-c_s} + \frac{\scrM_w}{\overline{V}_w} \jump{ \frac{\nb \cdot \taub\nb}{\rho} }\right)
    \,.
\end{equation}
The total volumetric flux in Eq.~\eqref{eq:modified-Starling-equation}, for an incompressible Newtonian bulk fluid, simplifies to 
\begin{equation}
    J_v =
    (\vb - \vb^{(m)}) \cdot \nb = \frac{L^{p}}{1 - 4 H \jump{\eta} L^{p}} \left(\jump{-p} - RT \jump{-c_s} \right) \,.
    \label{eq:modified-Starling-equation-ideal}
\end{equation}
Equation~\eqref{eq:modified-Starling-equation-ideal} serves as a boundary condition for the bulk fluid at an ideally selective membrane interface.


On the other hand, for a completely non-selective membrane, i.e., $\sigma =0$, the volume flux from the modified Starling equation can be seen to be 
\begin{align}
    J_v &= \frac{L^{p}}{1 - 4 H \jump{\eta} L^{p}} \jump{-p} \,,
    \label{eq:mech-law}
\end{align}
which is only dependent on the pressure drop $\jump{-p}$ across the membrane. In this case, $L^{pD} = 0$, which when combined with the assumption of negligible cross phenomena, i.e., $L^{sw}=0$, leads to the mobility $\omega = \frac{L^{ss}}{\LMavg{c_s}}$ and the following solute flux: 
\begin{align}
    \dot{N}_s^{(m)} = \omega RT \jump{-c_s} + \LMavg{c_s} J_v \,,
    \label{eq:Ns-non-selective}
\end{align}
where we have used Eqs.~\eqref{eq:to-KK-lin-alg},~\eqref{eq:KK-solute-flux}, and~\eqref{eq:omega-mobility-defn}. In Eq.~\eqref{eq:Ns-non-selective}, $\LMavg{c_s} J_v$ corresponds to the solute advected with the volume flux, while the first term on the right-hand side corresponds to the diffusive flux of the solute through the membrane. 

Lastly, the volume flux in Eq.~\eqref{eq:mech-law} also coincides with the mechanical permeability of a membrane, which is defined as the ability of the membrane to resist overall mass flux being forced through it by way of a pressure gradient, i.e., when $RT\jump{-c_s} \equiv 0$. The solute flux in this case is simply $\dot{N}_s^{(m)} = \LMavg{c_s} J_v $.
If the fluids on both sides of the membrane have the same viscosity, or if the membrane has no curvature, then the resistance to mechanically forcing volume flux through the membrane is inversely proportional to $L^{p}$, recovering the usual interpretation of $L^{p}$ as the filtration coefficient for a membrane~\cite{kedem1958thermodynamic}.

\section{Equations of Motion for a Permeable Lipid Bilayer Membrane}
\label{sec:EOM-membrane-bulk-system}

In this section, we summarize the governing equations for a permeable membrane in a bulk fluid, along with a set of constitutive relations 
and the boundary conditions to close the system of equations. 
To this end, consider a lipid membrane $\Pc$ immersed in a bulk fluid $\Omega$ containing $N$ chemical species. We assume that the system is isothermal, and that no chemical species can accumulate on the membrane. Most biological problems take place at low enough Reynolds numbers where the inertial terms in both the bulk fluid and the membrane may be neglected \cite{purcell1977life, sahu2020geometry}. Assuming that the bulk consists of an incompressible Newtonian fluid, we can use the equations for Stokes flow to model the bulk fluid:
\begin{gather}
    \divg\vb = 0 \,,\\
    -\grad p + \mu \divg\grad\vb = \zerob \,,\\
    \dot{c}_i = -\divg\Jb_i \,,
\end{gather}
where $\Jb_i$ is given by the generalized Fick's law (Eqs.~\eqref{eq:generalized-Ficks-law}), which in dilute solutions corresponds to Eq.~\eqref{eq:bulk-Ficks-law}.

For an isothermal, incompressible membrane, substituting the constitutive relations for the membrane stresses (Eq.~\eqref{eq:sigma-alpha-beta-incompressible}) and moments (Eq.~\eqref{eq:M-alpha-beta-incompressible}) into the membrane momentum balance (Eq.~\eqref{eq:local-membrane-linear-momentum-balance}), leads to the following in-plane and out-of-plane momentum equations: 
\begin{subequations}
\label{eq:membrane-EOM}
\begin{align}
    0 &=
    \pi^{\beta\alpha}_{;\beta} - 2k_b(H-C)C^{,\alpha} + \lambda^{,\alpha} + \jump{\ab^\alpha \cdot \taub\nb} \,, 
    \label{eq:membrane-EOM-in-plane} \\
    0 &= 
    \pi^{\alpha\beta}b_{\alpha\beta} - k_b \Delta_\text{S} (H - C) - 2k_b (H-C)(H^2 + HC - K) + 2\lambda H + \jump{-p} + \jump{\nb \cdot \taub\nb} \,,
    \label{eq:membrane-EOM-out-of-plane}
\end{align}
\end{subequations}
where the membrane viscous stresses $\pi^{\alpha\beta}$ are given by Eq.~\eqref{eq:pi-ab-constitutive-law} and
$\Delta_\text{S}(\,\bullet\,) := a^{\alpha\beta} (\,\bullet\,)_{;\alpha\beta}$ is the surface Laplacian \cite{Rangamani2013, sahu-mandadapu-pre-2017}. The mass balance (Eq.~\eqref{eq:local-membrane-mass-balance}) for an incompressible membrane with no accumulation yields 
\begin{align}
    \divg_\text{S} \vb^{(m)} = v_{;\alpha}^\alpha - 2vH &= 0
    \,.
\end{align}

For the boundary conditions at the membrane-bulk interface, we have Navier-slip for the tangential velocity components
\begin{equation}
    (\vb^{(\pm)} - \vb^{(m)}) \cdot \ab^\alpha = \beta_\text{sl}^{(\pm)} \left(\ab^{\alpha} \cdot \taub^{(\pm)}(\pm\nb)\right) \,,
    \label{eq:slip-law-2-copy}
\end{equation}
and the generalized permeability law (Eq.~\eqref{eq:general-permeability-linear-law-2}) for the normal velocity components 
\begin{equation}
    c_i \left(\vb_i - \vb^{(m)}\right) \cdot \nb = \sum_{j=1}^N L^{ij} \left( \overline{V}_j \jump{-p} - RT \jump{\ln(x_j \gamma_j)} + \scrM_j\jump{ \frac{\nb \cdot \taub\nb}{\rho} } \right) \,.
    \label{eq:general-permeability-linear-law-2-copy}
\end{equation}
For an ideally selective membrane with no tangential slip, one may repeat the procedure in Section~\ref{subsec:ideally-selective-non-selective} and simplify Eq.~\eqref{eq:general-permeability-linear-law-2-copy} to obtain the modified Starling equation
\begin{align}
    (\vb - \vb^{(m)}) \cdot \nb &= \frac{L^p}{1 - 4 H \jump{\eta} L^p} \left(\jump{-p} - RT \jump{-c_s} \right) \,,
    \label{eq:modified-Starling-many-solutes}
\end{align}
where $c_s$ is now the total molar concentration of the impermeable solutes. Equation~\eqref{eq:modified-Starling-many-solutes} is equivalent to a slip condition for the bulk fluid through the membrane. 
Equation~\eqref{eq:general-permeability-linear-law-2-copy} also provides the boundary condition for the species balance at the membrane-bulk interface, which for an ideally selective membrane is
\begin{align}
    \Jb_i \cdot \nb + c_i (\vb - \vb^{(m)}) \cdot \nb  &= 0
    \,,
\end{align}
with the second term on the left-hand side being given by the modified Starling equation~\eqref{eq:modified-Starling-many-solutes}.

Lastly, we may analyze the order of magnitude for the dimensionless quantity $4 H \jump{\eta}{L}^{p}$ arising from the additional viscous driving force for a lipid membrane. The membrane permeability is commonly measured in the literature as a permeability coefficient $P_f$, which is related to Kedem and Katchalsky's filtration coefficient through $L^p = \dfrac{P_f \overline{V}_w}{RT}$. For single-component lipid bilayer membranes, permeability coefficients on the order of $10^{-4}\ \si{m/s}$ have been reported\footnote{Note that the experiment in Ref.~\cite{mathai2008structural}, when viewed as the tensionless shrinking of a vesicle with no pressure jump, is compatible with the membrane and bulk equations of motion presented in this section.}~\cite{paula1996permeation,mathai2008structural}, which at physiological temperatures of $310\ \si{K}$ translates to an $L^p$ on the order of $10^{-12}\ \si{\frac{m}{s Pa}}$. With these measurements, we find that even extremely small vesicles on the order of $20\ \si{nm}$, and a viscosity contrast three orders greater than the viscosity of water at $1\ \si{Pa s}$ still yield $4 H \jump{\eta} {L}^{p}$ on the order of $10^{-5}$, which is a negligible modification to the permeability coefficient. Thus, the contribution from the viscous driving force $\jump{\dfrac{\nb\cdot\taub\nb}{\rho}}$ is negligible for biological membranes. For highly curved membranes with a much larger permeability to water, such as desalination membranes~\cite{geise2011water}, this modification may serve as a significant correction to the permeability coefficient.


\section{Conclusions} \label{sec:conclusion}

In this work, we developed an irreversible thermodynamic framework for a two-dimensional material surface immersed in a three-dimensional bulk fluid. 
Our approach combined an earlier treatment of standalone arbitrarily curved and deforming membranes~\cite{sahu-mandadapu-pre-2017} with Kedem and Katchalsky's~\cite{kedem1958thermodynamic} work on permeable membranes.
In order to account for finite jump discontinuities in bulk fields (such as the density, velocity, and stress) at the membrane-bulk interface, we derived the modified divergence and Reynolds transport theorems for material surfaces of discontinuity in Appendix~\ref{sec:integral-theorem-proofs}. In proving these modified integral theorems, we demonstrated that the surface Reynolds transport theorem can only be separated out if no mass accumulation takes place on the membrane.

The modified integral theorems enabled us to derive local balances of mass, linear momentum, angular momentum, energy, and entropy for both the bulk fluid and the membrane. Subsequently, by introducing functional forms for the Helmholtz free energy, we identified the equation for the internal entropy production at the membrane. 
Previous work by Sahu \etal{}~\cite{sahu-mandadapu-pre-2017} derived an equation for internal entropy production for a standalone curved and deforming lipid membrane, while Kedem and Katchalsky~\cite{kedem1958thermodynamic} presented an equation for the internal entropy production of a permeable membrane. In this work, we showed that the internal entropy production equation for a permeable, deforming membrane contains more terms than just the direct sum of these two equations. 
In particular, we identified the normal component of the deviatoric stress as a novel driving force for permeability. 
Though it may not be clear \textit{a priori} that one should include the deviatoric stresses in the equation for internal entropy production at the membrane, the formalism provided by the discontinuous integral theorems and non-equilibrium thermodynamics allows one to systematically identify the thermodynamic forces and the corresponding thermodynamic fluxes causing irreversibilities at the membrane-bulk fluid interfaces.

The jump in deviatoric stress can act as a driving force for both tangential and normal slip (i.e., permeability) 
at the membrane, which we incorporated into the generalized slip and permeability laws.
In the case of permeability, the normal component of the deviatoric stress is an independent driving force from the pressure and concentration jumps that Starling first identified~\cite{starling1896absorption}, and serves to enhance or impede the permeability of membranes. 
For a Newtonian fluid, we derived an expression that relates the normal deviatoric stress to the mean curvature of the membrane, the viscosity contrast between the two bulk fluid solutions, and the mass flux through the membrane. Furthermore, if the solution is a dilute, binary mixture we demonstrated that the normal deviatoric stress serves to modify the permeability coefficient in the Starling equation, resulting in a \textbf{Modified Starling Equation}.
This modification was ultimately found to be negligible for biological membranes via an order of magnitude estimate. 
Nevertheless, for highly curved membranes with a much larger filtration coefficient, this driving force may act as a significant modification to the permeability.

We provided a complete set of balance laws, constitutive relations, and boundary conditions to describe a viscous lipid membrane with curvature elasticity and a multi-component Newtonian bulk fluid system. These equations can be applied to study instabilities induced by osmotic shocks~\cite{Pullarkat2006,Datar2019TheAtrophy} in different geometries such as flat, spherical, and cylindrical~\cite{sahu2020geometry,tchoufag2022absolute,al2018hydro}, which we leave as future work. 
The set of coupled, non-linear partial differential equations presented here could also be solved using numerical methods~\cite{sauer2017stabilized, sahu-2020-ALE,torres2019modelling, reuther2020numerical, sahu2024arbitrary} to simulate the effects of osmotic gradients on biological membranes that underlie many biological processes~\cite{venkova2022mechano,rollin2023physical}.

We note that the same formalism used in this work can be extended to understand deposition processes such as binding/unbinding reactions at the membrane-fluid interfaces~\cite{sahu-mandadapu-pre-2017, stachowiak2012membrane}. 
Another possible extension is to address the role of permeability and osmosis in charged membrane-bulk systems.
Lastly, one could even apply this framework to study systems beyond lipid or biological membranes; any system that contains a deforming interface immersed in a fluid solution can be systematically studied using the framework of non-equilibrium thermodynamics and modified integral theorems presented here.

%
%

\section*{Acknowledgements}

We are grateful to Alison Lui, Amaresh Sahu, Dimitrios Fraggedakis, and Joel Tchoufag for useful discussions. A.M.A. is supported by the Natural Sciences and Engineering Research Council of Canada Postgraduate Scholarships---Doctoral program. This work was initiated from support by the Director, Office of Science, Office of Basic Energy Sciences, of the U.S. Department of Energy under contract No. DEAc02-05CH11231. A.M.A and K.K.M. also acknowledge the support of the Hellman Foundation Grant, and University of California, Berkeley.
\vspace{02pt}


\begin{appendices}

\renewcommand{\theequation}{\thesection.\arabic{equation}}
\setcounter{equation}{0}

\newpage
\section{Arguments for the Modified Integral Theorems} 
\label{sec:integral-theorem-proofs}

Various proofs for the modified integral theorems have been presented by continuum mechanicians studying shock waves and phase transitions \cite{Slattery2007,gupta2007balance,gurtin2010mechanics,casey2011derivation}. However, in these cases, the surfaces of discontinuity do not represent material interfaces, and the regularity requirements on the motion of the three-dimensional body required in order to capture osmotic transport phenomena are not explicitly considered. While these distinctions are unimportant for the modified divergence theorem, the ability of the bulk fluid to permeate through, slip on, or deposit onto the membrane requires modifications in the proof of the modified Reynolds transport theorem. As we will argue in Section~\ref{subsec:modified-Reynolds-transport-theorem}, the central issue with proving the modified Reynolds transport theorem for a system containing a material interface is that there does not exist a material motion for a region of bulk fluid surrounding the interface. As a consequence, even the definition of the material time derivative requires some elucidation. We will remedy these issues by utilizing the flows of smooth extensions of the bulk and membrane velocity fields to construct a hypothetical (or ``sampling'' \cite{casey2011derivation}) motion that coincides with the membrane and bulk's positions and velocities at the instant that the material time derivative is being evaluated. In doing so, we argue that the modified integral theorem used in this work (see Eq.~\eqref{eq:jump-reynolds-transport}) is valid even when the material membrane interface, the bulk fluid on top, and the bulk fluid on the bottom undergo relative slip to one another.

In what follows, consider a membrane surface of discontinuity $\Mc(t)$ immersed in a bulk fluid $\Bc(t)$ at some time $t$ (see Figure~\ref{fig:membrane-immersion}). Just as in Section~\ref{sec:mathematical-preliminaries}, we assume that the membrane has no boundary (i.e., $\partial\Mc(t) = \emptyset$). We will first outline the appropriate choice of membrane patch $\Pc(t) \subseteq \Mc(t)$ and bulk region $\Omega(t) \subseteq \Bc(t)$ for the purposes of applying the integral and localization theorems. To that end, consider an arbitrary membrane patch $\Pc(t)$ with boundary $\partial\Pc(t)$ that is contained in the membrane interior (i.e., $\partial\Pc(t) \subset \Mc(t)$). Since $\Mc(t)$ is an embedded surface in $\bbR^3$, there exists an open set $\Omega(t)$ in $\bbR^3$ such that $\Pc(t) = \Omega(t) \cap \Mc(t)$ (see Figure~\ref{fig:surface-of-discontinuity}). We call the open set $\Omega(t)$ the \textit{bulk fluid neighbourhood} surrounding $\Pc(t)$, and choose it to be small enough so that it is contained in the bulk fluid $\Bc(t)$. By construction, $\Omega(t)$ is partitioned by $\Pc(t)$ into two open sets $\Omega^{(+)}(t)$ and $\Omega^{(-)}(t)$, where $\Omega^{(+)}(t)$ is the region that the membrane normal $\nb$ points into and $\Omega^{(-)}(t)$ is the region that the normal $\nb$ points away from. An explicit construction for such a set $\Omega(t)$ can be made by extending the local membrane coordinates $\{\theta^1,\theta^2\}$ to local curvilinear bulk coordinates $\{\theta^1,\theta^2,\theta^3\}$ by defining 
\begin{equation}
    \xb(\theta^i,t) := \xb^{(m)}(\theta^1,\theta^2,t) + \theta^3 \nb \,,
    \label{eq:local-coordinate-extension}
\end{equation}
with $|\theta^3| < \delta(\xb^{(m)})$ for some positive smooth function $\delta \colon {\Pc}(t) \to \bbR$.
The range of $\delta$ is chosen to be small enough such that Eq.~\eqref{eq:local-coordinate-extension} is injective, with its image 
\begin{equation}
    \Omega(t) := \{ \xb^{(m)} + \theta^3 \nb \mid \xb^{(m)} \in \Pc(t) \text{ and } |\theta^3| < \delta(\xb^{(m)}) \} \,,
    \label{eq:tubular-nbhd}
\end{equation}
contained in the bulk $\Bc(t)$, thereby satisfying the desired properties for the bulk region\footnote{The set in Eq.~\eqref{eq:tubular-nbhd} is sometimes referred to as a \textit{tubular neighbourhood} \cite{lee2012smooth}.}. The injective nature of \eqref{eq:local-coordinate-extension} also guarantees a valid local curvilinear coordinate system, which is useful for enforcing boundary conditions on the bulk fluid in numerical and analytical explorations of the equations that appear in this work, thus in describing the membrane-bulk interactions. 
In what follows, we will prove the modified divergence theorem in Section~\ref{subsec:modified-divergence-theorem} and the modified Reynolds transport theorem in Section~\ref{subsec:modified-Reynolds-transport-theorem}.

\subsection{The Modified Divergence Theorem}
\label{subsec:modified-divergence-theorem}

For any continuously differentiable vector field $\fb(\xb,t)$ defined in the bulk fluid, consider an integral of the form 
\begin{equation}
    i(t) := \int_{\Omega(t)} \divg\fb \,dv \,. 
\end{equation}
As described before, recall that $\Pc(t) = \Omega(t) \cap \Mc(t)$ represents the part of the membrane $\Pc(t)$ encapsulated by $\Omega(t)$ at time $t$. Since $\Pc(t)$ has measure zero inside of $\Omega(t)$, we can write 
\begin{equation}
    \int_{\Omega(t)} \divg \fb \,dv = \int_{\Omega(t)\setminus \Pc(t)} \divg \fb \,dv \,.
\end{equation}
Recognizing that $\Omega(t) \setminus \Pc(t)$ is the disjoint union of $\Omega^{(+)}(t)$ and $\Omega^{(-)}(t)$, we now have that 
\begin{align}
    \int_{\Omega(t)} \divg \fb \,dv &= \int_{\Omega^{(+)}(t)} \divg \fb \,dv + \int_{\Omega^{(-)}(t)} \divg \fb \,dv \,. \label{eq:appA-1-intermediate-1}
\end{align}
Since $f$ is continuously differentiable inside each of $\Omega^{(+)}(t)$ and $\Omega^{(-)}(t)$, we can use the divergence theorem on each term of the right-hand side of Eq.~\eqref{eq:appA-1-intermediate-1} to write
\begin{align}
    \int_{\Omega(t)} \divg \fb \,dv &= \int_{\partial\Omega^{(+)}(t)} \fb\cdot\nb \,da + \int_{\partial\Omega^{(-)}(t)} \fb\cdot\nb \,da \\
    \begin{split}
        &= \left(\int_{\partial\Omega^{(+)}(t)\setminus \Pc(t)} \fb\cdot\nb \,da + \int_{\Pc(t)} \fb^{(+)}\cdot(-\nb) \,da\right) 
        \, \\ 
        & \hspace{0.5in} 
        + \left(\int_{\partial\Omega^{(-)}(t)\setminus \Pc(t)} \fb\cdot\nb \,da + \int_{\Pc(t)} \fb^{(-)}\cdot\nb \,da\right) \,.
    \end{split} \label{eq:appA-divthm-1}
\end{align}
Note that the surface normal $\nb$ is taken to be the outward-pointing normal when integrating over $\Omega^{(\pm)}(t)$, which leads to a negative sign in the second term of Eq.~\eqref{eq:appA-divthm-1}. Recognizing that the bulk boundary $\partial\Omega(t)$ can be written as the disjoint union 
\begin{equation}
    \partial\Omega(t) = \left( \partial\Omega^{(+)} \setminus (\Pc(t) \cup \partial\Pc(t)) \right)
    \bigsqcup \left(\partial\Omega^{(-)} \setminus (\Pc(t) \cup \partial\Pc(t)) \right)
    \bigsqcup \partial\Pc(t) \,,
\end{equation}
and that $\partial \Pc(t)$ has measure zero in $\partial\Omega(t)$, Eq.~\eqref{eq:appA-divthm-1} simplifies to 
\begin{equation}
    \int_{\Omega(t)} \divg \fb \,dv = \int_{\partial\Omega(t)} \fb\cdot\nb \,da - \int_{\Pc(t)} (\fb^{(+)} - \fb^{(-)})\cdot\nb \,da \,. \label{eq:appA-divthm-2}
\end{equation}
Equation~\eqref{eq:appA-divthm-2} can be rearranged to obtain
\begin{equation}
    \int_{\partial\Omega(t)} \fb\cdot\nb \,da = \int_{\Omega(t)} \divg\fb \,dv + \int_{\Pc(t)} \jump{\fb} \cdot \nb \,da \,,
    \label{eq:appA-jump-div-thm}
\end{equation}
which is the final form of the modified divergence theorem that appears in the main text.

Equation \eqref{eq:appA-jump-div-thm} can be extended to scalar and tensor fields using the fact that $\forall \ab,\bb \in \bbR^3$,   
\begin{equation}
    \ab = \bb 
    \quad \text{ if and only if } \quad 
    \cb \cdot \ab = \cb \cdot \bb \quad \forall \cb \in \bbR^3 \,.\label{eq:appA:vector-scalar-equivalence}
\end{equation}
To that end, observe that for a bulk scalar field $\varphi$ and constant vector $\cb \in \bbR^3$,
\begin{align}
    \cb \cdot \int_{\Omega(t)} \grad \varphi \,dv &= \int_{\Omega(t)} \divg (\cb \varphi)\,dv\\
    &= \int_{\partial\Omega(t)} (\cb\varphi) \cdot \nb \,da - \int_{\Pc(t)} \jump{\cb\varphi} \cdot \nb \,da \label{eq:appA-divthm-3}\\
    &= \cb \cdot \left(\int_{\partial\Omega(t)} \varphi\nb \,da - \int_{\Pc(t)} \jump{\varphi} \nb \,da\right) \,, \label{eq:appA-divthm-4}
\end{align}
where we have used Eq.~\eqref{eq:appA-jump-div-thm} in Eq.~\eqref{eq:appA-divthm-3}. Then with Eq.~\eqref{eq:appA:vector-scalar-equivalence}, we conclude from Eq.~\eqref{eq:appA-divthm-4} that 
\begin{equation}
    \int_{\Omega(t)} \grad \varphi\,dv = \int_{\partial\Omega(t)} \varphi\nb \,da - \int_{\Pc(t)} \jump{\varphi} \nb \,da \,.
    \label{eq:appA-jump-div-thm-scalar}
\end{equation}
Similarly, for a bulk tensor field $\Fb$ and a constant vector field $\cb \in \bbR^3$, using Eq.~\eqref{eq:appA-jump-div-thm},
\begin{align}
    \cb \cdot \left(\int_{\Omega(t)} \divg \Fb \,dv\right) &= \int_{\Omega(t)} \divg (\Fb^T\cb) \,dv\\
    &= \int_{\partial\Omega(t)} (\Fb^T\cb) \cdot \nb \,da - \int_{\Pc(t)} \jump{\Fb^T\cb} \cdot \nb \,da\\
    &= \cb \cdot  \left(\int_{\partial\Omega(t)} \Fb\nb \,da - \int_{\Pc(t)} \jump{\Fb}\nb \,da\right) \,,
\end{align}
which from Eq.~\eqref{eq:appA:vector-scalar-equivalence} leads to 
\begin{equation}
    \int_{\Omega(t)} \divg \Fb \,dv = \int_{\partial\Omega(t)} \Fb\nb \,da - \int_{\Pc(t)} \jump{\Fb}\nb \,da \,.
    \label{eq:appA-jump-div-thm-tensor}
\end{equation}
The modified divergence theorems in Eqs.~\eqref{eq:appA-jump-div-thm}, \eqref{eq:appA-jump-div-thm-scalar}, and \eqref{eq:appA-jump-div-thm-tensor} are equivalent to the form of the divergence theorems for systems with non-material surfaces of discontinuity. As we move on to proving the modified Reynolds transport theorem, however, such considerations of a permeable material surface will become more germane.

\subsection{The Modified Reynolds Transport Theorem}
\label{subsec:modified-Reynolds-transport-theorem}

The Reynolds transport theorem is used to evaluate the material rate of change of the integral of a quantity $\fb(\xb,t)$ over a region $\Omega(t)$ within a bulk fluid $\Bc(t)$. 
In a continuous system, the material time derivative of this integral is mathematically defined as its rate of change in time $t$ evaluated by tracking the motion of the set of material particles that make up the region $\Omega(t)$.
In systems containing a material surface of discontinuity $\Mc(t)$, however, this definition of the material time derivative may no longer be used, as material in the bulk fluid $\Omega(t)$ may accumulate (or deplete) onto (or from) the portion of the membrane $\Pc(t)$ enclosed by $\Omega(t)$. Additionally, the membrane can also act as a \textit{surface of discontinuity}, as the function $\fb(\xb,t)$ may be discontinuous at the membrane. The goal of this section is to derive a modified Reynolds transport theorem for regions of fluid $\Omega(t)$ that contain a \textit{material surface of discontinuity} $\Pc(t)$. In doing so, we will also argue that the surface Reynolds transport theorem remains unchanged, as this is not \textit{a priori} evident due to the accumulation/depletion of material.

Our starting point is to consider integrals of the form 
\begin{equation}
    \Ib(t) := \Ib_1(t) + \Ib_2(t) := \int_{\Omega(t)} \fb(\xb,t) \,dv + \int_{\Pc(t)} \fb^{(m)}(\xb^{(m)},t) \,da \,, 
    \label{eq:appA-I-intro}
\end{equation}
where $\Ib_1(t)$ and $\Ib_2(t)$ are, respectively, the first and second terms on the right-hand-side of Eq.~\eqref{eq:appA-I-intro}, 
$\fb(\xb,t)$ is a vector field defined in the bulk fluid $\Bc(t)$, and $\fb^{(m)}(\xb^{(m)},t)$ is a vector field of the associated quantity over the membrane $\Mc(t)$. 
Typically, $\fb(\xb,t)$ and $\fb^{(m)}(\xb^{(m)},t)$ are intensive quantities (e.g., mass, momentum, or energy density) so that $\Ib(t)$ is the associated extensive quantity over the composite bulk fluid and membrane system $\Omega(t) \oplus \Pc(t)$. 
Despite the exchange of material between the membrane and bulk fluid, the material time derivative of $\Ib(t)$, i.e., $\dfrac{d\Ib(t)}{dt}$, is a physically well-defined quantity when considering the composite system.
However, it is not \textit{a priori} clear what the appropriate definition for the material time derivative of right-hand side of Eq.~\eqref{eq:appA-I-intro} is in the continuum sense when there is an appearance/loss of material points or relative slip on either side of the membrane, as in such a case there is no material motion for the bulk fluid surrounding the membrane.
This problem can be addressed by constructing hypothetical motions that allow for taking the material time derivative, which will be discussed in the following sections. 
To that end, it is instructive to first consider the proof of the Generalized Reynolds transport theorem for a continuous system. In doing so, we will review the notions of a time-dependent flow and its associated infinitesimal generator in Section~\ref{subsubsec:flows-and-generalized-rtt}. Consequently, in Section~\ref{subsubsec:gen-rtt-piecewise-smooth-boundary} we use these notions to construct a one-parameter family of homeomorphisms (known as a hypothetical ``sampling motion'' \cite{casey2011derivation}) to derive a generalized version of the Reynolds transport theorem for a bulk region with a piecewise smooth velocity field on its boundary. Finally, we derive the modified Reynolds transport theorem for a system with a material surface of discontinuity in Section~\ref{subsubsec:proof-of-modified-rtt}.

\subsubsection{Time-dependent Flows and the Generalized Transport Theorem for a Continuous System}
\label{subsubsec:flows-and-generalized-rtt}

The Generalized transport theorem provides an expression for the time rate of change of the integral of a scalar, vector, or tensor field $\fb(\xb,t)$ over an open control volume $\Omega_\mathrm{cv}(t)$ whose boundary $\partial\Omega_\mathrm{cv}(t)$ evolves at some velocity $\vb_\brm$, and is given by \cite{truesdell1960classical}
\begin{equation}
    \frac{d}{dt} \int_{\Omega_\mathrm{cv}(t)} \fb(\xb,t) \,dv = \int_{\Omega_\mathrm{cv}(t)} \pder{\fb(\xb,t)}{t} \,dv + \int_{\partial\Omega_\mathrm{cv}(t)} \fb(\xb,t) \left( \vb_\brm \cdot \nb \right) \,da \,.
    \label{eq:generalized-transport-theorem}
\end{equation}
In general, the boundary velocity $\vb_\brm$ need not coincide with the material velocity of the fluid, though when it does the Generalized transport theorem reduces to the Reynolds transport theorem for a material region. 
Additionally, note that the derivative on the left-hand-side of Eq.~\eqref{eq:generalized-transport-theorem} need not correspond to the material time derivative. Some authors \cite{Slattery2007,truesdell1960classical} choose to decorate the time derivative as $\dfrac{d_{(\mathrm{cv})}}{dt}$ or $\dfrac{d_{\vb_\brm}}{dt}$ to avoid confusion. By contrast, in this work, we will differentiate between these notions of the time derivative by using different labels for the region of integration for each motion under consideration. In particular, for the derivation of the modified Reynolds Transport Theorem, we will take the bulk fluid region $\Omega(t)$ and the control volume $\Omega_\mathrm{cv}(t)$ to overlap in the same region of space at time $t$, but the boundary of the former region $\partial\Omega(t)$ is understood to move with the material velocity $\vb$ while the boundary of the latter region $\partial\Omega_\mathrm{cv}(t)$ moves with the prescribed velocity $\vb_\brm$.

In what follows, we will adopt the definitions of \cite{lee2012smooth, frankel2011geometry} to introduce the flow of a time-dependent vector field. Subsequently, we will utilize this flow to prove the Generalized transport theorem stated as Eq.~\eqref{eq:generalized-transport-theorem} in a manner similar to that presented in \cite{marsden1994mathematical,frankel2011geometry}.

\paragraph{Time-dependent Flows}\mbox{}

Let us begin by considering a smooth time-independent velocity field $\vb(\xb)$ defined within a region of interest $\Omega(t)$ at some fixed time $t$. From the existence, uniqueness, and smoothness of autonomous differential equations \cite{lee2012smooth}, there exists a one-parameter collection of maps $\thetab_s \colon \Omega(t) \to \bbR^3$, called the \textit{flow} of $\vb(\xb)$, such that the parameter $s$ belongs to some open interval $\Jc = (-\epsilon,\epsilon)$, and the flow $\thetab_s$ satisfies 
\begin{equation}
    \thetab_0(\xb) = \xb 
    \quad \text{and} \quad 
    \vb(\xb) = \left[ \frac{d}{ds} \thetab_s(\xb) \right|_{s=0} \quad \text{for all } \xb \in \Omega(t)\,. \label{eq:flow-static}
\end{equation}
Crucially, the flow satisfies the one-parameter group property $\thetab_s \circ \thetab_r = \thetab_{s + r}$ for all $r,s \in \Jc$. The map $s \mapsto \thetab_s(\xb)$ is also known as an \textit{integral curve} of $\vb(\xb)$ starting at $\xb$. Conversely, the velocity field $\vb(\xb)$ is said to be the \textit{infinitesimal generator} of the flow $\thetab_s$.

For a smooth time-dependent vector field $\vb(\xb,t)$ defined over a region $\Omega(t)$, there does not exist a flow of $\vb(\xb,t)$ in the usual sense as in Eq.\eqref{eq:flow-static}. Instead, we must view the moving bulk region $\Omega(t)$ as a $4$-dimensional sub-manifold of space-time $\bbR^3 \times \bbR$. This introduction of space-time coordinates enables the autonomization of the vector field $\vb(\xb,t)$ through the space-time vector field
\begin{equation}
    \Vb(\xb,t) := \vb(\xb,t) + \eb_t \,, \label{eq:4-vector-velocity}
\end{equation}
where $\eb_t$ is the unit vector in the time axis\footnote{Note that the notation in Eq.\eqref{eq:4-vector-velocity} views all vectors $\vb$ in $\bbR^3$ to also correspond to vectors in space-time with the time-component being zero, i.e., $(v_1, v_2, v_3,0)$ for instance in Cartesian coordinates.}.
Then there exists a flow $\thetab_s(\xb,t)$ for which $\Vb(\xb,t)$ is the infinitesimal generator in the ordinary sense. We may further project this to real space using the projection map $\pi_{\bbR^3}(\xb,t) = \xb$ to obtain a \textit{time-dependent flow} $\thetab_{s,t}(\xb) := \pi_{\bbR^3} \circ \thetab_s(\xb,t)$. The parameter $s$ can be shifted to being centred at time $t$ so that the time-dependent flow $\thetab_{s,t}$ satisfies 
\begin{equation}
    \thetab_{t,t}(\xb) = \xb 
    \quad \text{and} \quad 
    \vb(\xb,t) = \left[ \frac{d}{ds} \thetab_{s,t}(\xb) \right|_{s=t}
    \quad \text{for all } \xb \in \Omega(t) \,, \label{eq:time-dep-flow-properties}
\end{equation}
with $s \in \Jc := (t - \epsilon, t + \epsilon)$ for some $\epsilon > 0$.
In the special case where the vector field $\vb$ is time-independent, the time-dependent flow $\thetab_{s,t}$ depends only on the difference $s - t$, and coincides with the flow generated of $\vb$ obtained from Eq.\eqref{eq:flow-static}, i.e., $\thetab_{s-t} := \thetab_{s,t}$. 
In what follows, we will only consider time-dependent velocity fields, and for brevity we will refer to $\thetab_{s,t}$ as simply the \textit{flow}\footnote{In the strictest sense of the word, a ``time-dependent flow'' isn't a ``flow'', since it doesn't satisfy the one-parameter group property (see \cite{frankel2011geometry} for more detail).} of the velocity field $\vb(\xb,t)$.

To illustrate the physical meaning of the flow, consider a solid body $\Omega(t)$ undergoing a motion $\chib_t \colon \Omega(t_0) \to \bbR^3$, where $\Omega(t_0)$ is the reference configuration. In this case, $\thetab_{s,t} = \chib_{s} \circ \chib_t^{-1}$ indicating that the flow is exactly the motion of the solid body, but with the reference configuration shifted to time $s$. In the case where the body $\Omega(t)$ is a region of a fluid, however, the description of the velocity field is the primary object of interest, and so the aforementioned formalism of flows allows us to recover a motion $\thetab_{s,t}$ of the body for which the velocity $\vb(\xb,t)$ is the infinitesimal generator.

\paragraph{The Generalized Transport Theorem}\mbox{}

We now return to proving the Generalized transport theorem (Eq.~\eqref{eq:generalized-transport-theorem}). As before, let $\Omega_\mathrm{cv}(t)$ be a control volume with boundary velocity field $\vb_\brm(\xb_\brm,t)$. Note the motion of the material points in interior $\Omega_\mathrm{cv}(t)$ do not influence the rate of change of the integral quantity. This allows us to consider 
an arbitrary smooth extension of the boundary velocity field $\widetilde{\vb}_\brm(\xb,t)$\footnote{Note that here and henceforth, the notation $\widetilde{(\bullet)}$ is used to denote smooth extensions of fields. Additionally, smooth extensions of vector fields defined on closed subsets of $\bbR^3$ to vector fields defined on open open subsets of $\bbR^3$ always exist  (see for example \cite[Lemma~2.26]{lee2012smooth}).} defined over both the interior and the boundary of the control volume $\Omega_\mathrm{cv}(t) \cup \partial\Omega_\mathrm{cv}(t)$. Then using the flow $\thetab_{s,t}(\xb)$ of the velocity field $\widetilde{\vb}_\brm(\xb,t)$, the time-derivative in the left-hand-side of Eq.~\eqref{eq:generalized-transport-theorem} is given by 
\begin{align}
    \frac{d}{dt} \int_{\Omega_\mathrm{cv}(t)} \fb(\xb,t) \,dv 
    &= \left[ \frac{d}{ds} \int_{\thetab_{s,t}(\Omega_\mathrm{cv}(t))} \fb(\thetab_{s,t}(\xb), s) \,dv(s) \right|_{s=t} \,.
    \label{eq:appA-cv-derivative-defn}
\end{align}
Pulling the integral on the right-hand-side of Eq.~\eqref{eq:appA-cv-derivative-defn} back to time $t$ yields 
\begin{align}
    \frac{d}{dt} \int_{\Omega_\mathrm{cv}(t)} \fb(\xb,t) \,dv 
    &= \left[ \frac{d}{ds} \int_{\Omega_\mathrm{cv}(t)} \fb(\thetab_{s,t}(\xb), s) \det\left(\pder{\thetab_{s,t}(\xb)}{\xb}\right) \,dv \right|_{s=t} \,.
    \label{eq:appA-cv-derivative-pullback}
\end{align}
Defining the quantities
\begin{align}
    \xb(s) := \thetab_{s,t}(\xb)
    \quad \text{and} \quad 
    \Omega_\mathrm{cv}(s) := \thetab_{s,t}(\Omega_\mathrm{cv}(t))
    \,,
\end{align}
Eq.~\eqref{eq:appA-cv-derivative-pullback} can be evaluated as
\begin{align}
    \frac{d}{dt} \int_{\Omega_\mathrm{cv}(t)} \fb(\xb,t) \,dv 
    &= \left[ \frac{d}{ds} \int_{\Omega_\mathrm{cv}(t)} \fb(\xb(s), s) \det\left(\pder{\xb(s)}{\xb}\right) \,dv \right|_{s=t}\\
    \begin{split}
       &= \int_{\Omega_\mathrm{cv}(t)}  \Bigg[ \left( \frac{\partial}{\partial s}\fb(\xb(s), s) + \frac{\fb(\xb(s), s)}{\partial\xb(s)} \right) \det\left(\pder{\xb(s)}{\xb}\right) \\ 
       & \hspace{1.5in} + \fb(\xb(s), s)\frac{d}{ds}\det\left(\pder{\xb(s)}{\xb}\right) \Bigg]\Bigg|_{s=t} \,dv \,. \label{eq:appA-cv-derivative-intermediate}
    \end{split}
\end{align}
Using the kinematic relation $\dfrac{d}{ds}\det\left(\dfrac{\partial \xb(s)}{\partial \xb}\right) = \det\left(\dfrac{\partial \xb(s)}{\partial \xb}\right)\Bigg(\dfrac{\partial}{\partial\xb(s)} \cdot \dfrac{d\xb(s)}{ds}\Bigg)$, we can further reduce Eq.~\eqref{eq:appA-cv-derivative-intermediate} to
\begin{equation}
\begin{split}
    \frac{d}{dt} &\int_{\Omega_\mathrm{cv}(t)} \fb(\xb,t) \,dv \\
    &= \int_{\Omega_\mathrm{cv}(t)}  \left[ \left( \pder{}{s} \fb(\xb(s), s) + \frac{\partial}{\partial\xb(s)}\cdot\left(\fb(\xb(s), s) \otimes \frac{d\xb(s)}{ds} \right) \right) \det\left(\pder{\xb(s)}{\xb}\right) \right]\Bigg|_{s=t} \,dv \,.\label{eq:appA-cv-derivative-intermediate-1}
\end{split}
\end{equation}
Using the properties for the flow in Eq.~\eqref{eq:time-dep-flow-properties}, i.e., $\dfrac{\partial \xb(s)}{\partial \xb} \Bigg|_{s=t} =  \dfrac{\partial \theta_{t,t}(\partial \xb)}{\xb} = \Ib$ and $\dfrac{d\xb(s)}{ds} \Bigg|_{s=t} = \dfrac{d\thetab_{s,t}(\xb)}{ds} \Bigg|_{s=t} = \widetilde{\vb}_\brm(\xb,t)$, Eq.~\eqref{eq:appA-cv-derivative-intermediate-1} can be rewritten as   
\begin{align}
    \frac{d}{dt} \int_{\Omega_\mathrm{cv}(t)} \fb(\xb,t) \,dv 
    &= \int_{\Omega_\mathrm{cv}(t)} \left(\pder{\fb(\xb,t)}{t} + \divg( \fb(\xb,t) \otimes \widetilde{\vb}_\brm(\xb,t) \right) \,dv\\
    &= \int_{\Omega_\mathrm{cv}(t)} \pder{\fb(\xb,t)}{t} \,dv + \int_{\partial\Omega_\mathrm{cv}(t)} \fb(\xb,t) (\vb_\brm(\xb,t) \cdot \nb) \,da \,,
\end{align}
which establishes the Generalized transport theorem.

\subsubsection{Sampling Motions and the Generalized Transport Theorem for a Piecewise Smooth Boundary Velocity Field}
\label{subsubsec:gen-rtt-piecewise-smooth-boundary}

The derivation of the modified Reynolds transport theorem requires the development of Generalized transport theorems for the partitioned regions $\Omega_\mathrm{cv}^{(\pm)}(t)$. However, the assumption of a smooth boundary velocity field that was made in deriving Eq.~\eqref{eq:generalized-transport-theorem} is not valid when the bulk fluid can permeate through and slip on the membrane;
since the membrane and the bulk fluid can move independently of one another, the boundary velocity of each partition $\Omega_\mathrm{cv}^{(\pm)}(t)$ is only piecewise smooth (and possibly ill-defined at the membrane boundary $\partial\Pc(t)$). Furthermore, the relative motion between the bulk fluid and the membrane renders a description of the evolution of the enclosed region $\Omega^{(\pm)}_\mathrm{cv}(t)$ at later times unclear. 
In this section, we derive a Generalized transport theorem for each partitioned region $\Omega_\mathrm{cv}^{(\pm)}(t)$ by combining the formalism of flows introduced in Section~\ref{subsubsec:flows-and-generalized-rtt} with the technique of ``sampling motions'' introduced by Casey in Ref.~\cite{casey2011derivation}. In doing so, we argue that expressions of the form given by Eq.~\eqref{eq:generalized-transport-theorem} for the rates of change of integrals over the regions $\Omega^{(\pm)}_\mathrm{cv}(t)$ are valid even in the case where relative slip and mass deposition onto the membrane take place.

\begin{figure}
    \centering
    \includegraphics[width=0.5\textwidth]{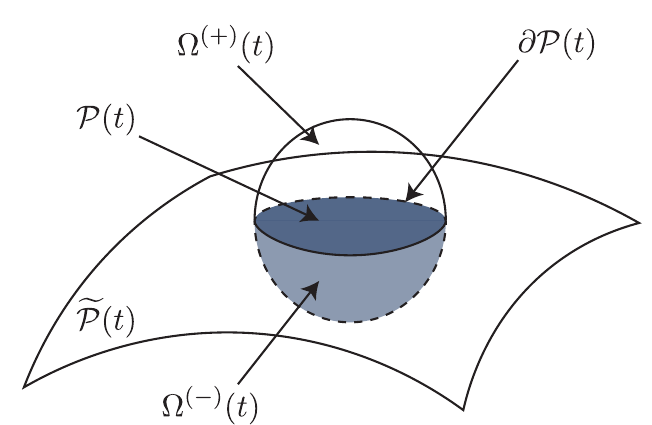}
    \caption{A bulk fluid neighbourhood $\Omega(t)$ transversely intersecting a membrane patch $\widetilde{\Pc}(t)$. The membrane patch intersection $\Pc(t) := \Omega(t) \cap \Mc(t)$ partitions the bulk fluid neighbourhood into two connected regions $\Omega^{(+)}(t)$ and $\Omega^{(-)}(t)$.}
    \label{fig:RTT-nbhds}
\end{figure}

\begin{figure}
    \centering
    \includegraphics[width=0.9\textwidth]{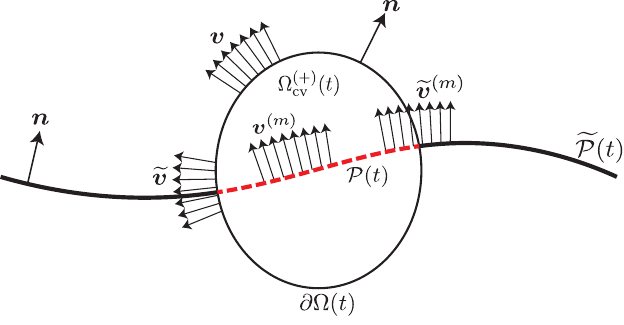}
    \caption{Smooth extension $\widetilde{\vb}$ of the bulk velocity field on $\partial\Omega^{(+)}\setminus\Pc(t)$ onto $\partial\Omega(t)$, and the natural extension $\vb^{(m)}$ of the velocity field on $\Pc(t)$ onto $\widetilde{\Pc}(t)$. Each body is evolved separately according to the flow of their respective velocity field. Note that $\Pc(t)$ is denoted by the dashed red line.}
    \label{fig:field-extension}
\end{figure}

\begin{figure}
	\centering
	\includegraphics[width=1.\textwidth]{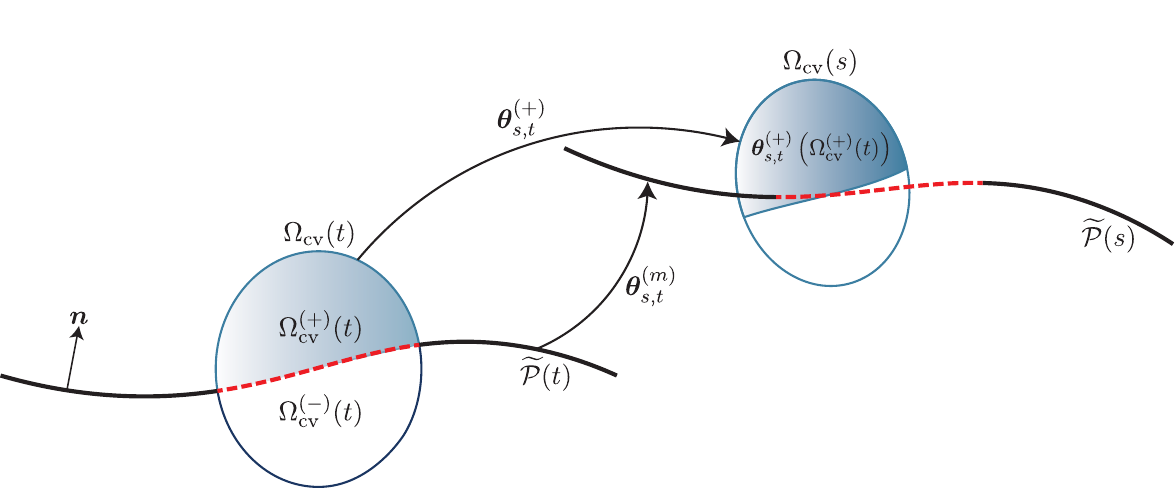}
	\caption{
        Visual depiction of the control volume $\Omega_\mathrm{cv}(t)$ mapped under the flow $\thetab_{s,t}^{(+)}$, and the extended membrane patch $\widetilde{\Pc}(t)$ mapped under the flow $\thetab_{s,t}^{(m)}$.
        Here, we see the need for the extended membrane patch, as $\Pc(s)$ does not partition $\Omega_\mathrm{cv}(s)$ into two regions, whereas $\widetilde{\Pc}(s)$ does. Additionally, the shaded region depicts the image of the upper region $\Omega_\mathrm{cv}^{(+)}(t)$ under the flow $\thetab_{s,t}^{(+)}$. 
	}
	\label{fig:flow-mapping}
\end{figure}

To this end, consider once again a scalar, vector, or tensor field $\fb(\xb,t)$ that is measurable within the region $\Omega_\mathrm{cv}(t)$ (e.g., any bounded function). The measurability condition will allow us to omit regions of measure zero in integrals of the quantity $\fb(\xb,t)$ without altering the value of the integral. We will first analyze only the upper region $\Omega_\mathrm{cv}^{(+)}(t)$, and an analogous result will follow  for the lower region $\Omega_\mathrm{cv}^{(-)}(t)$. 
To ensure that the bulk fluid region $\Omega_\mathrm{cv}(t)$ continues to be partitioned into an upper and a lower region $\Omega_\mathrm{cv}^{(\pm)}(t)$ as the membrane membrane $\Pc(t)$ and bulk fluid $\Omega_\mathrm{cv}(t)$ undergo independent motions, we will need to consider an open membrane patch $\widetilde{\Pc}(t)$ that is strictly bigger than the membrane patch $\Pc(t)$ (i.e., $\Pc(t) \subset \widetilde{\Pc}(t)$) as in Figure~\ref{fig:RTT-nbhds}. 
Now, consider an arbitrary smooth extension of the bulk fluid velocity field $\vb(\xb,t)$ defined in the upper fluid region $\Omega_\mathrm{cv}^{(+)}(t)$ to a velocity field $\widetilde{\vb}(\xb,t)$ defined over the entire control volume and boundary $\Omega_\mathrm{cv}(t) \cup \partial\Omega_\mathrm{cv}(t)$ (see Figure~\ref{fig:field-extension}). Note that in general, the velocity field $\widetilde{\vb}(\xb,t)$ will not agree with the material bulk velocity $\vb(\xb,t)$ within the lower region $\Omega^{(-)}(t)$ due to the presence of discontinuities. In addition, smoothly extend the membrane velocity field $\vb^{(m)}(\xb^{(m)},t)$ to a velocity field $\widetilde{\vb}^{(m)}(\xb,t)$ defined over the bulk fluid $\Bc(t)$, which includes $\widetilde{\Pc}(t)$. Then the extended bulk velocity field $\widetilde{\vb}(\xb,t)$ and the extended membrane velocity field $\widetilde{\vb}^{(m)}(\xb,t)$ are the infinitesimal generators of, respectively, the bulk flow $\thetab_{s,t}^{(+)}$ and the membrane flow $\thetab_{s,t}^{(m)}$ defined for $s \in \Jc := (t-\epsilon,t+\epsilon)$, with $\epsilon > 0$ (see Figure~\ref{fig:flow-mapping}).

\begin{figure}
	\centering
	\includegraphics[width=1.\textwidth]{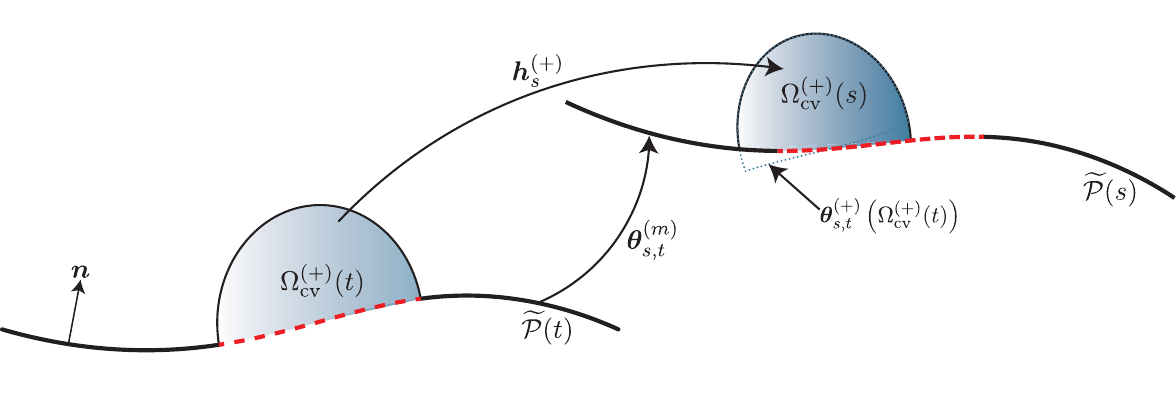}
	\caption{
		Schematic of a sampling motion:
        At the hypothetical time $s$, the image of the  region $\Omega_\mathrm{cv}^{(+)}(t)$ under the flow  $\thetab_{s,t}^{(+)}$, i.e. $\thetab_{s,t}^{(+)}\left( \Omega_\mathrm{cv}^{(+)}(t) \right)$, does not in general coincide with the enclosed upper region $\Omega_\mathrm{cv}^{(+)}(s)$. Also shown is a homeomorphic mapping $\hb_s^{(+)}$ between the enclosed upper bulk region $\Omega_\mathrm{cv}^{(+)}(t)$ at time $t$, and the enclosed upper bulk region $\Omega_\mathrm{cv}^{(+)}(s)$ at hypothetical time $s$. 
	}
	\label{fig:hs-mapping}
\end{figure}

Consider the flow of the regions 
\begin{align}
    \Omega_\mathrm{cv}(s) &:= \thetab_{s,t}^{(+)}\left(\Omega(t)\right) \,,\\
    \widetilde{\Pc}(s) &:= \thetab_{s,t}^{(m)}\left(\widetilde{\Pc}(t)\right) \,.
\end{align}
For sufficiently small $\epsilon$, the regions $\Omega_\mathrm{cv}(s)$ and $\widetilde{\Pc}(s)$ always enclose an upper region $\Omega_\mathrm{cv}^{(+)}(s)$ (Figure~\ref{fig:hs-mapping}). Note that the region $\Omega_\mathrm{cv}^{(+)}(s)$ is not material with respect to the flow $\thetab_{s,t}^{(+)}$, i.e., $\thetab_{s,t}^{(+)}\big(\Omega_\mathrm{cv}^{(+)}(t)\big) \neq \Omega_\mathrm{cv}^{(+)}(s)$. 
Yet, the two regions $\Omega_\mathrm{cv}^{(+)}(t)$ and $\Omega_\mathrm{cv}^{(+)}(s)$
are homeomorphic to each other since no topological changes have occurred. As indicated in Figure~\ref{fig:hs-mapping}, let $\hb_s^{(+)} \colon \Omega_\mathrm{cv}^{(+)}(t) \to \Omega_\mathrm{cv}^{(+)}(s)$ denote a one-parameter family of homeomorphisms that give the enclosed region $\Omega_\mathrm{cv}(s)$ for all $s \in \Jc$.
Since the movement of both $\Omega_\mathrm{cv}(s)$ and $\Pc(s)$ is smooth, we can choose this family of homeomorphisms $\hb_s^{(+)}(\xb)$ to be continuously differentiable with respect to both the parameter $s$ and the position coordinate $\xb$\footnote{More precisely, we require that $\hb_s^{(+)}(\xb)$ is continuously differentiable (i.e., in the class of $C^1$ functions) with $\det\left(\dfrac{\partial\hb_s^{(+)}(\xb)}{\partial\xb}\right) \ne 0$ at all points $\xb$ in $\Omega_\mathrm{cv}^{(+)}(t)$. It then follows from the inverse function theorem that $\hb_s^{(+)}(\xb)$ has an inverse that is also continuously differentiable (see \cite[Section~1.6]{marsden1994mathematical}); we call such functions diffeomorphisms. Furthermore, since we are interested in integrals involving $\hb_s^{(+)}(\xb)$, it is sufficient that $\hb_s^{(+)}(\xb)$ is a diffeomorphism almost everywhere in $\Omega_\mathrm{cv}^{(+)}(t)$.}. Additionally, we can choose $\hb_s^{(+)}(\xb)$ be the identity map at $s = t$. When these conditions are satisfied, such a family of homeomorphisms $\hb_s^{(+)}(\xb)$ is referred to as a ``sampling motion" \cite{casey2011derivation}. 

The sampling motion $\hb_s^{(+)}(\xb)$ satisfies the following properties:
\begin{align}
    \hb_t^{(+)}(\xb) = \xb &\quad \text{for all } \xb \in \Omega_\mathrm{cv}^{(+)}(t)\,,
    \label{eq:samp-motion-property-1}\\
    \left[\frac{d\hb_s^{(+)}(\xb)}{ds} \cdot \nb \right|_{s=t} = \vb_\brm(\xb,t) \cdot \nb 
    &\quad \text{for all } \xb \in \partial\Omega_\mathrm{cv}^{(+)}(t)\,,
    \label{eq:samp-motion-property-2}
\end{align}
where the boundary velocity field $\vb_\brm(\xb,t)$ is equal to the bulk velocity $\vb(\xb,t)$ on $\partial\Omega_\mathrm{cv}^{(+)}(t)\setminus\Pc(t)$ and the membrane velocity $\vb^{(m)}(\xb^{(m)},t)$ on $\Pc(t)$. These properties are similar to the properties of a flow described by Eq.~\eqref{eq:time-dep-flow-properties}, except Eq.~\eqref{eq:samp-motion-property-2} requires that only the normal components of the boundary velocity induced by the sampling motion $\hb_s^{(+)}(\xb)$ and the boundary velocity of the upper region $\Omega_\mathrm{cv}^{(+)}(s)$ agree at time $t$. 
This is because~$\hb_s^{(+)}\left( \Omega_\mathrm{cv}^{(+)}(t) \right)$ encloses the same region of space as~$\Omega_\mathrm{cv}^{(+)}(s)$ (i.e., $\hb_s^{(+)}\left(\Omega_\mathrm{cv}^{(+)}(t)\right) = \Omega_\mathrm{cv}^{(+)}(s)$ for all $s \in \Jc$), and so the geometric shape of the two regions is the same. Since, the shape of a region can only be changed by displacing its boundary in the direction of its normal $\nb$, it follows that the normal component of the boundary velocities for both $\hb_s^{(+)}\left( \Omega_\mathrm{cv}^{(+)}(t) \right)$ and $\Omega_\mathrm{cv}^{(+)}(s)$ must be the same.

The time rate of change of the bulk quantity $\fb(\xb,t)$ within the region $\Omega_\mathrm{cv}^{(+)}(t)$ can now be computed using the sampling motion $\hb_s^{(+)}(\xb)$ as 
\begin{align}
    \frac{d}{dt} \int_{\Omega_\mathrm{cv}^{(+)}(t)} \fb(\xb,t) \,dv 
    &= \left[ \frac{d}{ds} \int_{\hb_s^{(+)}(\Omega_\mathrm{cv}^{(+)}(t))} \fb(\hb_s^{(+)}(\xb), s) \,dv(s) \right|_{s=t} \,.
\end{align}
Pulling back this integral to time $t$ gives
\begin{align}
    \frac{d}{dt} \int_{\Omega_\mathrm{cv}^{(+)}(t)} \fb(\xb,t) \,dv 
    &= \left[ \frac{d}{ds} \int_{\Omega_\mathrm{cv}^{(+)}(t)} \fb(\hb_s^{(+)}(\xb), s) \det\left(\pder{\hb_s^{(+)}(\xb)}{\xb}\right) \,dv \right|_{s=t} \,.
\end{align}
Carrying out the differentiation in the same manner as Section~\ref{subsubsec:flows-and-generalized-rtt} and using Eqs.~\eqref{eq:samp-motion-property-1} and~\eqref{eq:samp-motion-property-2} yields
\begin{align}
    \frac{d}{dt} \int_{\Omega_\mathrm{cv}^{(+)}(t)} \fb(\xb,t) \,dv 
    &= \int_{\Omega_\mathrm{cv}^{(+)}(t)} \left(\pder{\fb(\xb,t)}{t} + \divg( \fb(\xb,t) \otimes \left[\frac{d\hb_s^{(+)}(\xb)}{ds}\right|_{s=t} \right) \,dv\\
    &= \int_{\Omega_\mathrm{cv}^{(+)}(t)} \pder{\fb(\xb,t)}{t} \,dv + \int_{\partial\Omega_\mathrm{cv}^{(+)}(t)} \fb^{(+)}(\xb,t) (\vb_\brm(\xb,t) \cdot \nb) \,da \,,
    \label{eq:gen-RTT-Omega-plus}
\end{align}
which establishes the generalized transport theorem for a piecewise smooth boundary. Note that when applying the divergence theorem to obtain Eq.~\eqref{eq:gen-RTT-Omega-plus}, the value of the bulk quantity $\fb(\xb,t)$ must take on its limiting value $\fb^{(+)}(\xb,t)$. 

In an analogous fashion, we can find a sampling motion $\hb_s^{(-)}(\xb)$ that satisfies 
\begin{align}
    \hb_t^{(-)}(\xb) = \xb &\quad \text{for all } \xb \in \Omega_\mathrm{cv}^{(-)}(t)\,,\\
    \left[\frac{d\hb_s^{(-)}(\xb)}{ds} \cdot \nb \right|_{s=t} = \vb_\brm(\xb,t) \cdot \nb 
    &\quad \text{for all } \xb \in \partial\Omega_\mathrm{cv}^{(-)}(t)\,,
\end{align}
from which it follows that 
\begin{align}
    \frac{d}{dt} \int_{\Omega_\mathrm{cv}^{(-)}(t)} \fb(\xb,t) \,dv 
    &= \left[ \frac{d}{ds} \int_{\hb_s^{(-)}(\Omega_\mathrm{cv}^{(-)}(t))} \fb(\hb_s^{(-)}(\xb), s) \,dv(s) \right|_{s=t}\\
    &= \int_{\Omega_\mathrm{cv}^{(-)}(t)} \pder{\fb(\xb,t)}{t} \,dv + \int_{\partial\Omega_\mathrm{cv}^{(-)}(t)} \fb^{(-)}(\xb,t) (\vb_\brm(\xb,t) \cdot \nb) \,da \,,
    \label{eq:gen-RTT-Omega-minus}
\end{align}
which is the Generalized transport theorem applied to the lower region $\Omega_\mathrm{cv}^{(-)}(t)$.

\subsubsection{Proof of the Modified Reynolds Transport Theorem}
\label{subsubsec:proof-of-modified-rtt}

\begin{figure}
\centering
	\includegraphics[width=0.8\textwidth]{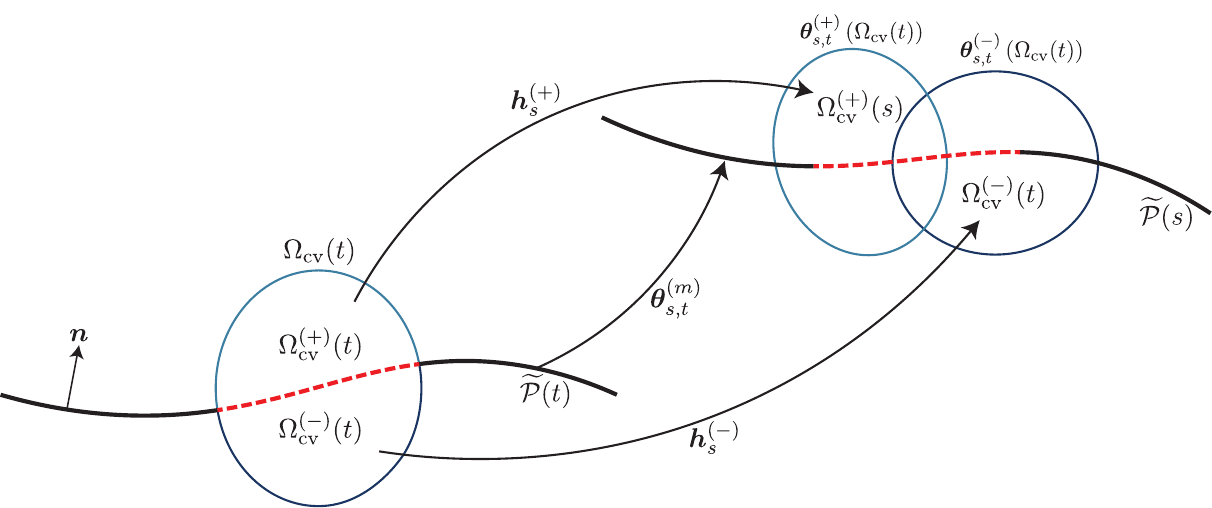}
	\caption{
		The three motions $\hb_s^{(+)}$, $\hb_s^{(-)}$, and $\thetab_{s,t}^{(m)}$ used to prove the modified Reynolds transport theorem. Since the bodies $\Omega_\mathrm{cv}^{(+)}(t)$, $\Omega_\mathrm{cv}^{(-)}(t)$, and $\widetilde{\Pc}(t)$ can undergo relative slip to one another, their boundaries at the hypothetical time $s$ may not coincide. Note that the red dashed line on the left depicts the membrane patch $\Pc(t)$, while the red dashed line on the right indicates the deformed membrane $\Pc(s) := \thetab_{s,t}^{(m)}(\Pc(t))$.}
	\label{fig:theta-mapping}
\end{figure}

We now return to analyzing the material time derivative of Eq.~\eqref{eq:appA-I-intro}. Since the membrane surface $\Pc(t)$ has measure zero inside the bulk region $\Omega(t)$, we can equivalently write Eq.~\eqref{eq:appA-I-intro} as 
\begin{align}
    \Ib(t) &= \int_{\Omega_\mathrm{cv}(t)\setminus\Pc(t)} \fb(\xb,t) \,dv + \int_{\Pc(t)} \fb^{(m)}(\xb^{(m)},t) \,da\\
    &= \int_{\Omega_\mathrm{cv}^{(+)}(t)} \fb(\xb,t) \,dv + \int_{\Omega_\mathrm{cv}^{(-)}(t)} \fb(\xb,t) \,dv + \int_{\Pc(t)} \fb^{(m)}(\xb^{(m)},t) \,da \,.
    \label{eq:appA-I-2}
\end{align}
In the presence of a material surface of discontinuity, bulk fluid particles can accumulate onto the membrane, effectively disappearing from the bulk system. This conversion of material from the bulk fluid to the membrane can be thought of as a chemical reaction, and could be postulated as such at the stage where we write the balance laws. Here, however, we would like to automatically incorporate this exchange into the Reynolds transport theorem. Although we cannot give a Lagrangian description of the bulk fluid since material particles can ``disappear'' onto the membrane, it is sufficient to consider \textit{any} hypothetical motion that \textit{instantaneously} has the same tendency to evolve as the combined bulk and membrane system. In other words, by choosing the motion of the control volumes $\Omega_\mathrm{cv}^{(+)}(t)$ and $\Omega_\mathrm{cv}^{(-)}(t)$ in Eq.~\eqref{eq:appA-I-2} appropriately, the mathematical time rate change of each region can be made to be equal to the material time derivative $\dot{\Ib}(t)$ of the extensive quantity $\Ib(t)$. We require only that the normal component of the boundary velocity $\vb_\brm \cdot \nb$ of $\partial\Omega_\mathrm{cv}(t)$ to \textit{instantaneously} match the normal component of the bulk fluid velocity $\vb \cdot \nb$ at time $t$, and for the motion of the common boundary between the two partitions $\partial\Omega_\mathrm{cv}^{(+)}(t) \cap \partial \Omega_\mathrm{cv}^{(-)}(t) = \Pc(t)$ to be the same. Using the flow $\thetab_{s,t}^{(m)}$ and the sampling motions $\hb_s^{(\pm)}$ developed in Section~\ref{subsubsec:gen-rtt-piecewise-smooth-boundary} (see Figure~\ref{fig:theta-mapping}), it follows that the material time derivative is given by 
\begin{align}
    \begin{split}
    \dot{\Ib}(t) &= \left[ \frac{d}{ds} \int_{\hb_s^{(+)}(\Omega_\mathrm{cv}^{(+)}(t))} \fb(\hb_s^{(+)}(\xb), s) \,dv(s) \right|_{s=t}
    \  \\
    &\hspace{0.5in} + \left[ \frac{d}{ds} \int_{\hb_s^{(-)}(\Omega_\mathrm{cv}^{(-)}(t))} \fb(\hb_s^{(-)}(\xb), s) \,dv(s) \right|_{s=t} 
    \ \\
    &\hspace{1in} + \left[ \frac{d}{ds} \int_{\thetab_{s,t}^{(m)}(\Pc(t))} \fb^{(m)}(\thetab_{s,t}^{(m)}(\xb^{(m)}),s) \,da(s) \right|_{s=t} \,.
    \end{split}
    \label{eq:appA-I-dot}
\end{align}

First, note that since $\thetab_{s,t}^{(m)}$ is the true material motion of the membrane, the third term on the right-hand-side of Eq.~\eqref{eq:appA-I-dot} gives rise to the surface transport theorem (Eq.~\eqref{eq:surface-reynolds-transport}). For the first and second terms, we can substitute in Eqs.~\eqref{eq:gen-RTT-Omega-plus} and~\eqref{eq:gen-RTT-Omega-minus}, respectively, to obtain 
\begin{align}
    \begin{split}
    \dot{\Ib}(t) &=  \int_{\Omega_\mathrm{cv}^{(+)}(t)} \pder{\fb(\xb,t)}{t} \,dv + \int_{\partial\Omega_\mathrm{cv}^{(+)}(t)} \fb^{(+)}(\xb,t) (\vb_\brm(\xb,t) \cdot \nb) \,da \\
    &\hspace{0.5in} + \int_{\Omega_\mathrm{cv}^{(-)}(t)} \pder{\fb(\xb,t)}{t} \,dv + \int_{\partial\Omega_\mathrm{cv}^{(-)}(t)} \fb^{(-)}(\xb,t) (\vb_\brm(\xb,t) \cdot \nb) \,da \\
    &\hspace{1in} + \int_{\Pc(t)} \left(\dot{\fb}^{(m)}(\xb^{(m)},t) + \fb^{(m)}(\xb^{(m)},t)\divg_\text{S} \vb^{(m)}(\xb^{(m)},t)\right) \,da \,.
    \end{split}
    \label{eq:appA-I-dot-2}
\end{align}
From here on, we will omit the arguments of the functions being integrated. Since the membrane boundary $\partial\Pc(t)$ has measure zero in $\partial\Omega_\mathrm{cv}^{(\pm)}(t)$, we can split the boundary integral terms in Eq.~\eqref{eq:appA-I-dot-2} as 
\begin{align}
    \begin{split}
    \dot{\Ib}(t) &=  \int_{\Omega_\mathrm{cv}^{(+)}(t)} \pder{\fb}{t} \,dv + \int_{\partial\Omega_\mathrm{cv}^{(+)}(t)\setminus\Pc(t)} \fb (\vb_\brm \cdot \nb) \,da + \int_{\Pc(t)} \fb^{(+)} (\vb^{(m)} \cdot (-\nb)) \,da \\
    &\hspace{0.5in} + \int_{\Omega_\mathrm{cv}^{(-)}(t)} \pder{\fb}{t} \,dv + \int_{\partial\Omega_\mathrm{cv}^{(-)}(t)\setminus\Pc(t)} \fb (\vb_\brm \cdot \nb) \,da + \int_{\Pc(t)} \fb^{(-)} (\vb^{(m)} \cdot \nb) \,da \\
    &\hspace{1in} + \int_{\Pc(t)} \left(\dot{\fb}^{(m)} + \fb^{(m)}\divg_\text{S} \vb^{(m)}\right) \,da \,.
    \end{split}
    \label{eq:appA-I-dot-3}
\end{align}
Since the membrane boundary $\partial\Pc(t)$ has measure zero in $\partial\Omega_\mathrm{cv}(t)$, the regions $\Omega^{(\pm)}(\tau)$ are disjoint, and $\Omega^{(+)}(t) \cup \Omega^{(-)}(t) = \Omega(t) \setminus \Pc(t)$, we can further rearrange Eq.~\eqref{eq:appA-I-dot-3} to obtain   
\begin{align}
    \begin{split}
    \dot{\Ib}(t) &= \int_{\Omega_\mathrm{cv}(t)} \pder{\fb}{t} \,dv + \int_{\partial\Omega_\mathrm{cv}(t)} \fb (\vb \cdot \nb) \,da - \int_{\Pc(t)} \left(\jump{\fb} \otimes \vb^{(m)} \right) \nb \,da \\
    &\hspace{0.5in} + \int_{\Pc(t)} \left(\dot{\fb}^{(m)} + \fb^{(m)}\divg_\text{S} \vb^{(m)}\right) \,da \,.
    \end{split}
    \label{eq:appA-I-dot-4}
\end{align}
Applying the modified divergence theorem (Eq.~\eqref{eq:appA-jump-div-thm-tensor}) to the second term of Eq.~\eqref{eq:appA-I-dot-4} then yields 
\begin{align}
    \begin{split}
    \dot{\Ib}(t) &=  \int_{\Omega_\mathrm{cv}(t)} \left[ \pder{\fb}{t} + \divg\left(\fb \otimes \vb\right) \right] \,dv + \int_{\Pc(t)} \left(\jump{\fb \otimes \left(\vb - \vb^{(m)}\right) } \right) \nb \,da \\
    &\hspace{0.5in} + \int_{\Pc(t)} \left(\dot{\fb}^{(m)} + \fb^{(m)}\divg_\text{S} \vb^{(m)}\right) \,da \,.
    \end{split}
    \label{eq:appA-I-dot-5}
\end{align}
Since the control volume $\Omega_\mathrm{cv}(t)$ overlaps with the material region $\Omega(t)$ at time $t$, we can integrate over the latter region instead and rewrite Eq.~\eqref{eq:appA-I-dot-5} as 
\begin{align}
    \begin{split}
    \dot{\Ib}(t) &=  \int_{\Omega(t)} \left(\dot{\fb} + \fb\divg(\vb) \right)\,dv
    \\
    &\hspace{0.5in} + \int_{\Pc(t)} \left(\dot{\fb}^{(m)} + \fb^{(m)}\divg_\text{S} \vb^{(m)} + \jump{\fb \otimes \left(\vb - \vb^{(m)}\right) } \nb\right) \,da \,.
    \end{split}
    \label{eq:appA-I-dot-6}
\end{align}
Equation~\eqref{eq:appA-I-dot-6} gives the modified Reynolds transport theorem.

Similar to the modified divergence theorem, the modified Reynolds transport theorem \eqref{eq:appA-I-dot-6} is easily extended to scalar and Tensor fields. Observe that for some scalar field $\varphi$ and a constant vector $\cb \in \bbR^3$,
\begin{align}
    \cb \cdot \frac{d}{dt} \Bigg( &\int_{\Omega(t)} \varphi \,dv + \int_{\Pc(t)} \varphi^{(m)} \,da \Bigg)\\
    &=  \frac{d}{dt} \Bigg( \int_{\Omega(t)} \cb\varphi \,dv + \int_{\Pc(t)} \cb\varphi^{(m)} \,da \Bigg)\\
    \begin{split}
        &= \int_{\Omega(t)} \left(\frac{d(\cb \varphi)}{dt} + (\cb \varphi)\divg(\vb) \right)\,dv\\
        &\hspace{0.5in} + \int_{\Pc(t)} \left( \frac{d(\cb\varphi^{(m)})}{dt} + \cb\varphi^{(m)}\divg_\text{S} \vb^{(m)} + \jump{(\cb \varphi) \otimes (\vb - \vb^{(m)})} \nb \right) \,da
    \end{split}\\
    \begin{split}
        &= \cb \cdot \Bigg(\int_{\Omega(t)} (\dot{\varphi} + \varphi\divg\vb) \,dv\\
        &\hspace{0.75in} + \int_{\Pc(t)} \left(\dot{\varphi}^{(m)} + \varphi^{(m)}\divg_\text{S} \vb^{(m)} + \jump{\varphi(\vb - \vb^{(m)})} \cdot \nb \right) \,da\Bigg)
    \end{split}\\
    \begin{split}
        \iff \frac{d}{dt}\Bigg( &\int_{\Omega(t)} \varphi \,dv + \int_{\Pc(t)} \varphi^{(m)} \,da \Bigg)
        = \int_{\Omega(t)} (\dot{\varphi} + \varphi\divg\vb) \,dv\\
        &\hspace{0.75in} + \int_{\Pc(t)} \left(\dot{\varphi}^{(m)} + \varphi^{(m)}\divg_\text{S} \vb^{(m)} + \jump{\varphi(\vb - \vb^{(m)})} \cdot \nb \right) \,da \, ,
    \end{split}
    \label{eq:appA-rtt-scalar}
\end{align}
where we used the modified Reynolds transport theorem for vectors in \eqref{eq:appA-I-dot-6}. 
For tensor fields $\Fb$, make use of the fact that for all tensor fields $\Ab,\Bb \in \text{Lin}(\bbR^3, \bbR^3)$, 
\begin{equation}
    \Ab = \Bb \iff \Ab\cb = \Bb\cb \quad \forall \cb \in \bbR^3 \,.
\end{equation}
Observe,
\begin{align}
    \frac{d}{dt} \Bigg( &\int_{\Omega(t)} \Fb \,dv + \int_{\Pc(t)} \Fb^{(m)} \,da \Bigg) \cb\\
    &=  \frac{d}{dt} \Bigg( \int_{\Omega(t)} \Fb\cb \,dv + \int_{\Pc(t)} \Fb^{(m)}\cb \,da \Bigg)\\
    \begin{split}
        &= \int_{\Omega(t)} \left(\frac{d(\Fb\cb)}{dt} + (\Fb\cb)\divg(\vb) \right)\,dv\\
        &\hspace{0.5in} + \int_{\Pc(t)} \left( \frac{d(\Fb^{(m)}\cb)}{dt} + (\Fb^{(m)}\cb)\divg_\text{S} \vb^{(m)} + \jump{( \Fb\cb) \otimes (\vb - \vb^{(m)})} \nb \right) \,da
    \end{split}\\
    \begin{split}
        &= \Bigg(\int_{\Omega(t)} (\dot{\Fb} + \Fb\divg\vb) \,dv\\
        &\hspace{0.75in} + \int_{\Pc(t)} \left(\dot{\Fb}^{(m)} + \Fb^{(m)}\divg_\text{S} \vb^{(m)} + \jump{\left[(\vb - \vb^{(m)}) \cdot \nb\right] \Fb} \cdot \nb \right) \,da\Bigg) \cb \,,
    \end{split}\\
    \begin{split}
        \iff \frac{d}{dt}\Bigg( &\int_{\Omega(t)} \Fb \,dv + \int_{\Pc(t)} \Fb^{(m)} \,da \Bigg)
        = \int_{\Omega(t)} (\dot{\Fb} + \Fb\divg\vb) \,dv\\
        &\hspace{0.75in} + \int_{\Pc(t)} \left(\dot{\Fb}^{(m)} + \Fb^{(m)}\divg_\text{S} \vb^{(m)} + \jump{\Fb \otimes (\vb - \vb^{(m)})} \nb \right) \,da \,.
    \end{split}
    \label{eq:appA-rtt-tensor}
\end{align}
Equations~\eqref{eq:appA-I-dot-6},~\eqref{eq:appA-rtt-scalar}, and~\eqref{eq:appA-rtt-tensor} establish the modified Reynolds transport theorems for vector, scalar, and tensor fields, respectively.

\paragraph{Remark.} Due to the exchange of material between the membrane and the bulk fluid, the material rate of change of an extensive quantity $\Ib(t) = \Ib_1(t) + \Ib_2(t)$ is evaluated on the composite membrane and bulk system $\Omega(t) \oplus \Pc(t)$ (see Eq.~\eqref{eq:appA-I-intro} and the discussion in the paragraph proceeding it). When there is no deposition onto the membrane, however, it is possible to separate Eq.~\eqref{eq:appA-I-dot-6} into a bulk modified transport theorem and a surface Reynolds transport theorem. That is, when no exchange of material takes place between $\Omega(t)$ and $\Pc(t)$, the material nature of the membrane need not be taken into account so that we can set $\fb^{(m)} \equiv 0$ in Eq.~\eqref{eq:appA-I-dot-6}. 
This yields separate expressions for the material time derivative of the bulk integral $\Ib_1(t)$, and the membrane integral $\Ib_2(t)$:
\begin{align}
    \dot{\Ib}_1(t) &= \frac{d}{dt}\int_{\Omega(t)} \fb \,dv = \int_{\Omega(t)} \left(\dot{\fb} + \fb\divg(\vb) \right)\,dv + \int_{\Pc(t)} \jump{\fb \otimes (\vb - \vb^{(m)})} \nb \,da \,, \label{eq:appA-rtt-separated-I1}\\
    \dot{\Ib}_2(t) &= \frac{d}{dt} \int_{\Pc(t)} \fb^{(m)} \,da = \int_{\Pc(t)} \left(\dot{\fb}^{(m)} + \fb^{(m)}\divg_\text{S} \vb^{(m)}\right) \,da \,.
\end{align}
Equation~\eqref{eq:appA-rtt-separated-I1} gives the form of the modified transport theorem that appears in previous works \cite{casey2011derivation} on non-material interfaces, though here it remains valid even when tangential slip between the membrane and the bulk takes place. Equation~\eqref{eq:appA-rtt-separated-I1} gives the form of the surface Reynolds transport theorem that appears in previous work that analyzes standalone membranes \cite{sahu-mandadapu-pre-2017}.

\section{Thermodynamic Relations for Curved Surfaces} 
\label{sec:thermodynamic-relations-curved-surfaces}
In this section, we present a derivation of the Euler equation for a curved surface. The extensive Helmholtz energy $\Ft^{(m)}$ of a surface patch $\Pc$ is specified by its area $A$, shape as determined by the metric tensor $a_{\alpha\beta}$ and curvature tensor $b_{\alpha\beta}$, temperature $T$, and molar composition $n_1^{(m)},\dots, n_N^{(m)}$, i.e., 
\begin{align}
    \Ft^{(m)} &= \Ft^{(m)}(A, a_{\alpha\beta}, b_{\alpha\beta}, T, n_1^{(m)},\dots, n_N^{(m)}) \,.
    \label{eq:extensive-membrane-Helmholtz-free-energy}
\end{align}
In writing Eq.~\eqref{eq:extensive-membrane-Helmholtz-free-energy}, note that two surfaces with the same first fundamental form $\mathrm{I} := d\xb^{(m)} \cdot d\xb^{(m)} = a_{\alpha\beta} \, d\theta^\alpha d\theta^\beta$ and second fundamental form $\mathrm{I\!I} := -d\xb^{(m)} \cdot d\nb = b_{\alpha\beta} \,d\theta^\alpha d\theta^\beta$ are identical up to rotations and translations \cite[pp. 247-249]{pressley2010elementary}. 
Moreover, if the first and second fundamental forms satisfy the Codazzi-Mainardi and Gauss equations~\cite{ciarlet2002recovery}, then there exists a surface patch $\Pc$ with $a_{\alpha\beta}$ and $b_{\alpha\beta}$ as its metric and curvature tensors. This is sometimes referred to as the Fundamental Theorem of Surface Theory or Bonnet's theorem \cite[p. 236]{carmo1976differential}.
This means that although~$a_{\alpha\beta}$ and~$b_{\alpha\beta}$ are in general not completely independent of one another, we can treat them as independent variables in the functional form of $\Ft^{(m)}$ that determines the free energy.

With the local equilibrium assumption in place, one can imagine that a given membrane patch may be made up of many smaller patches, each of which is an equilibrium system, and endowed with constant metric tensor $a_{\alpha\beta}$ and curvature tensor $b_{\alpha\beta}$. Then the differential of Eq.~\eqref{eq:extensive-membrane-Helmholtz-free-energy} is 
\begin{align}
\begin{split}
    d\Ft^{(m)} &= \pder{\Ft^{(m)}}{A} \,dA + \frac{1}{2}\left( \pder{\Ft^{(m)}}{a_{\alpha\beta}} + \pder{\Ft^{(m)}}{a_{\beta\alpha}} \right) da_{\alpha\beta} + \frac{1}{2}\left( \pder{\Ft^{(m)}}{b_{\alpha\beta}} + \pder{\Ft^{(m)}}{b_{\beta\alpha}} \right) db_{\alpha\beta} \\
    &\qquad \hspace{1in} + \pder{\Ft^{(m)}}{T} \,dT + \sum_{i} \pder{\Ft^{(m)}}{n_i^{(m)}} \,dn_i^{(m)} \,.
    \label{eq:extensive-membrane-Helmholtz-differential}
\end{split}
\end{align}
Applying the local equilibrium assumption, we have that 
\begin{align}
    \gamma^{(m)} := \pder{\Ft^{(m)}}{A}
    \,, \quad 
    S^{(m)} := -\pder{\Ft^{(m)}}{T}
    \,, \quad \text{and} \quad 
    \mu_i^{(m)} := \pder{\Ft^{(m)}}{n_i^{(m)}} 
    \,,
\end{align}
where $\gamma^{(m)}$ is the surface tension, $S^{(m)}$ is the extensive entropy of the surface, and $\mu_i^{(m)}$ the chemical potential of surface species ``$i$''. 

Defining the intensive Helmholtz free-energy per unit area as $\ft^{(m)} := \Ft^{(m)} / A$, Eq.~\eqref{eq:extensive-membrane-Helmholtz-differential} becomes 
\begin{align}
    d\Ft^{(m)} &= \ft^{(m)} \,dA + A \,d\ft^{(m)}\\
\begin{split}
    &= \gamma^{(m)} \,dA + \frac{1}{2}A\left( \pder{\ft^{(m)}}{a_{\alpha\beta}} + \pder{\ft^{(m)}}{a_{\beta\alpha}} \right) da_{\alpha\beta} + \frac{1}{2}A\left( \pder{\ft^{(m)}}{b_{\alpha\beta}} + \pder{\ft^{(m)}}{b_{\beta\alpha}} \right) db_{\alpha\beta} \\
    &\qquad \hspace{0.5in} - A \rho^{(m)} s^{(m)} \,dT + \sum_{i} \mu_i^{(m)} \,\left(A\,dc_i^{(m)} + c_i^{(m)} \,dA\right) \,.
\end{split}
\end{align}
Collecting all the terms in $A$ and $dA$, we have  
\begin{multline}
    A \Bigg[d\ft^{(m)} - \frac{1}{2}\left( \pder{\ft^{(m)}}{a_{\alpha\beta}} + \pder{\ft^{(m)}}{a_{\beta\alpha}} \right) da_{\alpha\beta} - \frac{1}{2}\left( \pder{\ft^{(m)}}{b_{\alpha\beta}} + \pder{\ft^{(m)}}{b_{\beta\alpha}} \right) db_{\alpha\beta} \\ 
    + \rho^{(m)} s^{(m)} \,dT - \sum_{i} \mu_i^{(m)} \,dc_i^{(m)} \Bigg]
    + dA \left[\ft^{(m)} - \gamma^{(m)} - \sum_{i} \mu_i^{(m)}  c_i^{(m)} \right] = 0 \,.
    \label{eq:surface-area-differential-collected}
\end{multline}
As we shrink the area of the surface patch $\Pc$ (i.e., as we send $A \to 0$), Eq.~\eqref{eq:surface-area-differential-collected} yields the Euler relation 
\begin{align}
    \ft^{(m)} &= \gamma^{(m)} + \sum_{i} \mu_i^{(m)}  c_i^{(m)} \,,
    \label{eq:surface-Euler-relation}
\end{align}
as well as the fundamental form 
\begin{align}
    d\ft^{(m)} &= \frac{1}{2}\left( \pder{\ft^{(m)}}{a_{\alpha\beta}} + \pder{\ft^{(m)}}{a_{\beta\alpha}} \right) da_{\alpha\beta} + \frac{1}{2}\left( \pder{\ft^{(m)}}{b_{\alpha\beta}} + \pder{\ft^{(m)}}{b_{\beta\alpha}} \right) db_{\alpha\beta} - \rho^{(m)} s^{(m)} \,dT + \sum_{i} \mu_i^{(m)} \,dc_i^{(m)} \,.
    \label{eq:surface-fundamental-form}
\end{align}
Taking the differential of Eq.~\eqref{eq:surface-Euler-relation} and subtracting Eq.~\eqref{eq:surface-fundamental-form}, we obtain a Gibbs-Duhem relation for a curved surface: 
\begin{multline}
    d\gamma^{(m)} - \frac{1}{2}\left( \pder{\ft^{(m)}}{a_{\alpha\beta}} + \pder{\ft^{(m)}}{a_{\beta\alpha}} \right) da_{\alpha\beta} - \frac{1}{2}\left( \pder{\ft^{(m)}}{b_{\alpha\beta}} + \pder{\ft^{(m)}}{b_{\beta\alpha}} \right) db_{\alpha\beta}\\ 
    + \rho^{(m)} s^{(m)} \,dT + \sum_{i} c_i^{(m)} d\mu_i^{(m)} = 0 \,.
    \hspace{0.5in}
\end{multline}

Using the constitutive relations from Eqs.~\eqref{eq:sigma-ab-constitutive-law} and~\eqref{eq:M-ab-constitutive-law}, we can also rewrite the fundamental form (Eq.~\eqref{eq:surface-fundamental-form})---sometimes also called the Gibbs equation---in terms of the membrane stresses and moments. Since Eq.~\eqref{eq:surface-fundamental-form} assumes that a reversible path has been taken, we must also assume that the membrane deforms in such a way that the stresses that lead to viscous dissipation are zero (i.e., $\pi^{\alpha\beta} \equiv 0$). In this case, the in-plane stresses that arise are those related to the elastic (reversible) deformation of the membrane, which we denote by
\begin{align}
    \left(\sigma^{\alpha\beta}\right)^{\mathrm{el}} &:= \left(\dfrac{\partial \ft^{(m)}}{\partial a_{\alpha\beta}} + \dfrac{\partial \ft^{(m)}}{\partial a_{\beta\alpha}}\right) + \gamma^{(m)} a^{\alpha\beta}
    \,.
    \label{eq:sigma-ab-elastic-constitutive-law}
\end{align}
Additionally, for a membrane with no accumulation from the bulk, the total mass balance $\dot{\rho}^{(m)} = -\rho^{(m)}\divg_\text{S}\cdot\vb^{(m)}$ (Eq.~\eqref{eq:local-membrane-mass-balance-spec-i-no-acc}) can be combined with the kinematic identity given by Eq.~\eqref{eq:J-dot} to obtain the differential form 
\begin{align}
    d\rho^{(m)} &= -\frac{1}{2}\rho^{(m)} a^{\alpha\beta} \,da_{\alpha\beta} \,.
    \label{eq:membrane-mass-bal-no-acc-differential-form}
\end{align}
Substituting Eqs.~\eqref{eq:sigma-ab-elastic-constitutive-law} and~\eqref{eq:M-ab-constitutive-law} into Eq.~\eqref{eq:surface-fundamental-form}, and incorporating Eq.~\eqref{eq:membrane-mass-bal-no-acc-differential-form}, we obtain the Gibbs Equation 
\begin{align}
\begin{split}
     \rho^{(m)} \,du^{(m)} &= \rho^{(m)} T ds^{(m)} + \left[\frac{1}{2}\left(\sigma^{\alpha\beta}\right)^\mathrm{el} \,da_{\alpha\beta} + M^{\alpha\beta} \,db_{\alpha\beta}\right] \\
     &\qquad \hspace{1in} 
     + \sum_{i=1}^N \frac{\mu_i^{(m)}}{\scrM_i} \left( d\rho_i^{(m)} + \frac{1}{2}\rho_i^{(m)} a^{\alpha\beta} \,da_{\alpha\beta} \right)
     \,.
     \label{eq:surface-Gibbs-equation}
\end{split}
\end{align}
Lastly, Eq.~\eqref{eq:surface-Gibbs-equation} is analogous to the Gibbs relation $dU = T \,dS - p \,dV + \mu \,dN$ for a macroscopic three-dimensional bulk system.

\end{appendices}
\vspace{45pt}


\phantomsection
\addcontentsline{toc}{section}{References}
\bibliography{refs}
\bibliographystyle{bibStyle}

\end{document}